\documentclass[letter,11pt]{article}
\pdfoutput=1

\usepackage{jheppub}
\usepackage{graphicx}
\usepackage{amsmath}
\usepackage{bm}
\usepackage{cancel}
\usepackage{booktabs}
\usepackage[small]{subfigure}
\usepackage{url}
\usepackage[utf8]{inputenc}
\graphicspath{{./plots/}}
\usepackage{xcolor, colortbl}
\usepackage{placeins}
\usepackage{comment}
\usepackage{hhline}

\newcommand{\eq}[1]{eq.~\eqref{eq:#1}}
\newcommand{\eqs}[2]{eqs.~\eqref{eq:#1} and \eqref{eq:#2}}
\renewcommand{\sec}[1]{section~\ref{sec:#1}}

\newcommand{\app}[1]{Appendix~\ref{app:#1}}
\newcommand{\fig}[1]{figure~\ref{fig:#1}}
\newcommand{\figs}[2]{figures~\ref{fig:#1} and \ref{fig:#2}}
\newcommand{\mycite}[1]{ref.~\cite{#1}}
\newcommand{\mycites}[1]{refs.~\cite{#1}}
\newcommand{\tab}[1]{table~\ref{tab:#1}}
\newcommand{\tabs}[2]{tables~\ref{tab:#1} and \ref{tab:#2}}

\newcommand{\as}{\alpha_s}
\newcommand{\mt}{m_t}
\newcommand{\mtt}{m_{t\bar{t}}}
\newcommand{\mttu}{m_{t\bar{t}}^u}
\newcommand{\yt}{y_t}
\newcommand{\ytt}{y_{t\bar{t}}}
\newcommand{\pT}{p^t_T}
\newcommand{\chisqmin}{\chi^2_{\rm min}}
\definecolor{Gray}{gray}{0.8}

\title{Simultaneous extraction of $\as$ and $\mt$ from LHC $t\bar{t}$ differential distributions}

\author[a]{Amanda M.~Cooper-Sarkar,}
\author[b]{Michal Czakon,}
\author[c]{Matthew A.~Lim,}
\author[d]{Alexander Mitov}
\author[d,e]{and Andrew S.~Papanastasiou}
\affiliation[a]{Department of Physics, Oxford University, Keble Road, Oxford, OX1 3RH, UK}
\affiliation[b]{Institut  f{\" u}r  Theoretische  Teilchenphysik  und  Kosmologie, RWTH  Aachen  University,  D-52056  Aachen,  Germany}
\affiliation[c]{Universit{\` a} degli Studi di Milano-Bicocca, Piazza della Scienza 3, 20126, Milan, Italy}
\affiliation[d]{Cavendish Laboratory, University of Cambridge, J.J. Thomson Avenue, CB3 0HE, Cambridge, UK}
\affiliation[e]{MRC Institute of Genetics and Molecular Medicine, University of Edinburgh, Crewe Road, Edinburgh EH4 2XU, UK}

\emailAdd{Amanda.Cooper-Sarkar@physics.ox.ac.uk}
\emailAdd{mczakon@physik.rwth-aachen.de}
\emailAdd{matthew.lim@unimib.it}
\emailAdd{adm74@cam.ac.uk}
\emailAdd{andrewp@hep.phy.cam.ac.uk}

\abstract{
  We present a joint extraction of the strong coupling $\alpha_s$ and the top-quark pole mass $m_t$ from measurements of top-quark pair production
  performed by the ATLAS and CMS experiments at the 8 TeV LHC. For the first time, differential NNLO theory predictions for different
  values of the top-quark mass are utilised for four kinematic distributions: the average transverse momentum of the top-quark, its
  average rapidity and the pair invariant mass and rapidity. The use of \texttt{fastNLO} tables for these distributions allows rapid
  evaluation of the differential theory predictions for different PDF sets. We consider the single differential distributions from the
  experiments both separately and in combination in order to obtain the best fit to theory. Our final values are $\as=0.1159^{+0.0013}_{-0.0014}$ and $\mt=173.8^{+0.8}_{-0.8} ~\mathrm{GeV}$ which are compatible with previous extractions using top-quark measurements. In the case of $\mt$, our value is also compatible with the world average value collated by the Particle Data Group.
}


\preprint{
\begin{flushright}
CAVENDISH-HEP-20/12 \\
TTK-20-31\\
P3H-20-051
\end{flushright}
}

\begin{document}
\maketitle
\flushbottom

\section{Introduction}
\label{sec:intro}

The strong coupling constant $\alpha_s$ and the top quark mass $m_t$ are among the most important parameters of the Standard Model (SM). The former is present in every perturbative calculation as the expansion parameter while the latter plays an important r\^{o}le in governing the stability of the electroweak vacuum~\cite{Degrassi:2012ry,Branchina:2014usa,Bezrukov:2014ina,Markkanen:2018pdo} and enters into calculations of important SM backgrounds to LHC processes such as single and pair top-quark production. Despite its importance, the value of $\alpha_s$ is one of the least well-constrained parameters in the SM and is a dominant uncertainty in many precision calculations. The measured value of the top-quark mass is also accompanied by a large uncertainty and is subject to assumptions made by event generators.

In order to minimise the size of the uncertainties on these parameters, it is vital that extractions use theoretical predictions at the highest available accuracy -- for many important processes, this means including next-to-next-to-leading order (NNLO) QCD corrections. Several extractions of $\as$ have been performed using NNLO theory already. The CMS collaboration has performed an extraction using measurements of the total $t\bar{t}$ cross section at 7 TeV~\cite{Chatrchyan:2013haa} and 13 TeV~\cite{Sirunyan:2018goh}, while ATLAS has taken a similar approach using 13 TeV measurements in ref.~\cite{Aad:2019hzw}. In ref.~\cite{Klijnsma:2017eqp}, the authors considered a variety of measurements of $\sigma_{t\bar{t}}$ at the LHC and the Tevatron compared to the NNLO prediction in order to obtain their value. NNLO calculations of deep inelastic scattering have also been compared to HERA data to extract $\as$ in refs.~\cite{Andreev:2017vxu,Britzger:2019kkb}, while CMS have used measurements of the inclusive Drell-Yan process in refs.~\cite{dEnterria:2019aat,Sirunyan:2019crt}.

Measurements of $\mt$, on the other hand, may proceed either by direct reconstruction of the top-quark decay products or by examination of the total cross section or kinematic distributions for processes featuring top-quarks in the final state. Differential/total cross section extractions require NNLO calculations with $m_t$ dependence which are readily available for the total inclusive cross section~\cite{Czakon:2011xx} but are not yet accessible for differential distributions. For this reason, in the past only the NNLO cross section has been utilised. One of the goals of this paper is to produce differential predictions with sufficiently flexible $m_t$ dependence.

Extractions using measurements of $\sigma_{t\bar{t}}$ have been performed in refs.~\cite{Sirunyan:2018goh,Aad:2014kva,Khachatryan:2016mqs}. In addition, NLO extractions using measurements of the invariant mass distribution $\mtt$ by CDF, D0, ATLAS and CMS appear in refs.~\cite{Tevatron:2014cka,Aaboud:2018zbu,Sirunyan:2018mlv}. An approach which does use NNLO differential information has been presented in ref.~\cite{Yuan:2020nzd}, where single top-quark production has been studied.

In this work, we perform a simultaneous extraction of the parameters $\as$ and $\mt$ using measurements of $t\bar{t}$ production from ATLAS and CMS.
Such an extraction cannot be performed via knowledge of the total inclusive cross section alone, since an increase in the value of
$\as$ can be compensated by an increase in the value of $\mt$. It is therefore impossible to meaningfully constrain both
parameters simultaneously. However, the non-trivial dependence on $\as$ and $\mt$ of the shapes of kinematic distributions can provide the additional information required to make this feasible. In ref.~\cite{Sirunyan:2019zvx}, the CMS collaboration exploit a triple differential $t\bar{t}$ distribution measured at 13 TeV to simultaneously fit $\as$ and $\mt$ by comparing to NLO predictions. A simultaneous fit of $\as$, $\mt$ and PDFs is also performed. We take a similar approach here, exploiting measurements of single differential distributions by ATLAS and CMS at 8 TeV but for the first time performing the extraction using differential NNLO QCD predictions. In this way we obtain full information about possible correlations between the parameters.

The paper is arranged as follows. In \sec{diff-ttbar} we detail the experimental measurements and theoretical calculations which we have used as input to this analysis. In \sec{methodology}, we describe how we extract the parameters from the measured data, considering both single differential distributions and combinations thereof. We present our results in \sec{results} and discuss them in detail in \sec{discussion}. Finally, we present some general conclusions and opportunities for future improvement in \sec{conclusions}.

\section{Differential $t\bar{t}$ production at 8 TeV LHC}\label{sec:diff-ttbar}

\subsection{Experimental measurements}

In this work we consider ATLAS and CMS $t\bar{t}$ measurements at 8 TeV of the total cross section~\cite{Aad:2014kva,Khachatryan:2016mqs} as well as of kinematic distributions~\cite{Aad:2015mbv,Khachatryan:2015oqa}. The differential measurements we exploit are those of the invariant mass and rapidity of the top-quark pair system, $\mtt$ and $\ytt$, and the average transverse momentum and average rapidity of the top or antitop-quark, $p^t_T$ and $\yt$. ATLAS and CMS used a set of common bin edges, catalogued in \tab{binnings}, for their measurements of these distributions. The experimental results include both statistical and systematic uncertainties and they enter our definition of the $\chi^2$ objective (see \sec{methodology}). We make use of the covariance matrices provided by ATLAS and CMS~\cite{Aad:2015mbv,Khachatryan:2015oqa}, as per the approach taken in ref.~\cite{Czakon:2016olj}.

\begin{table}[h!]
\centering
\begin{tabular}{ |c|c| }
\hline
Observable & Bin edges \\
\hline
$\pT$  & \{0, 60, 100, 150, 200, 260, 320, 400, 500\}~GeV \\
$\mtt$ & \{345, 400, 470, 550, 650, 800, 1100, 1600\}~GeV \\
$\yt$  & \{-2.5, -1.6, -1.2, -0.8, -0.4, 0.0, 0.4, 0.8, 1.2, 1.6, 2.5\} \\
$\ytt$ & \{-2.5, -1.3, -0.9, -0.6, -0.3, 0.0, 0.3, 0.6, 0.9, 1.3, 2.5\} \\
\hline
\end{tabular}
\caption{ATLAS and CMS common bin edges for measurements of $\pT$, $\mtt$, $\yt$ and $\ytt$ at 8 TeV.}
\label{tab:binnings}
\end{table}

\subsection{Theoretical predictions}

To quantitatively compare against experimental measurements and extract parameters, theoretical predictions of the measured quantities are required at as high precision as possible.
For the total $t\bar{t}$ cross section we use $\sigma^{\rm theory}(\as,\mt) = \sigma^{\rm NNLO+NNLL}_{t\bar{t}}(\as,\mt)$, i.e. the NNLO-QCD fixed-order prediction supplemented with soft-gluon resummation~\cite{Cacciari:2011hy,Baernreuther:2012ws,Czakon:2012zr,Czakon:2012pz,Czakon:2013goa}. The fast computation of this is provided through \texttt{top++}~\cite{Czakon:2011xx}. For the total cross section we set the renormalisation and factorisation scales to the common scale $\mu=\mu_R=\mu_F=\mt$. The differential cross section for this process is known up to complete NLO including NNLO QCD and resummation effects~\cite{Czakon:2017wor,Czakon:2019txp} -- in this work we use only the NNLO QCD predictions $d\sigma^{\rm theory}(\as,\mt) = d\sigma^{\rm NNLO}_{t\bar{t}}(\as,\mt)$~\cite{Czakon:2015owf,Czakon:2016ckf,Czakon:2016dgf} computed using the \texttt{Stripper} framework~\cite{Czakon:2010td,Czakon:2014oma,Czakon:2019tmo}.

Throughout this work we define $m_t$ in the pole mass scheme. NNLO differential predictions for top-quark pair production in the $\overline{\mathrm{MS}}$ mass scheme have recently been made available in ref.~\cite{Catani:2020tko}, following the earlier implementation of the process at NNLO in the $q_T$ subtraction framework~\cite{Catani:2019hip}. The renormalisation and factorisation scales chosen are those found to be optimal in ref.~\cite{Czakon:2016dgf}, namely,
\begin{align}
\mu_R&=\mu_F=H_T/4, \;\; \text{for} \;\; \mtt,\, \yt, \, \ytt \,, \\
\mu_R&=\mu_F=M_T/2, \;\; \text{for} \;\; \pT \,.
\end{align}
Normalised distributions were obtained from the corresponding absolute distributions by dividing the weight of each bin by the sum of weights in all bins.

Finding the best-fit values for $\as$ and $\mt$ requires theoretical predictions made with many input values of these fundamental parameters. Furthermore, since the value of $\as$ is typically taken from the associated input PDF set used to evaluate cross sections, these predictions must be made with multiple PDF sets. To allow for the fast computation of the relevant differential quantities, we have made
use of the \texttt{fastNLO} interface~\cite{Kluge:2006xs,Britzger:2012bs,Czakon:2017dip}, producing tables that allow, for a fixed set of observables and binnings and a fixed input top-quark mass, $\mathcal{O}(1\;{\rm second})$ evaluations of the NNLO differential cross section with any desired PDF set. We have produced \texttt{fastNLO} tables for the distributions and binnings given in \tab{binnings} and for the following values of the top-quark mass:
\begin{align}
\mt=\{169.0,171.0,172.5,173.3,175.0\} \;\text{GeV}.
\end{align}
The \texttt{fastNLO} tables for these values of $\mt$ are available for download from ref.~\cite{website}. Predictions for different input $\as$ values are obtained by convolving the \texttt{fastNLO} tables with PDF sets of the same family which have been fit using different values of $\as$.

\subsubsection{Choice of input PDFs}

In order to make an extraction of $\as$ we require a PDF family to have PDF sets that have been fit with different values of $\as$. To construct a robust fit of the theoretical predictions (see \sec{interpolation}) it is preferable that the PDF family contains at least three values of $\as$. The PDFs considered in this study (all NNLO fits) are the CT14~\cite{Dulat:2015mca}, NNPDF3.0~\cite{Ball:2014uwa} and NNPDF3.1~\cite{Ball:2017nwa} sets, which come with 13, 5 and 11 values of $\as$ respectively.

Of these sets, only CT14 was fit without top-quark data. NNPDF3.0 has used measurements of the total top-pair cross section at the LHC, whilst NNPDF3.1 additionally includes differential measurements, namely the ATLAS $\yt$ and CMS $\ytt$ distributions at 8 TeV.

One might be concerned that the presence of top-quark data in the PDF may bias the extraction of $\mt$. In order to estimate the impact this may have we have considered various PDF sets containing varying degrees of top-quark data. As we shall see, it appears that at least at present this is not a very large effect in most cases. We shall, however, comment in some detail on the inclusion of differential top-quark data in NNPDF3.1 in \sec{discussion}. Even if no top-quark data were included in the PDFs at all, one should recall that certain bias is still present in the experimental data where modelling with Monte Carlo codes introduces some $\mt$ dependence. A remedy to this problem may be to produce PDF sets with various values of $\mt$, treating it in a manner similar to $\as$, but such sets are not available at present. Similarly, there are subtleties in the extraction of $\as$ from PDF sets with fixed values of $\as$. This has been discussed in refs.~\cite{Ball:2018iqk,Forte:2020pyp}.

We have also considered PDF sets from the MMHT family~\cite{Harland-Lang:2014zoa}. However, since the sets do not pass our internal consistency checks when fitting theory predictions (discussed in the following section), we do not present the corresponding extractions.

\subsubsection{Two-dimensional interpolation in $\as$ and $\mt$} \label{sec:interpolation}

To aid the parameter extraction via minimisation of a $\chi^2$ objective function, for each PDF family (CT14, NNPDF3.0, NNPDF3.1) it is convenient to move from a discrete grid of predictions in $\as$ and $\mt$ to a smooth two-dimensional surface. This is achieved by fitting functions to the available sets of $(\as,\mt)$ points, both for the total cross section as well as for the weight of each bin of every distribution. In the one-dimensional case we fit functions in $\as$ or in $\mt$, while in the two-dimensional case the fit takes a factorised form as the product of a function of $\as$ with a function of $\mt$.

The total cross section is fit as a function of $\alpha_s$ using polynomials of order 1, 2 and 3 for the NNPDF3.0, CT14 and NNPDF3.1 sets respectively. We use the functional form of mass dependence recommended in ref.~\cite{Czakon:2013goa}
\begin{align}
\sigma^{\rm theory}_i(m)=\sigma^{\rm theory}_i(m_{\mathrm{ref}})
\left( \frac{m_{\mathrm{ref}}}{m} \right)^4  \left(1+a_1\frac{(m-m_{\mathrm{ref}})}{m_{\mathrm{ref}}}+a_2\frac{(m-m_{\mathrm{ref}})^2}{m_{\mathrm{ref}}^2}\right)
\,,
\end{align}
where we take $m_{\rm ref} = 172.5$~GeV as a reference value for the top-quark mass.

The functional forms of the fits to the differential distributions are chosen based on the number of available points in $\mt$ and $\alpha_s$ (the latter depending on the PDF set used) and based on the quality and stability of the interpolation and extrapolation. To parametrise the $\alpha_s$ dependence of each bin, we choose a polynomial of order 4 for the CT14 and NNPDF3.1 sets and a linear fit for the NNPDF3.0 set.
\footnote{Since only five values of $\alpha_s$ are available for NNPDF3.0, we impose a linear dependence to avoid overfitting. In this way, we can fit the predictions to within $3\sigma$ for the rapidity distributions and to within $2\sigma$ for the $\pT$ and $\mtt$ distributions, where $\sigma$ refers to the MC error of the NNLO calculation. Note that the NNLO MC error is negligible compared to the NNLO scale uncertainty.}
For the mass dependences of the distributions, we use the forms
\begin{align}
w_{\mtt}^{i}(m)&=w_{\mtt}^{i}(m_{\mathrm{ref}}) \left( \frac{m_{\mathrm{ref}}}{m} \right)^4
     \left(1+a_1\frac{(m-m_{\mathrm{ref}})}{m_{\mathrm{ref}}}+a_2\frac{(m-m_{\mathrm{ref}})^2}{m_{\mathrm{ref}}^2} \right. \nonumber \\
     & \left. \hspace{7.07cm} +a_3\Big(1-\frac{1}{\cosh^6(m-m_{\mathrm{ref}})}\Big) \right)\,, \\
w_{\pT}^{i}(m)&=w_{\pT}^{i}(m_{\mathrm{ref}}) \left( \frac{m_{\mathrm{ref}}}{m} \right)^4  \left(1+a_1\frac{(m-m_{\mathrm{ref}})}{m_{\mathrm{ref}}}+\frac{a_2}{m}\right)\,, \\
w_{y}^{i}(m)&=w_{y}^{i}(m_{\mathrm{ref}})\left(1+a_1 e^{-a_2(m-m_{\mathrm{ref}})}\right)\,,
\end{align}
where $w^i(m)$ is the weight of the $i^{\mathrm{th}}$ bin and we again take a reference mass $m_{\mathrm{ref}}=172.5$~GeV. The same functional form is used for both rapidity distributions $\yt$ and $\ytt$. The coefficients $a_j$ are fit independently for each distribution and PDF set.

\begin{figure}
\includegraphics[trim=0.4cm 0.5cm 1.5cm 1.0cm,clip,width=0.245\textwidth]{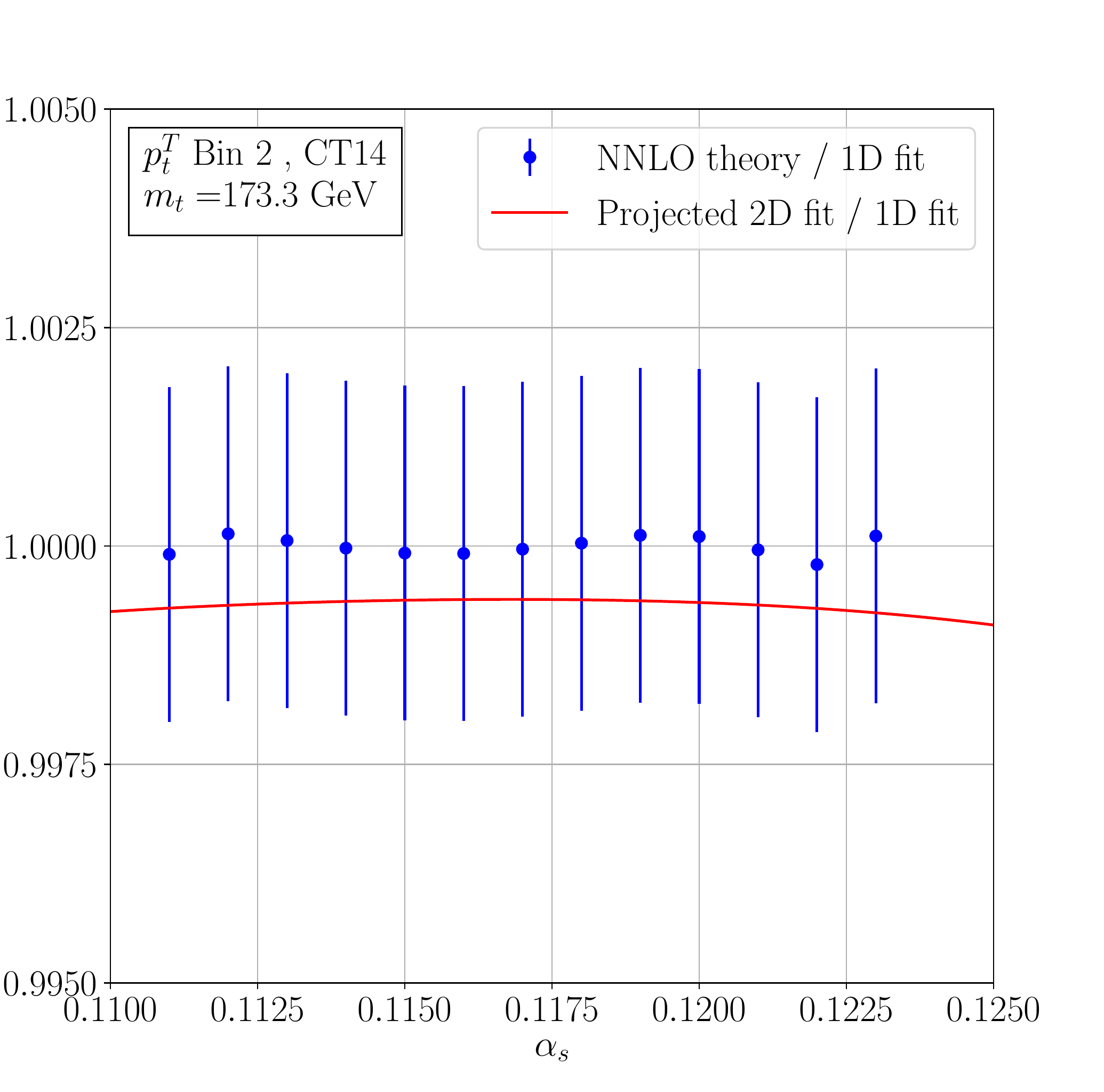}
\includegraphics[trim=0.4cm 0.5cm 1.5cm 1.0cm,clip,width=0.245\textwidth]{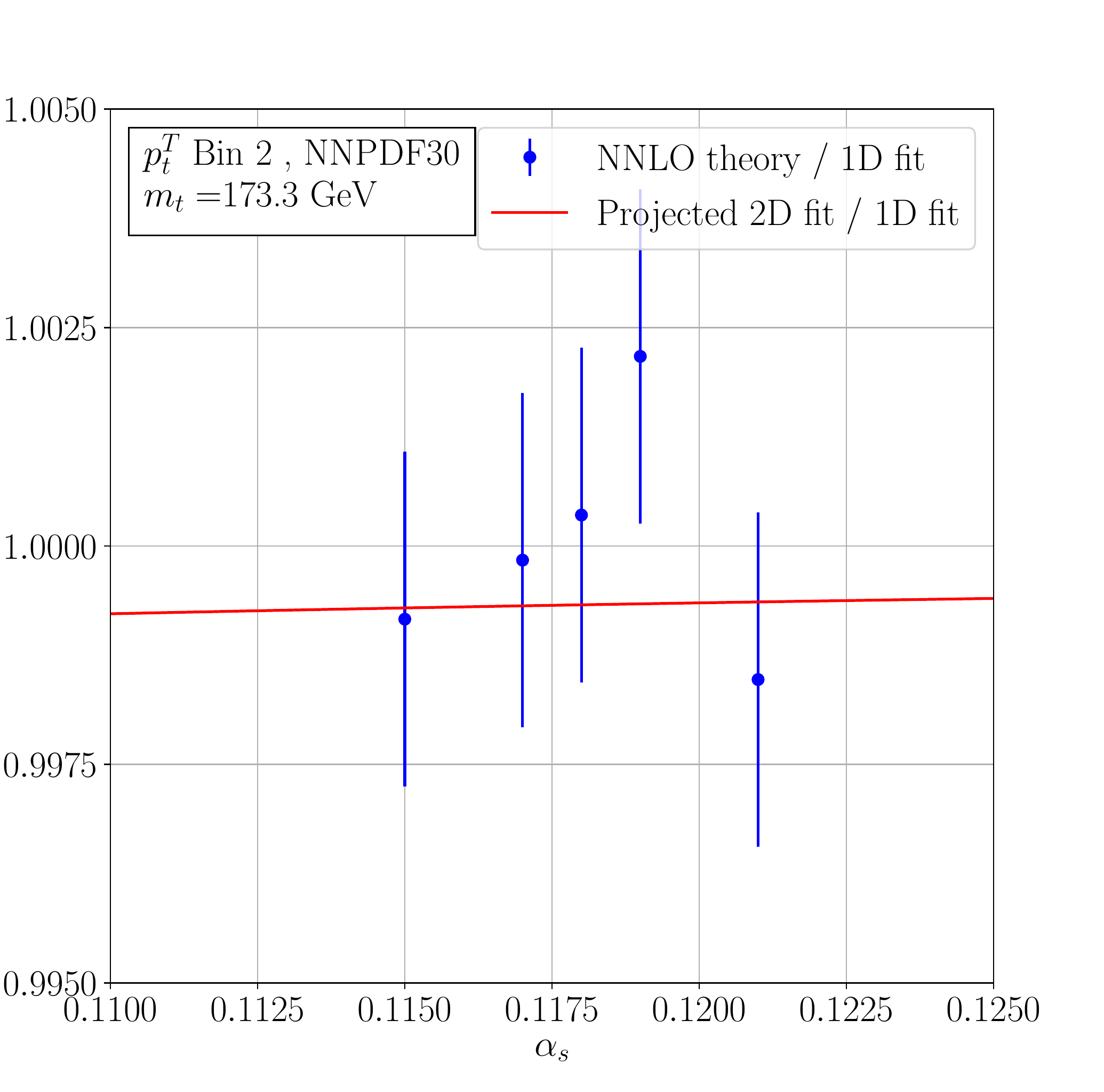}
\includegraphics[trim=0.4cm 0.5cm 1.5cm 1.0cm,clip,width=0.245\textwidth]{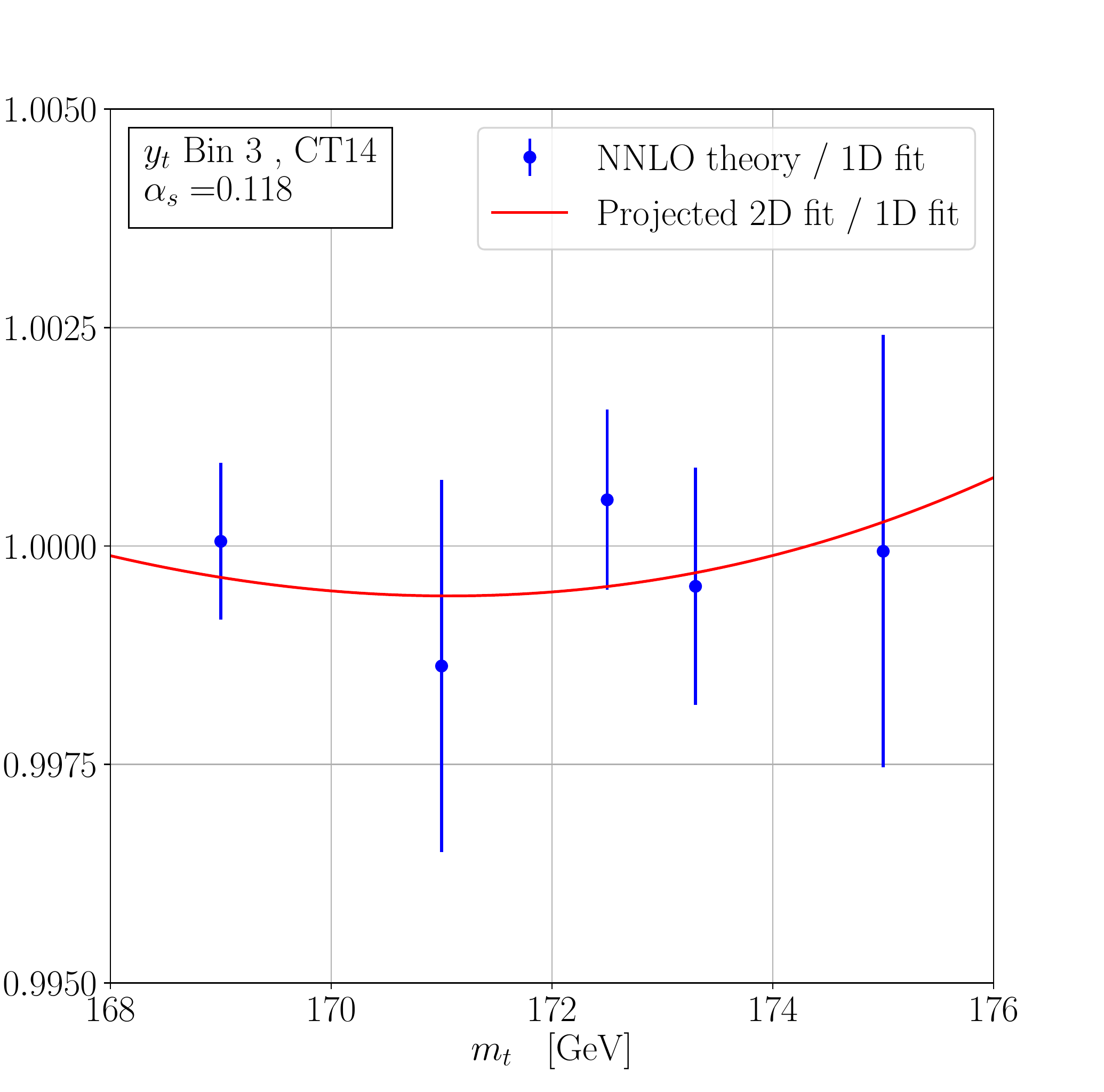}
\includegraphics[trim=0.4cm 0.5cm 1.5cm 1.0cm,clip,width=0.245\textwidth]{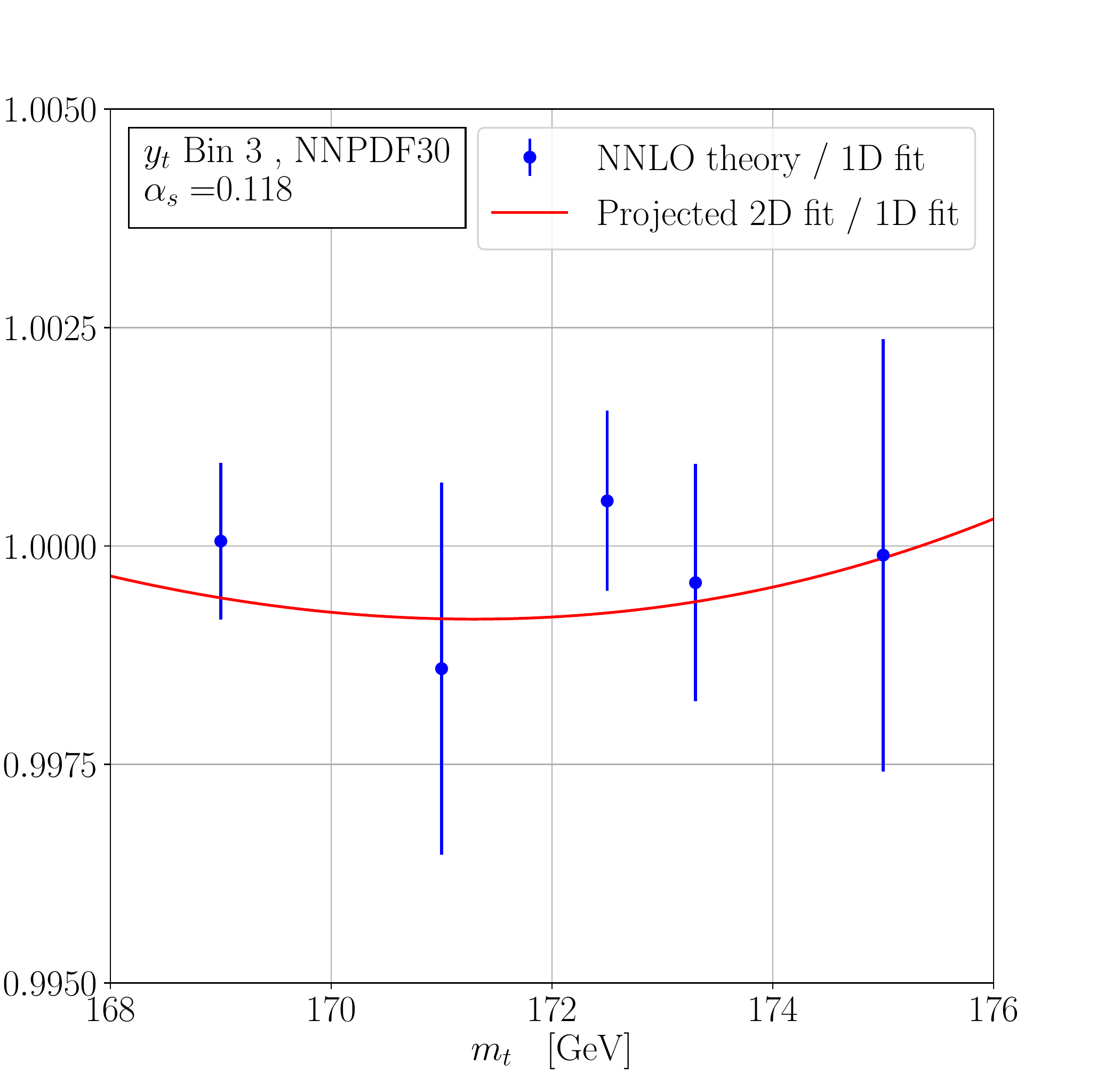}
\caption{Example plots showing the interpolations in $\as$ and $\mt$. The first and second (third and fourth) plots show the interpolations of theoretical predictions in $\as$ ($\mt$) for the second bin of $\pT$ (third bin of $\yt$), using the CT14 or NNPDF3.0 PDF sets. The blue points display the ratio of the computed NNLO predictions to the one-dimensional interpolation. The vertical bars indicate the MC error of the NNLO calculation. The red line is the projection of the corresponding two-dimensional fit in $\as$ or $\mt$.}
\label{fig:interpolation-checks}
\end{figure}

Although for the extractions in \sec{results} we have consistently used the two-dimensional fits (projecting out one dimension as required for the one-dimensional extractions of $\as$ or $\mt$), performing also the one-dimensional fits independently acts as a consistency check on our fitting procedure. In this way we can verify that, by fixing either $\mt$ or $\as$ to some value, the resulting projections of the two-dimensional fits onto one dimension are equivalent to fitting in one dimension to within the Monte Carlo error of the theory predictions. Some examples of the quality of the fits can be seen in \fig{interpolation-checks}. The requirement of good quality fits and extrapolations is one of the aspects which determined which PDF sets we have chosen to use -- our results for the CT14, NNPDF3.0 and NNPDF3.1 sets all show very good agreement, as a function of $\as$ and $\mt$, between the bin weight from the fit and the discrete set of the calculated points to within Monte Carlo error.

\subsubsection{Dependence of differential observables on $\as$ and $\mt$}

\begin{figure}[t]
\centering
\includegraphics[trim=0.7cm 0.2cm 1.5cm 2.0cm,clip,width=0.24\textwidth]{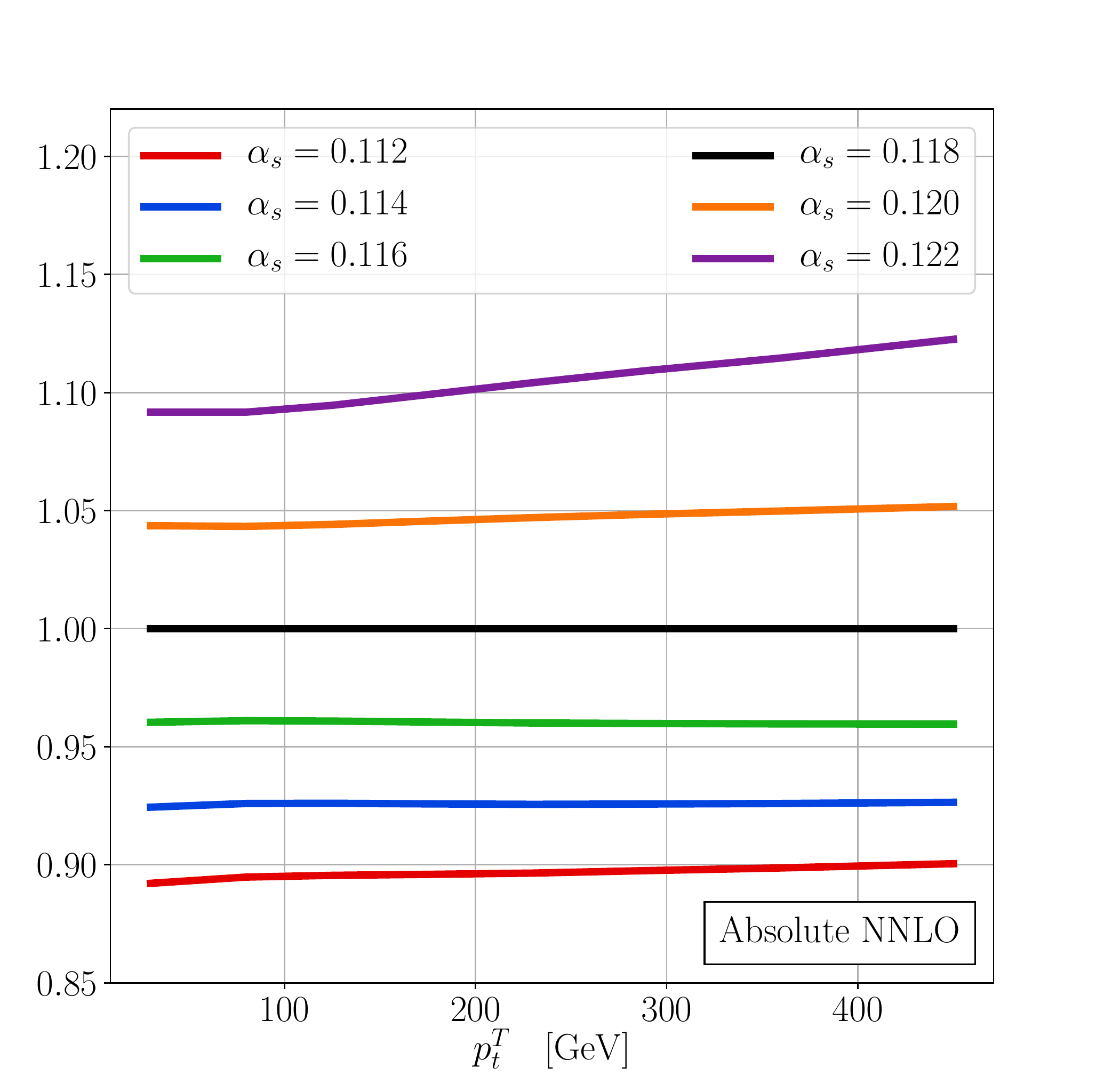}
\includegraphics[trim=0.7cm 0.2cm 1.5cm 1.5cm,clip,width=0.24\textwidth]{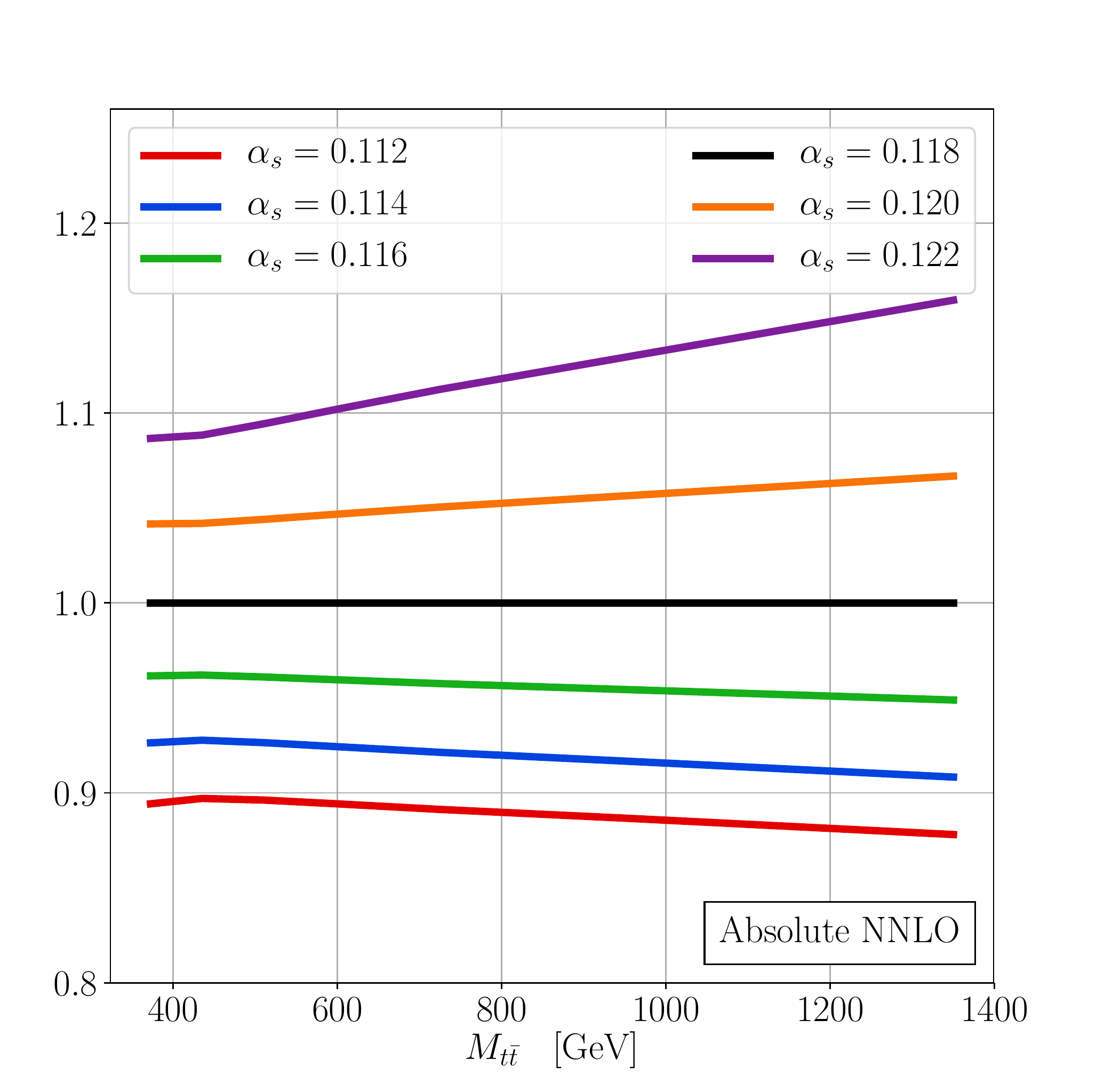}
\includegraphics[trim=0.7cm 0.2cm 1.5cm 1.5cm,clip,width=0.24\textwidth]{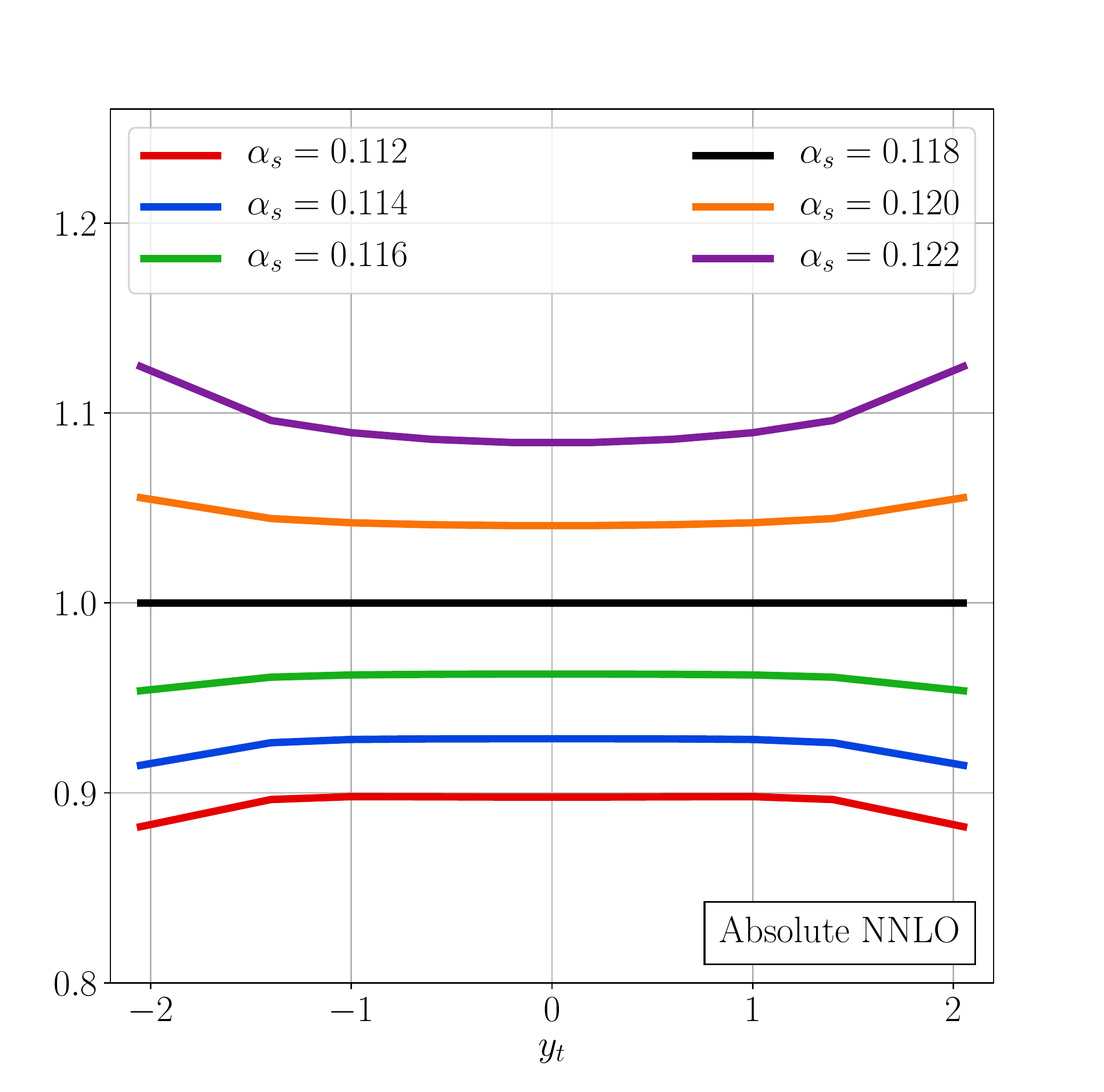}
\includegraphics[trim=0.7cm 0.2cm 1.5cm 1.5cm,clip,width=0.24\textwidth]{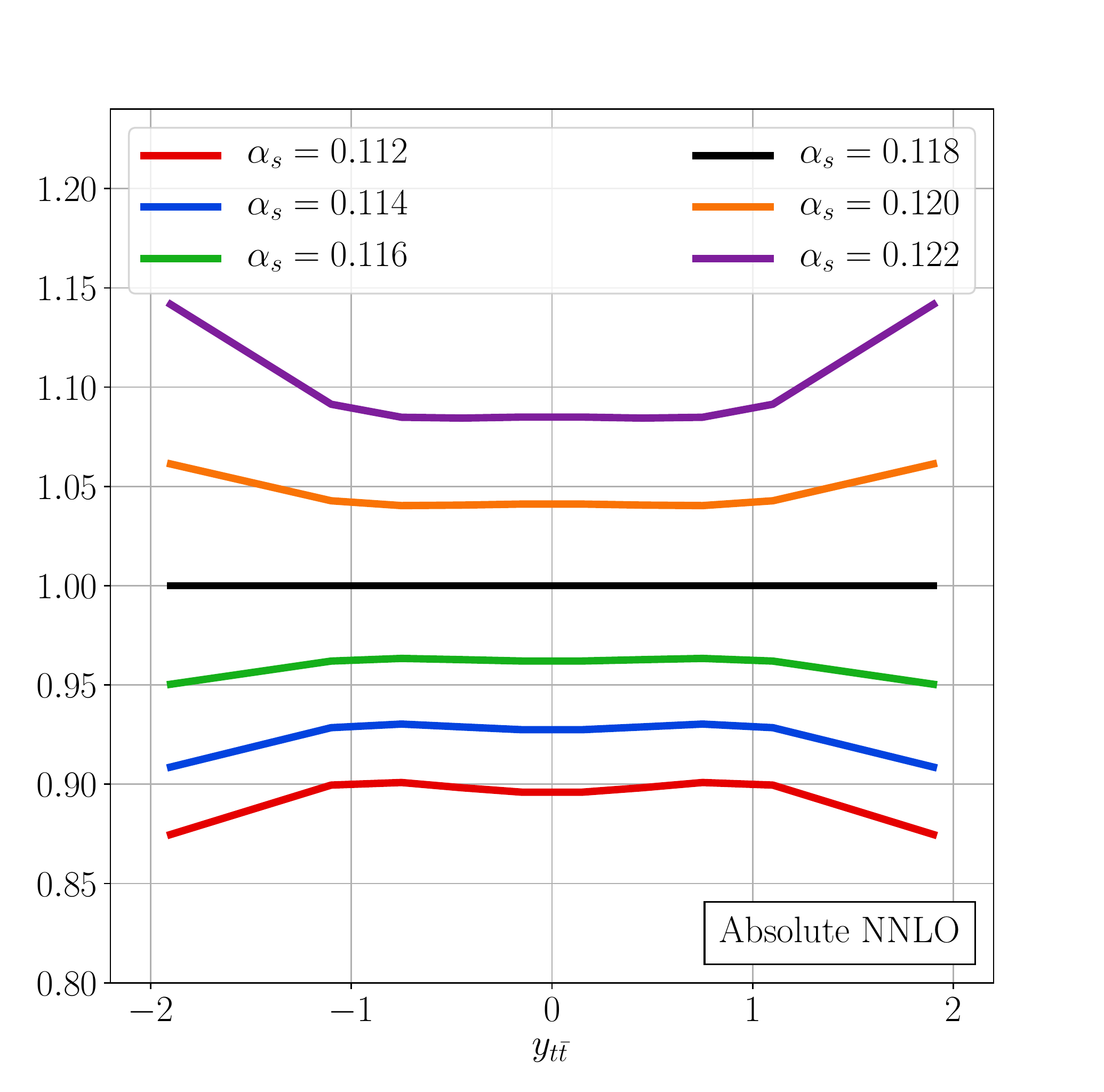}\\
\includegraphics[trim=0.7cm 0.2cm 1.5cm 2.0cm,clip,width=0.24\textwidth]{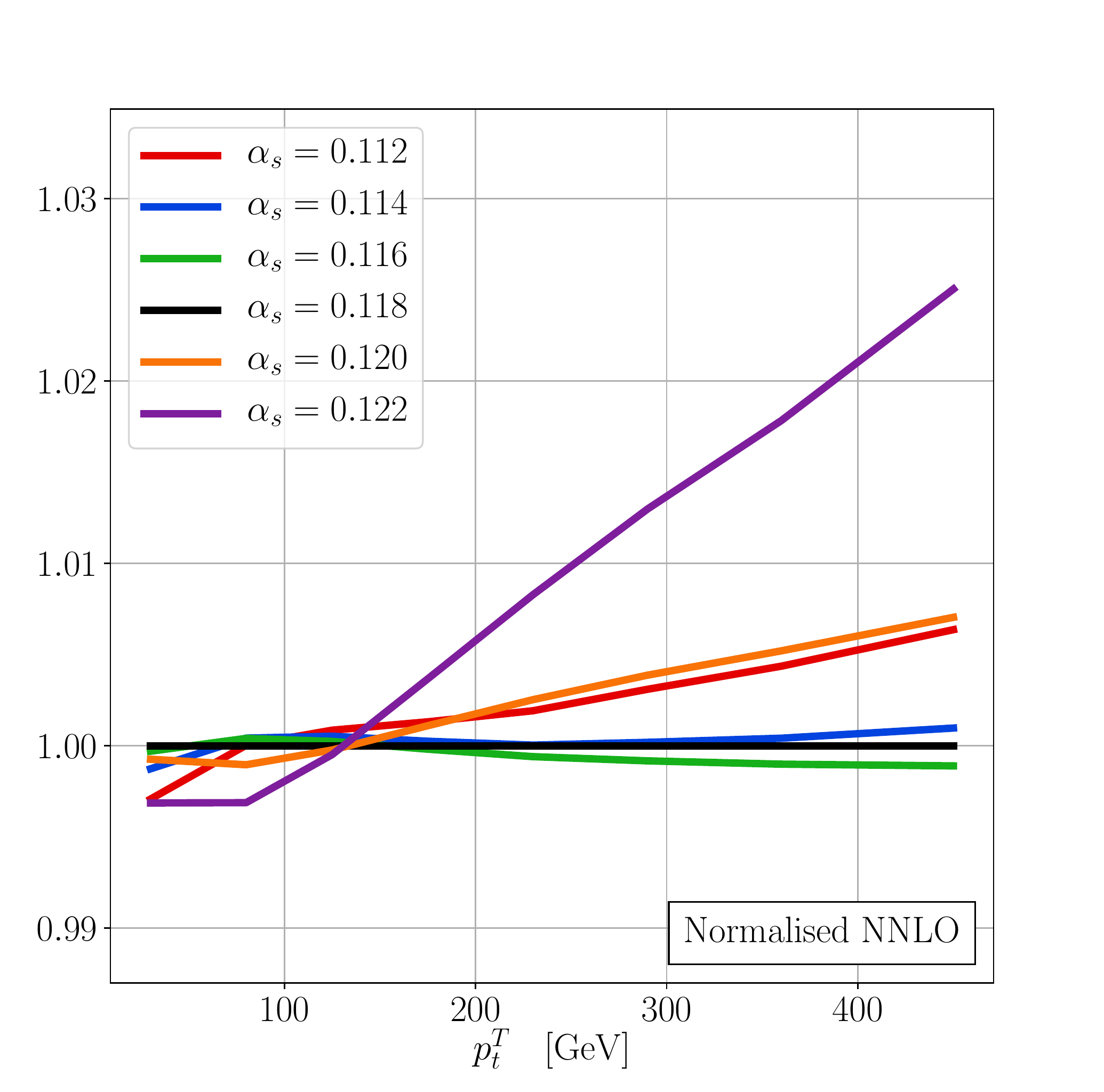}
\includegraphics[trim=0.7cm 0.2cm 1.5cm 1.5cm,clip,width=0.24\textwidth]{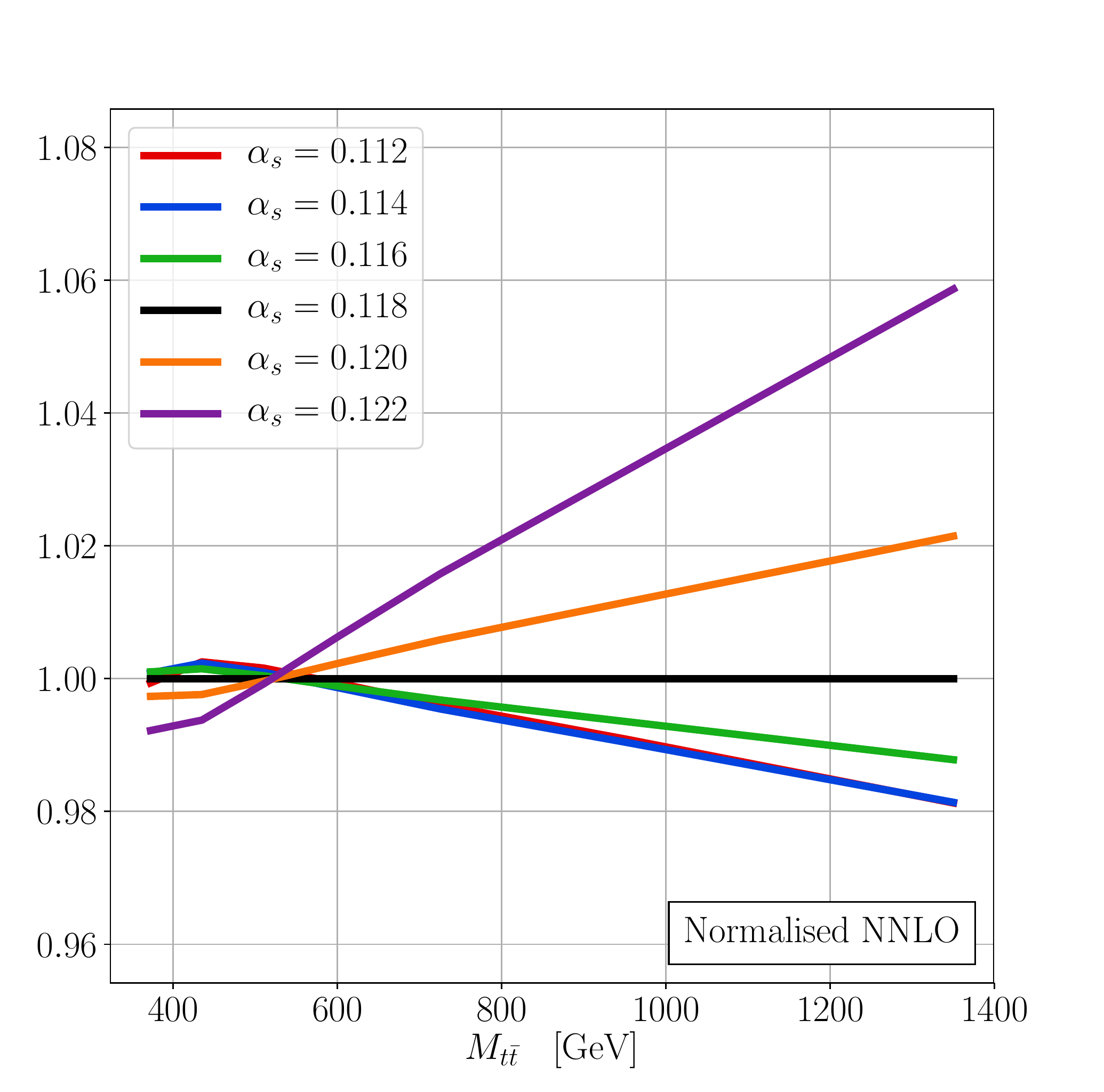}
\includegraphics[trim=0.7cm 0.2cm 1.5cm 1.5cm,clip,width=0.24\textwidth]{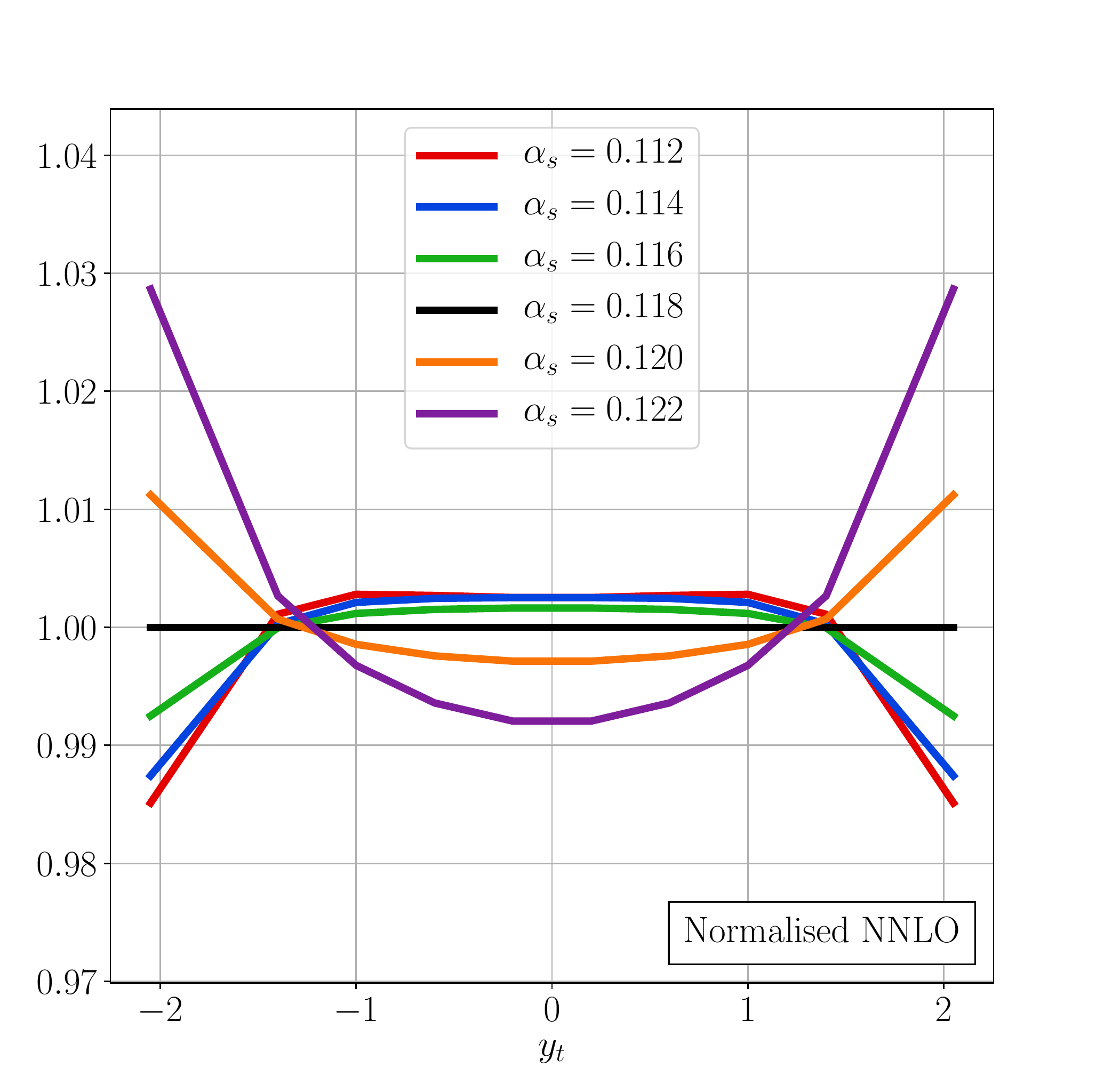}
\includegraphics[trim=0.7cm 0.2cm 1.5cm 1.5cm,clip,width=0.24\textwidth]{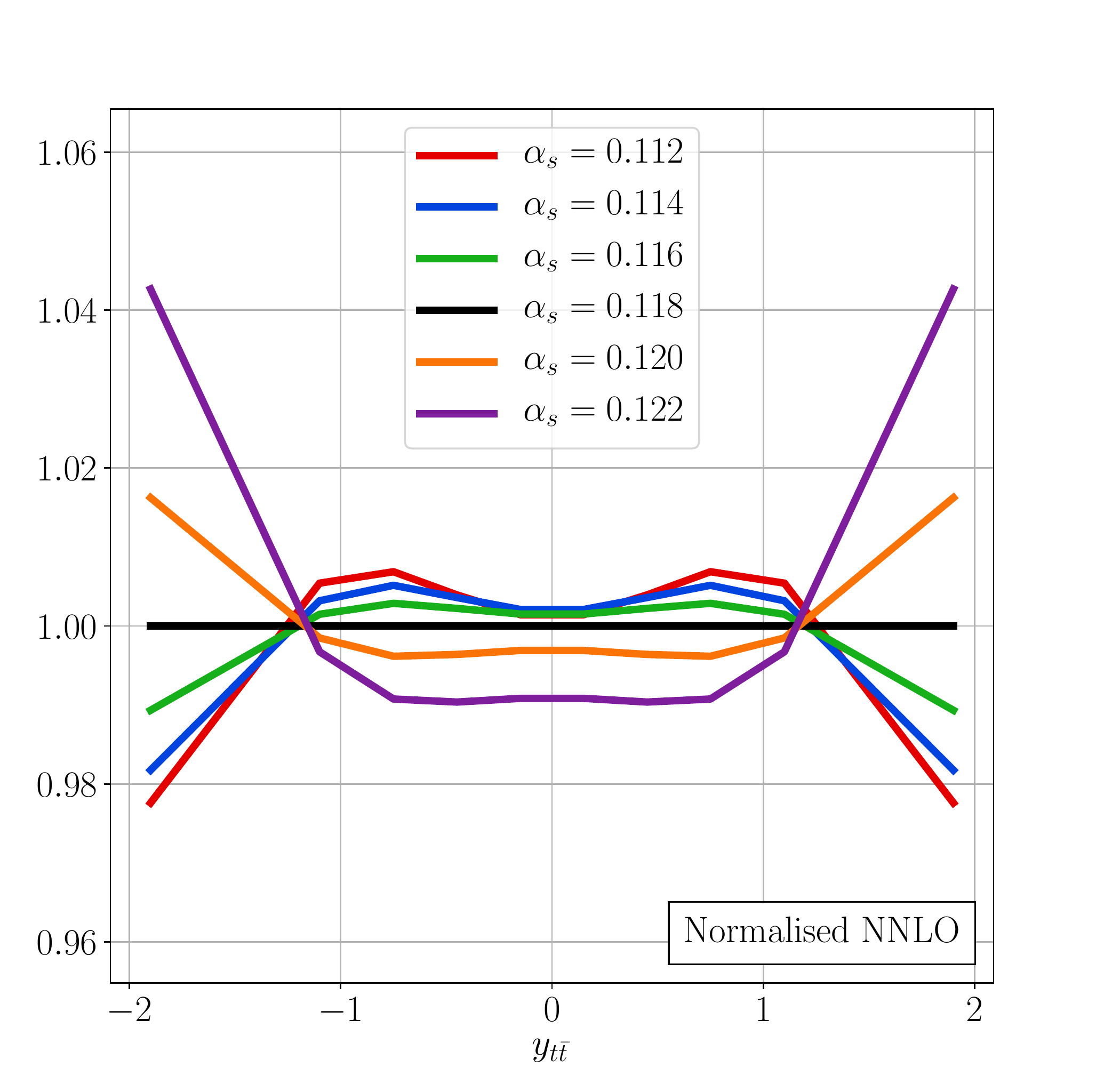}
\caption{The sensitivity on $\as$ of the $\pT$ (column 1), $\mtt$ (column 2), $\yt$ (column 3) and $\ytt$ (column 4) of the absolute (top) and normalised (bottom) distributions in NNLO QCD. All curves are relative to the world average values of $\as=0.118$ and $\mt=173.3$~GeV. The PDF set CT14 has been used to produce these curves. The pattern for other PDF sets is similar.}
\label{fig:as-sensitivity}
\end{figure}

In \figs{as-sensitivity}{mt-sensitivity} we show the sensitivity on $\as$ and $\mt$ of the absolute and normalised distributions respectively, at NNLO QCD accuracy. It is the non-trivial shape of these dependences that will allow us to perform simultaneous extractions of $\as$ and $\mt$. The plots have been produced using the CT14 PDF set -- however, very similar patterns are observed when using other PDF sets, in particular the size of the effects of varying $\as$ and $\mt$. These sensitivities have previously been discussed at NLO in \mycites{Czakon:2016vfr,Czakon:2016olj}.

\begin{figure}[t]
\centering
\includegraphics[trim=0.7cm 0.2cm 1.5cm 2.0cm,clip,width=0.24\textwidth]{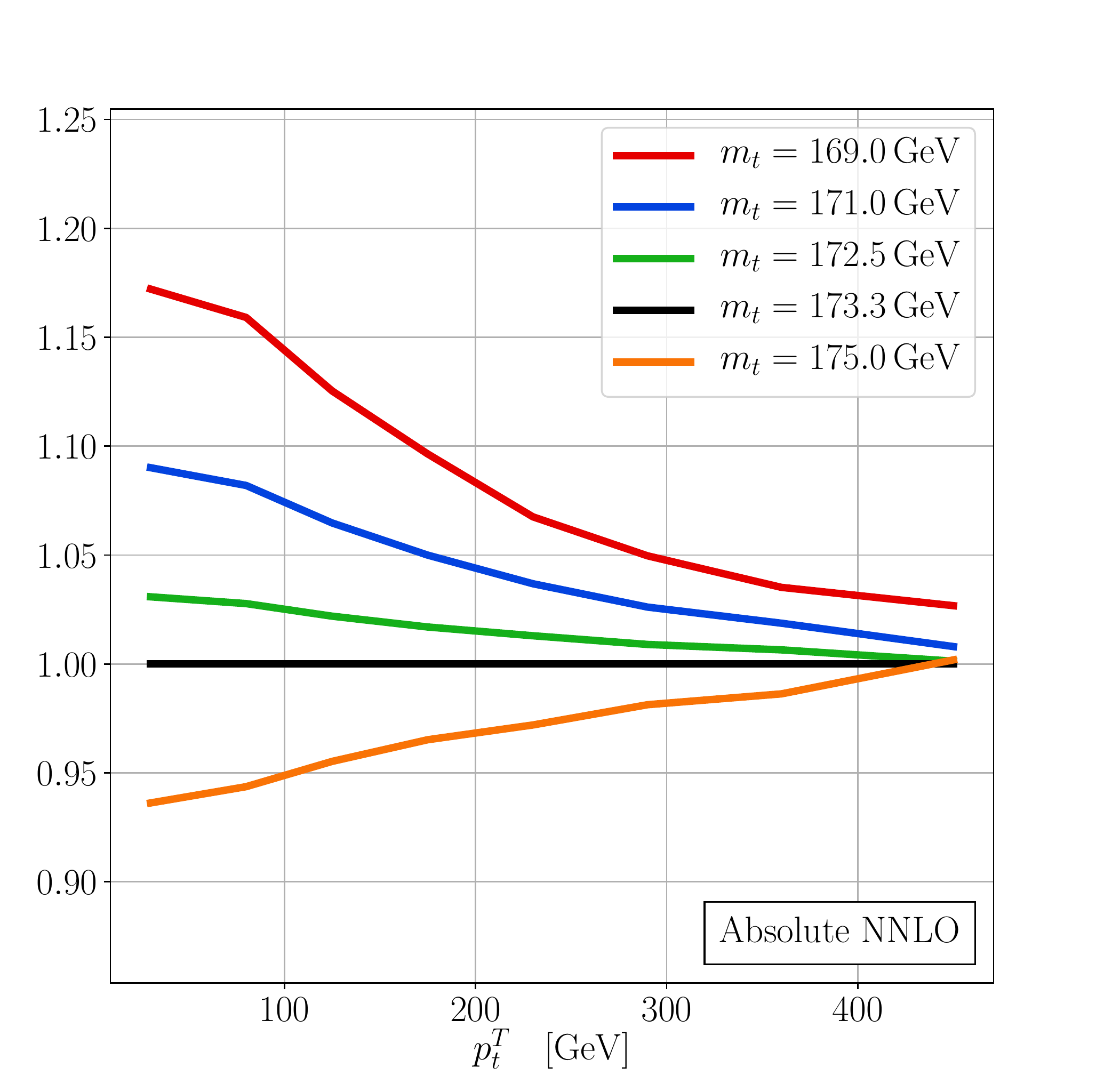}
\includegraphics[trim=0.7cm 0.2cm 1.5cm 1.5cm,clip,width=0.24\textwidth]{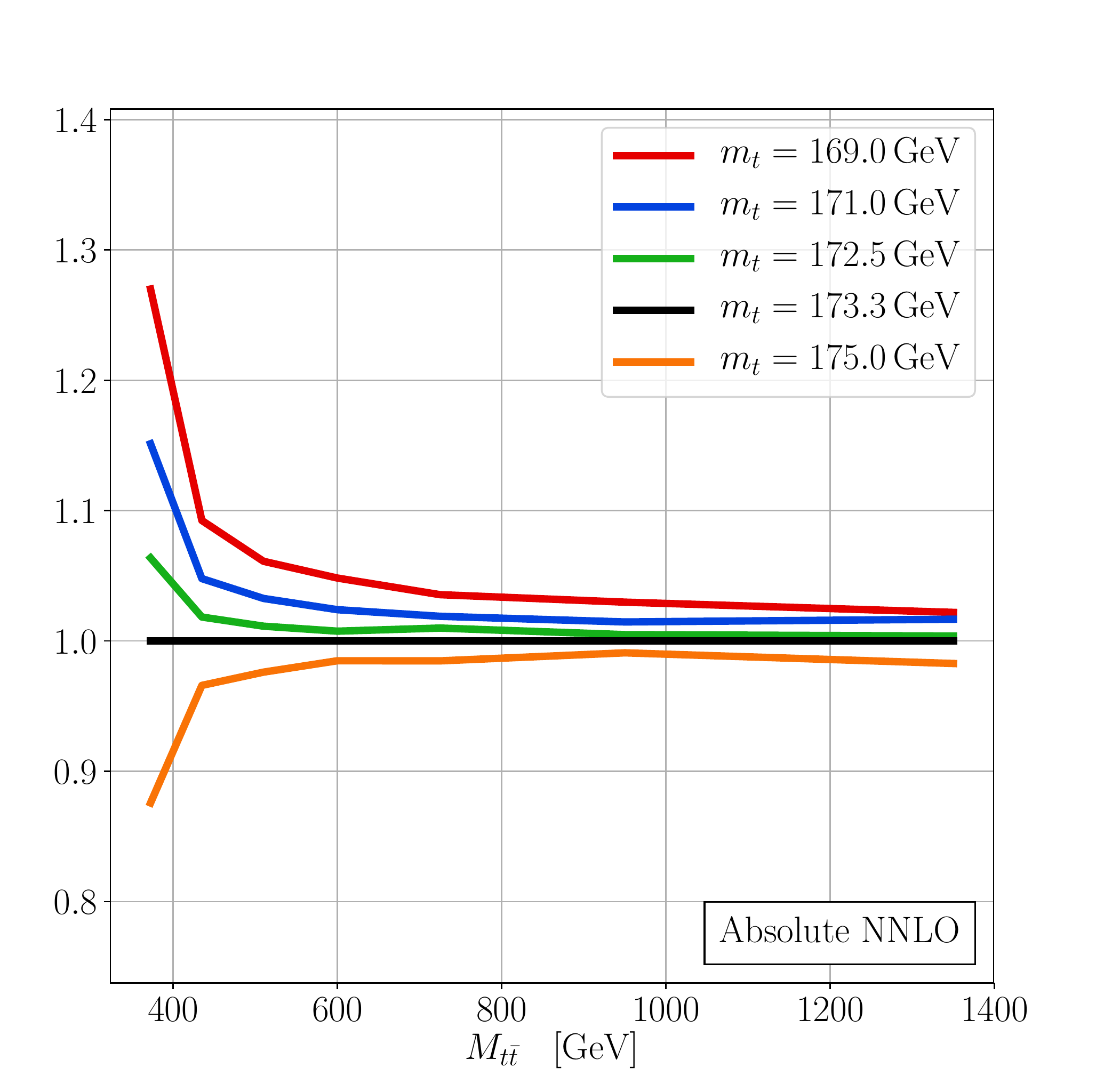}
\includegraphics[trim=0.7cm 0.2cm 1.5cm 1.5cm,clip,width=0.24\textwidth]{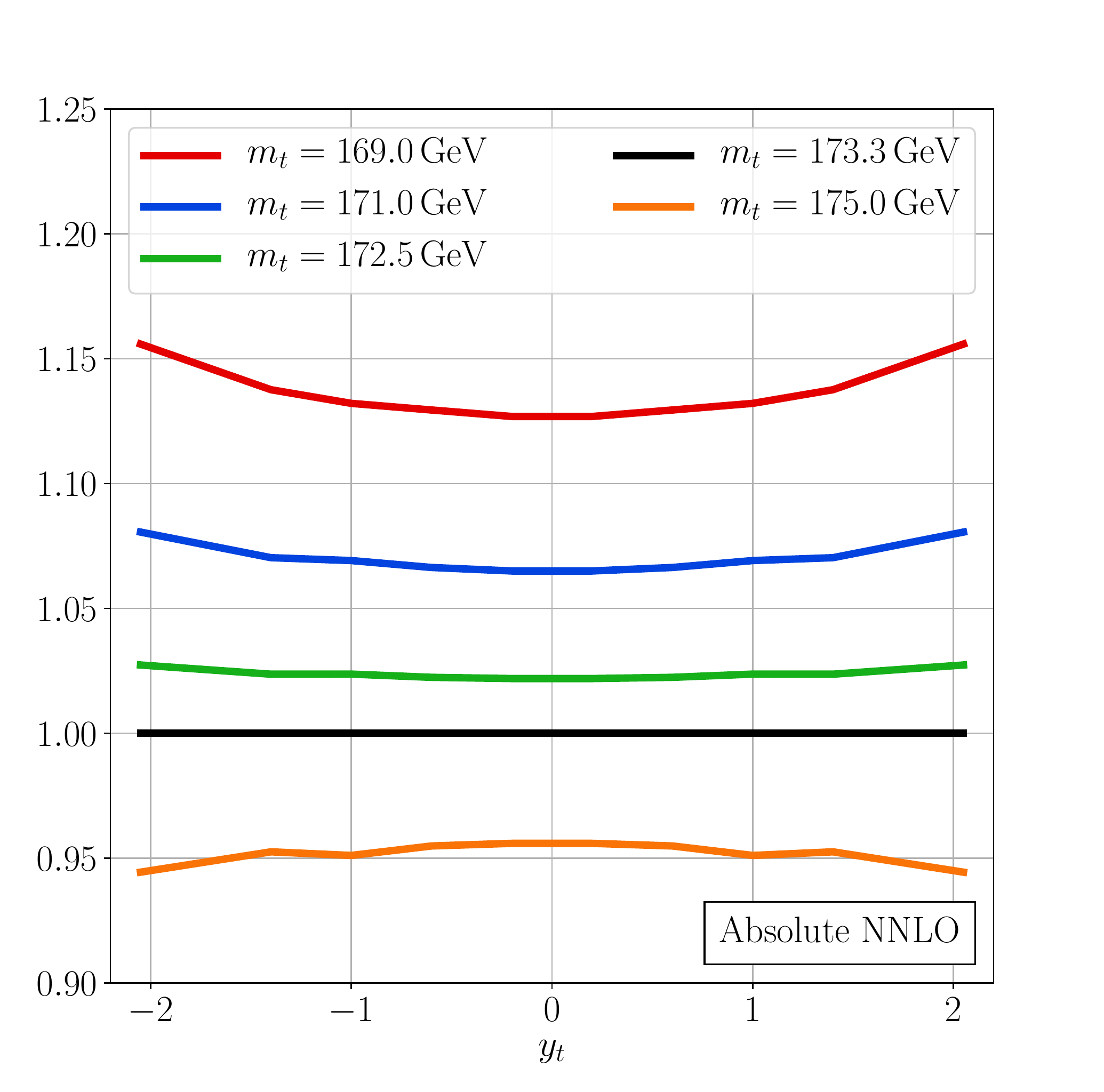}
\includegraphics[trim=0.7cm 0.2cm 1.5cm 1.5cm,clip,width=0.24\textwidth]{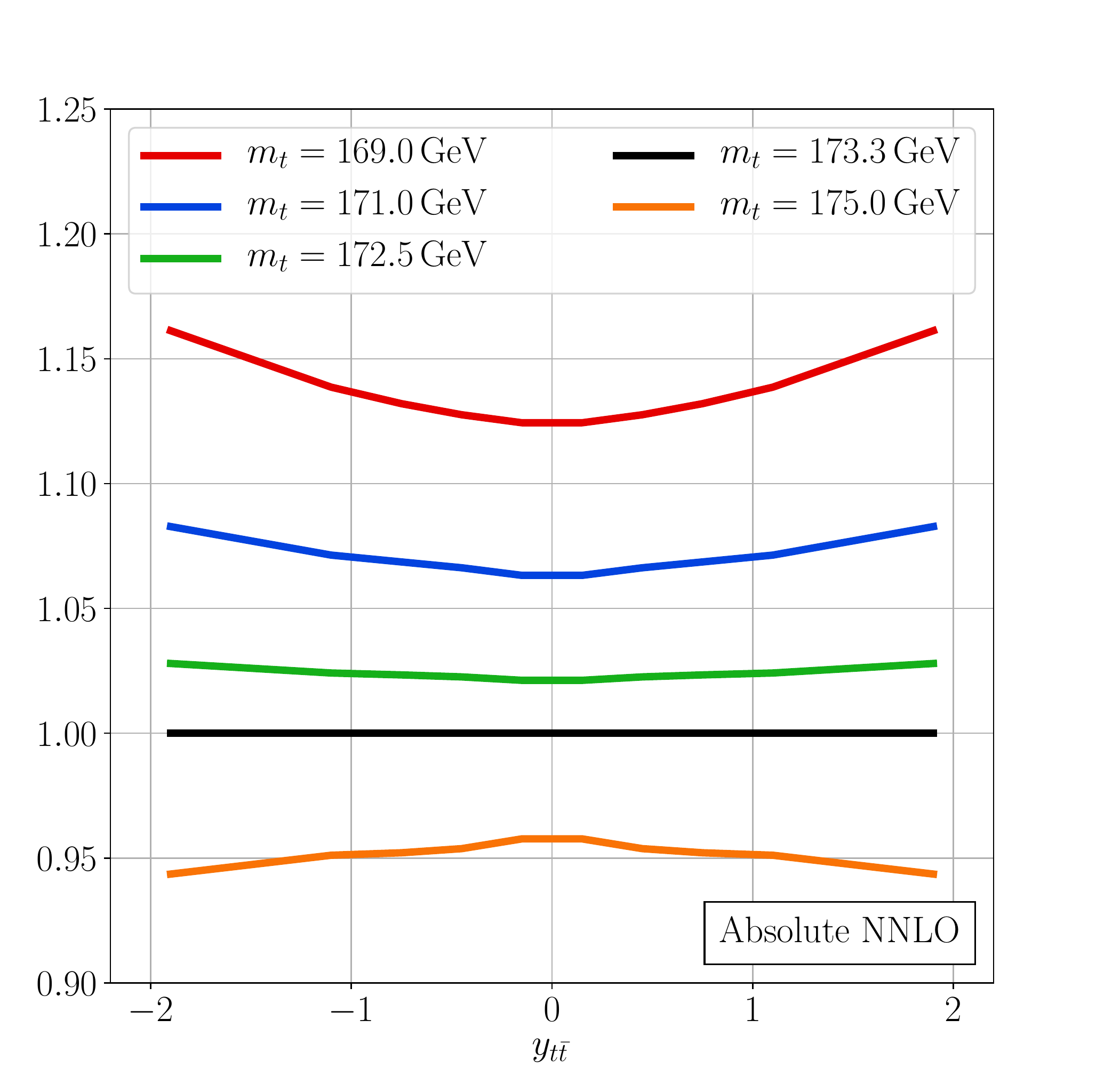}\\
\includegraphics[trim=0.7cm 0.2cm 1.5cm 2.0cm,clip,width=0.24\textwidth]{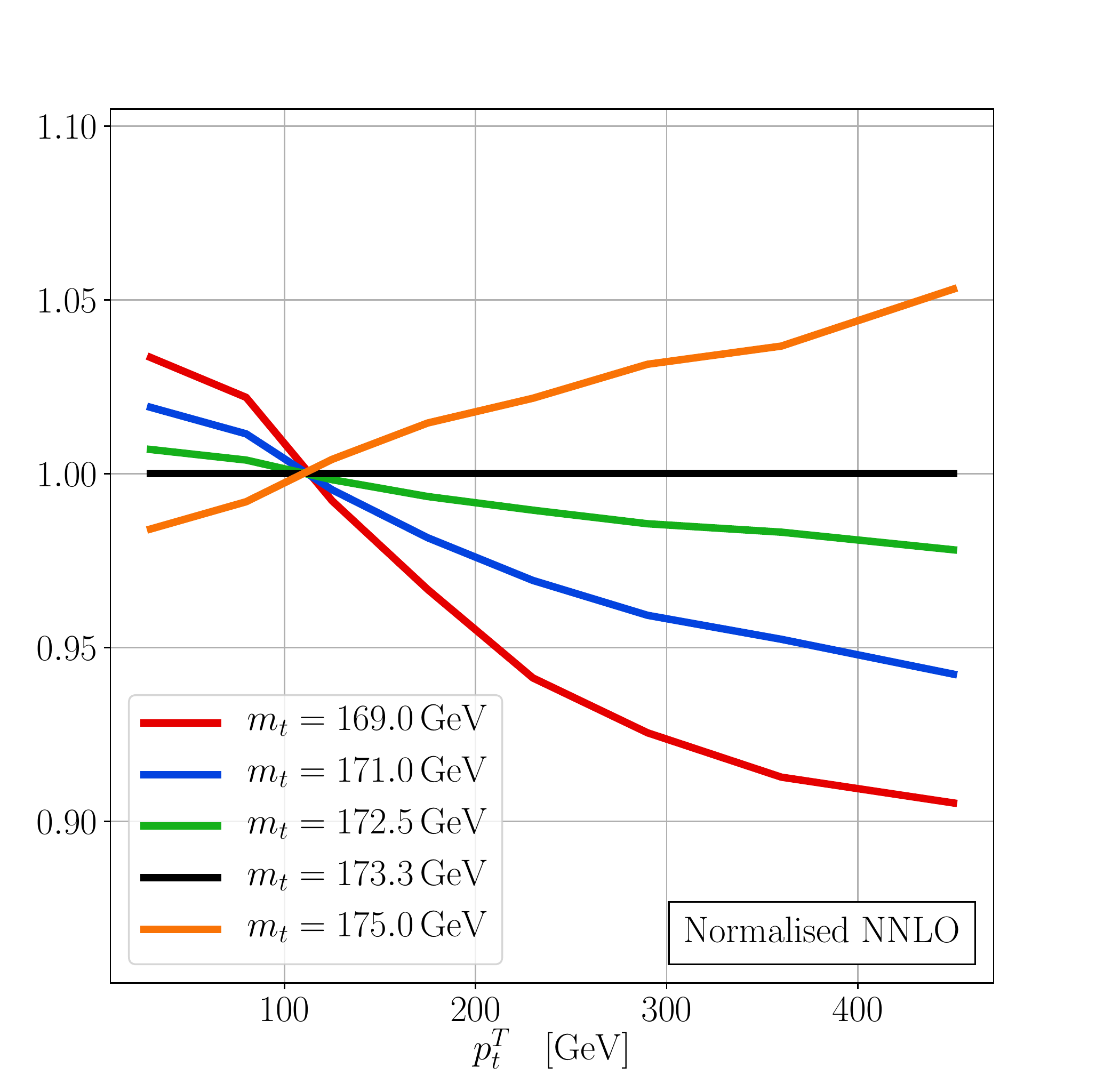}
\includegraphics[trim=0.7cm 0.2cm 1.5cm 1.5cm,clip,width=0.24\textwidth]{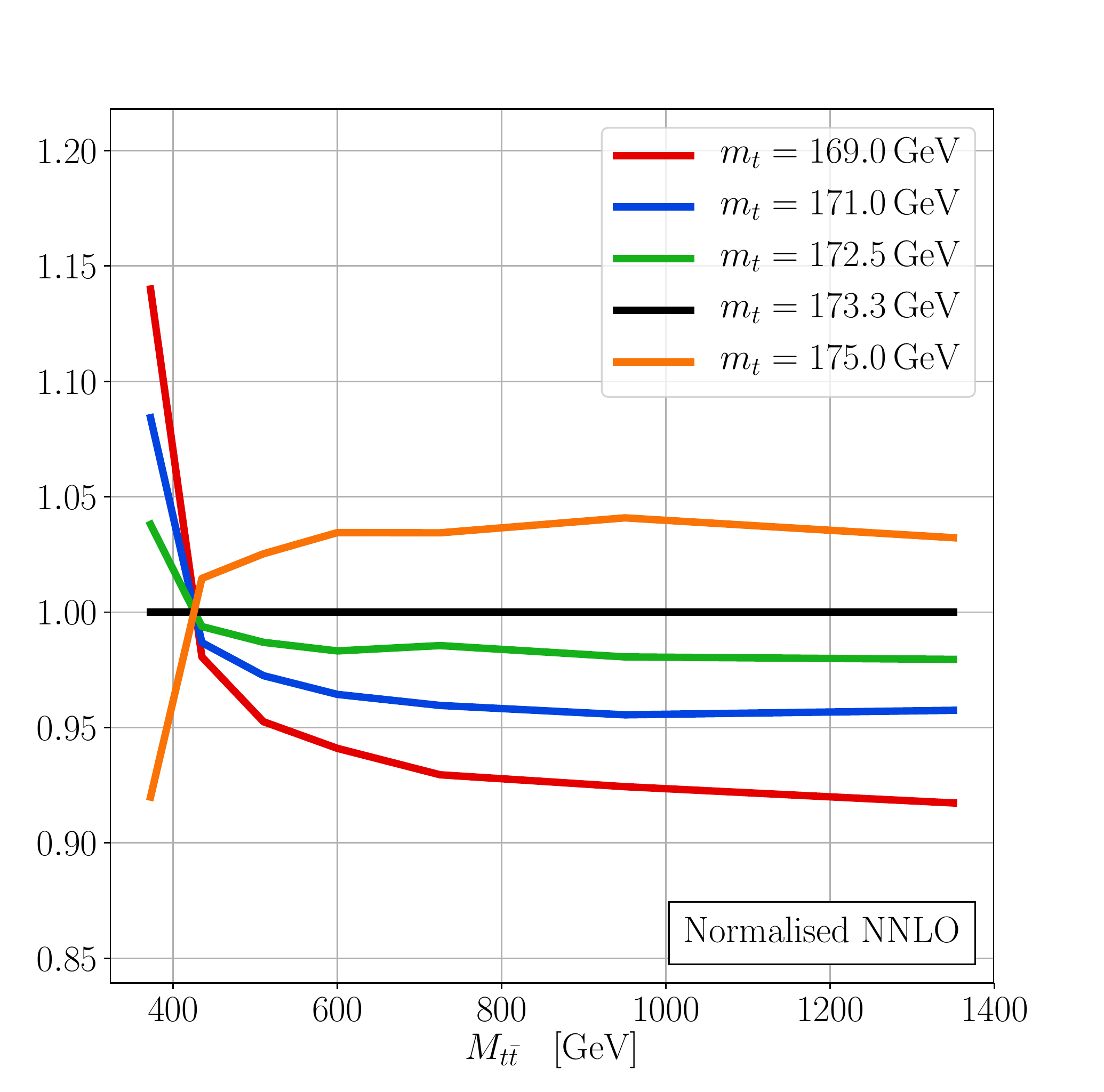}
\includegraphics[trim=0.7cm 0.2cm 1.5cm 1.5cm,clip,width=0.24\textwidth]{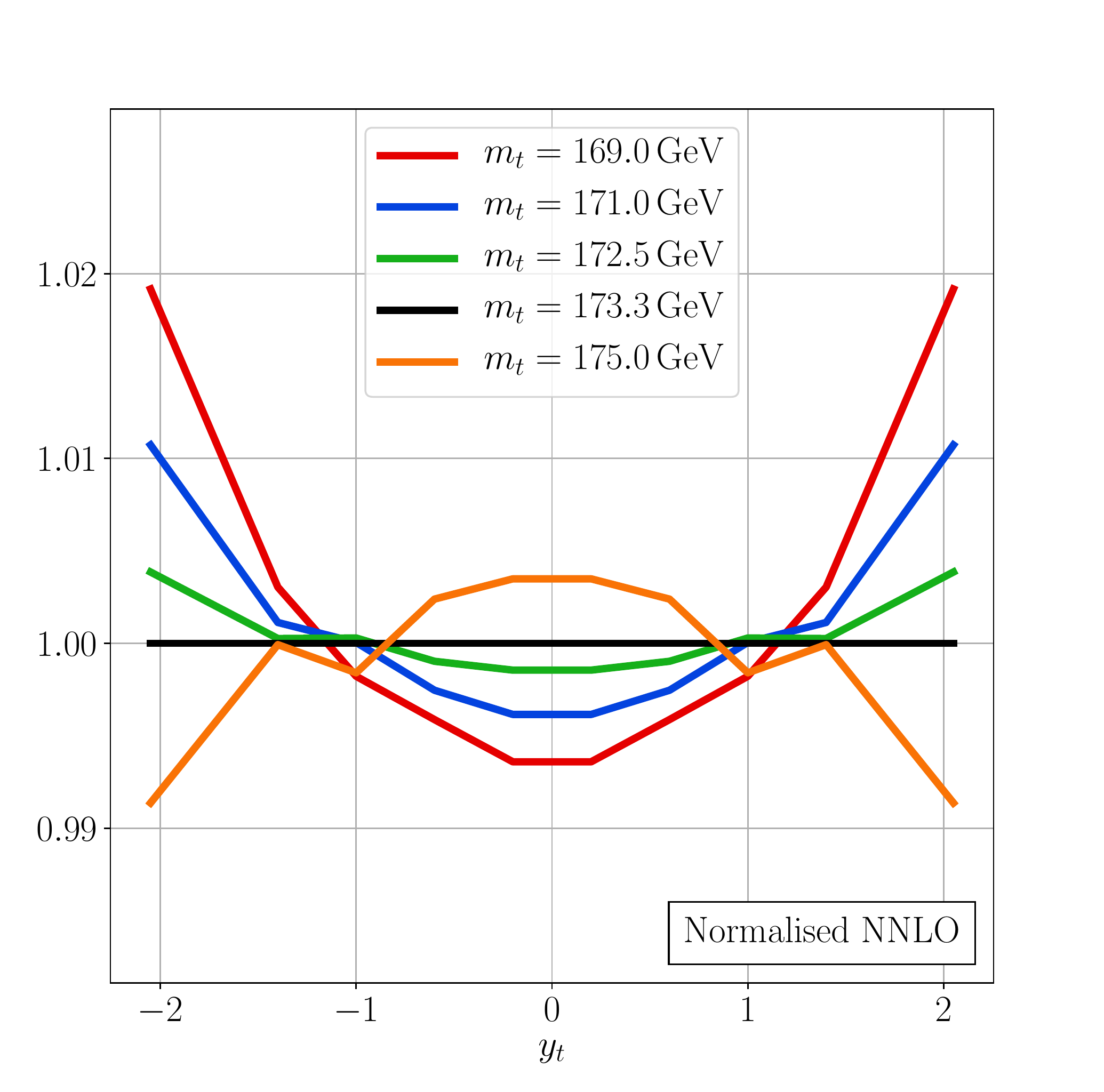}
\includegraphics[trim=0.7cm 0.2cm 1.5cm 1.5cm,clip,width=0.24\textwidth]{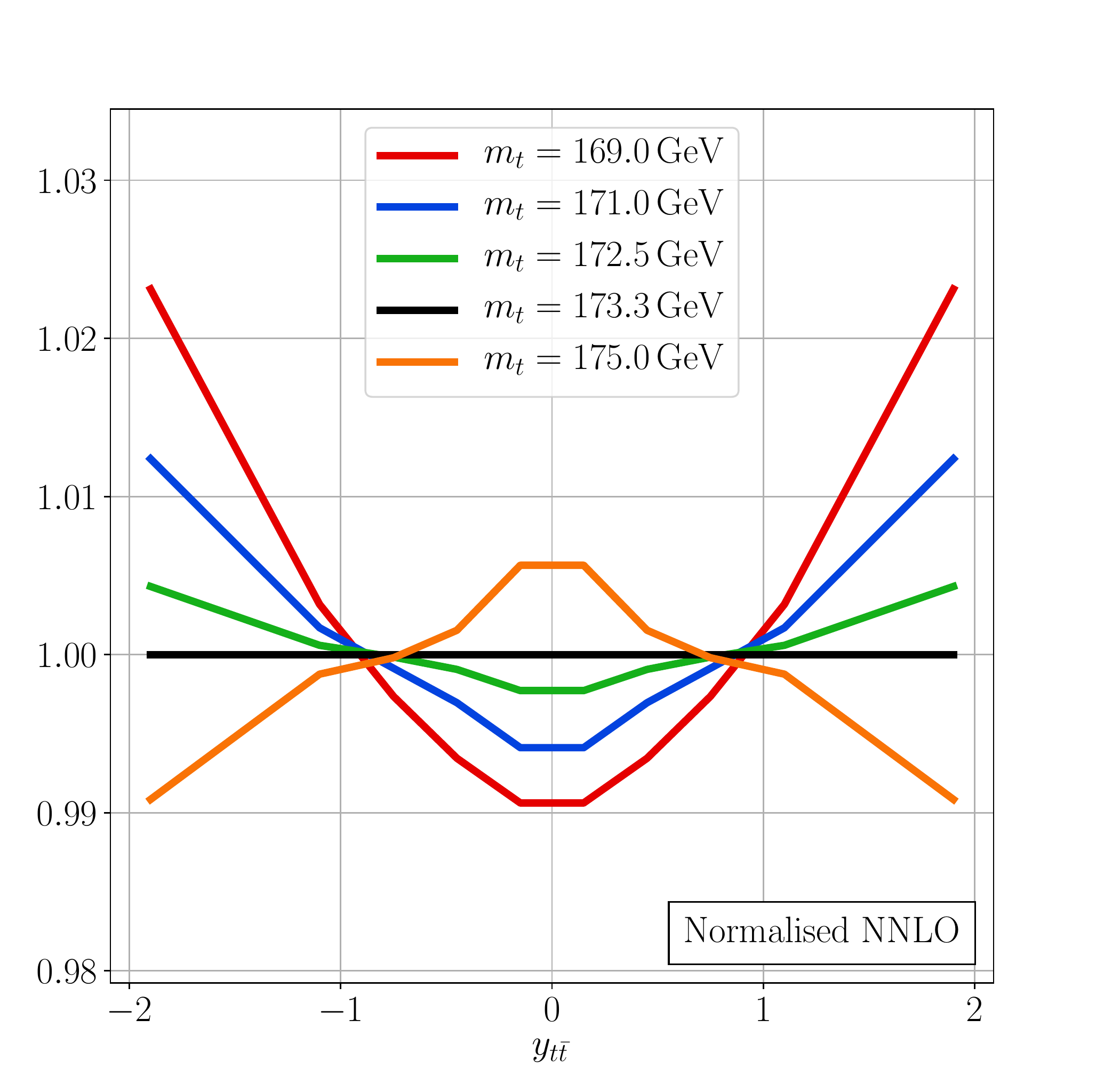}
\caption{As in fig.~\ref{fig:as-sensitivity} but for the $\mt$ sensitivity.}
\label{fig:mt-sensitivity}
\end{figure}

For the absolute distributions, we observe that the sensitivity on $\as$ closely follows the $\as$-dependence of the total cross section $\sim\mathcal{O}(\as^2)$. For all absolute distributions, the sensitivity increases in the tails of distributions, where corrections beyond the Born-level in the cross section will have larger effects. A similar pattern can be observed in the $\mt$ sensitivity of the rapidity distributions, $\yt$ and $\ytt$, ie the behaviour of the weights of each bin roughly follow the $\sim\mathcal{O}(\mt^{-4})$ dependence of the total cross section, and once again the tails see slightly increased sensitivity. On the other hand, the $\pT$ and $\mtt$ distributions display very different sensitivities with respect to the $\as$ case. The bins most sensitive to changes in $\mt$ are the lowest bins in the ranges where the bulk of the cross section lies. In the tails, however, the sensitivity is dramatically reduced -- this can be understood by the fact that in the tails the top-quark is effectively massless (the finite-mass effects being suppressed by powers of $\pT$ or $\mtt$).

In the case of normalised distributions, the sensitivity plots shown in \figs{as-sensitivity}{mt-sensitivity} show a far more involved dependence on $\as$ and $\mt$. These patterns arise because the bin-weights of the absolute distributions are divided by the cross section over all bins. In general we observe that, except in the tails of distributions, the effect of varying $\as$ by $\pm 0.003$ around the baseline value of $\as=0.118$ is at the $\sim 1$\%-level or less. On the other hand, variations of $\mt$ by $\pm (2-3)$~GeV around the value $\mt=173.3$~GeV can result in much larger effects: for $\pT$ the variations in $\mt$ lead to changes of $\sim 2\%$ and for $\mtt$ these are $\gtrsim 0-5$\% in the bins with largest weight. For $\yt$ and $\ytt$ the effects of varying $\mt$ are much smaller and fall below $0.5\%$ in the bins with largest weight. If experimental uncertainties were negligible, then we would expect these patterns to be reflected in the uncertainties on the extracted values of $\as$ and $\mt$. That is, we would expect that since the magnitude of the dependence on $\as$ is roughly the same across all distributions, this should translate to similar uncertainties on values of $\as$ extracted from each distribution. In contrast, we would expect that the uncertainties on the extracted values of $\mt$ would be largest when using the normalised $\yt$ and $\ytt$ data (and significantly larger than the uncertainties obtained when using the normalised $\pT$ and particularly $\mtt$ data), due to the relative insensitivity on the mass of these distributions. We indeed find evidence of such behaviour in \sec{results}.

\subsubsection{Top mass systematics and the lower edge of $\mtt$}\label{sec:lower-edge-mtt}

Experimental measurements of the cross section and of distributions depend on the value of the top-quark mass. This dependence arises because inclusive measurements of stable top quarks are actually extrapolations from measurements of top-quark decay products in fiducial regions, and these extrapolations are derived from simulated Monte Carlo data which explicitly carries a dependence on $\mt$.\footnote{The default choice of input $\mt$ used by both ATLAS and CMS is $\mt=172.5$~GeV.} Typically, this `experimental response' to the input value of $\mt$ is known to constitute
a $\sim 1-2\%$ effect at the level of the inclusive cross section. For generic bins of a kinematic distribution this is a similarly small effect.
However, there are kinematic regions where these effects could be significantly larger and are perhaps currently an underestimated systematic effect. This is particularly the case for the lowest bin in $\mtt$ -- for simulated stable top data, the lowest possible value of $\mtt$ is $2\mt$ and therefore when using data simulated with $\mt=172.5$~GeV, one can never obtain values $\mtt<345$~GeV.

Considering the $\mtt$ distribution, if the lowest value of $\mtt$ considered is $345$~GeV (as is done here, see \tab{binnings}), then for values of the top-quark mass $\mt < 172.5$~GeV, the theoretical predictions will contain events with kinematics such that $\mtt < 345$~GeV.
In our extractions above, and consistently with what is done by the two experiments, these `underflow' events are not included.
Since this lowest bin of $\mtt$ is the most sensitive to the value of $\mt$, this issue deserves a careful examination.
If in the extrapolation to stable-top quarks, some underflow events do leak into the first bin of $\mtt$, then this could have serious effects on the extraction of $\mt$ from this distribution. In order to study this further, we have considered how our extractions react to the addition of
underflow events to this lowest bin. For the discrete set of values of $\mt$ considered, it is only the predictions for
$\mt=169.0$~GeV and $\mt=171.0$~GeV that contain underflow events, but nevertheless, these do alter the fits of the theory predictions. In appendix~\ref{app:C} we present extractions using the invariant mass distribution supplemented with underflow events $\mttu$ and compare these with the results obtained using simply $\mtt$.

\section{Extraction Methodology}\label{sec:methodology}

In order to quantitatively compare experimental measurements to theoretical predictions we use a least-squares method. We consider two sets of measurements: (a) measurements of normalised distributions supplemented with measurements of the total cross section, or (b) measurements of absolute distributions. In both cases we first include data only from a single experiment in the $\chi^2$ objective (we comment later on using data from both ATLAS and CMS). The definitions of the $\chi^2$ variables we use are
\begin{align}
\chi^2_{\rm{norm}}&= \frac{1}{N_\mathrm{data}}\left[\sum_{i,j=1}^{N_\mathrm{data}-1}\zeta_iC_{ij}^{-1}\zeta_j \quad
      + \frac{(\sigma_{\rm{theory}}-\sigma_{\mathrm{data}})^2}{\delta\sigma_{\mathrm{data}}^2}\right] \,, \label{eq:chisq-norm} \\
\chi^2_{\rm{abs}}&=\frac{1}{N_\mathrm{data}}\sum_{i,j=1}^{N_\mathrm{data}}\zeta_iC_{ij}^{-1}\zeta_j \,, \label{eq:chisq-abs}
\end{align}
where $\zeta_i=\zeta_i^{\mathrm{data}}-\zeta_i^{\mathrm{theory}}$ is the difference between the measured and predicted weights in bin $i$ and $C_{ij}$ a covariance matrix encoding experimental sources of error and correlations between bins $i$ and $j$ of the distributions. Our covariance matrices are obtained from the correlation matrices made publicly available by the experiments. In the case of the ATLAS experiment, they contain information about all sources of systematic and statistical correlation. We were unable to obtain the corresponding matrices from CMS---for extractions from these data (and in cases where ATLAS and CMS data are combined) we use the so-called experimental definition of the covariance matrix, mirroring the analysis performed in section~3.3 of \mycite{Czakon:2016olj}. This definition takes into account correlated systematic uncertainties but is diagonal in the statistical uncertainties.

In the cases where normalised shape distributions have been used the final bin of each distribution has been removed in order to render the covariance matrix invertible
\footnote{The choice of bin to be removed does not affect the value of the $\chi^2$.}{}.
A possible advantage of using normalised distributions rather than absolute distributions is the improved control of systematic errors on the experimental measurements of the total cross section. There is also a partial cancellation of uncertainties such as the luminosity which takes place in normalising the distributions.

The sources of uncertainty included in the covariance matrices, and hence in our extractions, are at present purely experimental. These experimental uncertainties are statistical and systematic, also including uncertainties due to luminosity and beam energy and the bin-by-bin correlations are either fully (in the case of the ATLAS experiment) or partially (in the CMS case) retained.

In a more complete treatment of uncertainties, sources of theoretical uncertainty such as those due to scale variation and PDF uncertainties should also be included. The fact that we have not included the theoretical uncertainties in our extraction of $\mt$ and $\as$ deserves clarification. We justify this omission by the fact that these uncertainties, both due to scales and PDF, are not dominant for the observables and data sets we study here. Indeed, since the scale uncertainty is greatly reduced at NNLO when considering normalised (compared to absolute) distributions, we do not expect that including these would increase the uncertainties on the extracted parameters dramatically. On the other hand, in one-dimensional extractions of $\as$, the scale uncertainty on the cross section is known to contribute a significant fraction of the overall uncertainty on the best-fit $\as$~\cite{Klijnsma:2017eqp}, and therefore we would expect this to also play a r\^{o}le for our study here.

Despite the fact that the scale error itself has no direct statistical meaning, extensive experience with collider processes shows that scale variation could reasonably be interpreted as a proxy for a hypothetical error estimator which is statistical in nature. One could then imagine a modified $\chi^2$ function where the experimental and theoretical errors are added in quadrature. A subtlety arises when one tries to generalise the full covariance matrices entering the $\chi^2$ functions we study. The accumulated experience in estimating theory scale errors does not automatically translate from scale variation to the correlation between scale variations. As we observe here, the effect of correlations between different bins or different distributions play a very significant role and for the sake of being conservative we have decided not to include such theory correlations. The subject has recently been explored in ref.~\cite{AbdulKhalek:2019ihb} which may represent a starting point for improved future studies. An alternative interpretation of the scale uncertainty in a Bayesian fashion has also been given in ref.~\cite{Bonvini:2020xeo} which may also shed some light on this issue.

We consider individual extractions of the parameters in which $\mt$ is fixed at the world average $\mt=173.3$~GeV and we extract $\alpha_s$ or, conversely, fits where we fix $\as=0.118$ and extract $\mt$. In both of these cases, the resulting $\chi^2$ is one-dimensional and the best-fit value of $\mt$ or $\as$ is that which minimises the $\chi^2$ objective. The uncertainties on the extracted values of the parameters are estimated through the standard condition
\begin{equation}
\Delta\chi^2\equiv N_{\mathrm{data}}(\chi^2 - \chisqmin)= 1\,,
\label{eq:deltachi2}
\end{equation}
which naturally corresponds to a 1$\sigma$ deviation from the best-fit value.

While the above procedure is considered standard practice, this could potentially lead to a bias in the extracted value of $\mt$ (or $\as$) or yield extracted parameters that do not minimise the $\chi^2$ objective in the full $(\mt,\as)$ plane. We therefore extend the one-dimensional studies to the case where both parameters, $\mt$ \emph{and} $\as$, are allowed to vary independently resulting in a two-dimensional $\chi^2$. Once again the best-fit value is taken to be that which minimises this two-dimensional $\chi^2$. Estimating the uncertainties on the extracted values is slightly more involved in this case, since the condition \eq{deltachi2} traces out a contour in the $(\mt,\as)$ plane (in the ideal case, an ellipse). The estimate of the uncertainties on the extracted values of $\as$ and $\mt$ are derived from these contours by setting $\mt$ and $\as$ respectively to their best-fit value and deducing the range corresponding to $\Delta\chi^2= 1$ once this choice has been made.

Hitherto we have only discussed extractions from a single distribution measured by a single experiment. Given that the shape of the dependence on $\as$ or $\mt$ is distribution-dependent, combining information from different distributions in an extraction could provide complementary constraints. Indeed, the combination of several observables has been advocated in ref.~\cite{Frixione:2014ala} as an effective way of suppressing possible theoretical biases in the determination of $\mt$. Therefore, by combining distributions, it may be possible to obtain more reliable best-fit parameters as well as smaller overall uncertainties. To this end we perform extractions based on combinations of distributions from the ATLAS experiment, making use of the covariance matrices provided by the collaboration to include information about the correlations between distributions. It was not possible to perform such an extraction in the CMS case, since such information was not made publicly available and its omission would undoubtedly bias the results and underestimate the extraction uncertainties.
Accordingly, we make a simple extension of our $\chi^2$ definitions in \eqs{chisq-norm}{chisq-abs},
\begin{align}
\chi^2_{\mathrm{norm}}&= \frac{1}{(N-1)}\Bigg[\sum_{i,j=1}^{N-2}\zeta_{i,\textsc{atlas}}C_{ij,\textsc{atlas}}^{-1}\zeta_{j,\textsc{atlas}} + \frac{(\sigma_{\mathrm{NNLO}}-\sigma_{\textsc{atlas}})^2}{\delta\sigma_{\textsc{atlas}}^2}\Bigg] \,, \label{eq:chi2-norm-1exp-2dists} \\
\chi^2_{\mathrm{abs}}&=\frac{1}{N}\Bigg[\sum_{i,j=1}^{N}\zeta_{i,\textsc{atlas}}C_{ij,\textsc{atlas}}^{-1}\zeta_{j,\textsc{atlas}} \Bigg]\,,
\label{eq:chi2-abs-1exp-2dists}
\end{align}
where $N=N_\textsc{atlas,1}+N_\textsc{atlas,2}$. The vector $\zeta_{i,\textsc{atlas}}$ now contains data from both distributions and the covariance matrices $C_{ij}$ contain off-diagonal entries (see also \sec{twoatlas}).

Finally, since it is realistic to assume that to a large extent measurements performed by \emph{different} experiments are independent (and therefore that correlations between the experiments are negligible),
\footnote{The assumption of negligible correlated uncertainties does not hold for the luminosity uncertainty. Although non-negligible, this uncertainty is not dominant in this analysis.}
we also consider the combination of a single distribution from ATLAS with one from CMS. For these extractions we define further straightforward extensions of our $\chi^2$ definitions,
\begin{align}
\chi^2_{\mathrm{norm}}&= \frac{1}{(N_\textsc{atlas}+N_\textsc{cms})}\Bigg[\sum_{i,j=1}^{N_\textsc{atlas}-1}\zeta_{i,\textsc{atlas}}C_{ij,\textsc{atlas}}^{-1}\zeta_{j,\textsc{atlas}} +\sum_{i,j=1}^{N_\textsc{cms}-1}\zeta_{i,\textsc{cms}}C_{ij,\textsc{cms}}^{-1}\zeta_{j,\textsc{cms}} \nonumber \\
&\qquad\qquad+ \frac{(\sigma_{\mathrm{NNLO}}-\sigma_{\textsc{atlas}})^2}{\delta\sigma_{\textsc{atlas}}^2} +\frac{(\sigma_{\mathrm{NNLO}}-\sigma_{\textsc{cms}})^2}{\delta\sigma_{\textsc{cms}}^2}\Bigg] \,, \label{eq:chi2-norm-2exps} \\
\chi^2_{\mathrm{abs}}&=\frac{1}{(N_\textsc{atlas}+N_\textsc{cms})} \Bigg[\sum_{i,j=1}^{N_\textsc{atlas}}\zeta_{i,\textsc{atlas}}C_{ij,\textsc{atlas}}^{-1}\zeta_{j,\textsc{atlas}} +\sum_{i,j=1}^{N_\textsc{cms}}\zeta_{i,\textsc{cms}}C_{ij,\textsc{cms}}^{-1}\zeta_{j,\textsc{cms}} \Bigg] \,.
\label{eq:chi2-abe-2exps}
\end{align}

\section{Results}\label{sec:results}

In this section we present and discuss the results of our extractions of $\as$ and $\mt$, both separately from one-dimensional $\chi^2$ functions, as well as simultaneously letting both parameters vary. We focus on the extractions using measurements of normalised distributions supplemented with information from the total cross sections. Extractions using experimental results from absolute distributions can be found in \app{B}. We do not consider the rapidity distributions using NNPDF3.1, for reasons that will be discussed in \sec{nnpdf31}.

\subsection{Independent (one-dimensional) extractions of $\as$ and $\mt$}

\subsubsection{Extraction from complete distributions}

The first set of extractions we present are the separate extractions of $\as$ and $\mt$ using
the one-dimensional form of our $\chi^2$ objective in \eq{chisq-norm}.
The results of the $\as$ extraction, where the value of top mass has been fixed to $\mt=173.3$~GeV,
are shown in \tab{as-extr-1d}.
The results of the $\mt$ extraction, where the value of the strong-coupling constant has been
fixed to $\as=0.118$, are shown in \tab{mt-extr-1d}.

\begin{table}[h!]
\centering

\begin{tabular}{ |c|c c|c c|c c| }
\hline
\multicolumn{1}{|c|}{} & \multicolumn{6}{|c|}{ATLAS}\\
\hline
& \multicolumn{2}{|c|}{CT14} & \multicolumn{2}{|c|}{NNPDF30} & \multicolumn{2}{|c|}{NNPDF31} \\
& $\as$ & $\chisqmin$ & $\as$ & $\chisqmin$ & $\as$ & $\chisqmin$ \\
\hline
$\pT$ &\cellcolor{Gray}$0.1157^{+0.0020}_{-0.0022}$ \
&\cellcolor{Gray}0.42&\cellcolor{Gray}$0.1162^{+0.0027}_{-0.0027}$ \
&\cellcolor{Gray}0.50&\cellcolor{Gray}$0.1171^{+0.0022}_{-0.0022}$ \
&\cellcolor{Gray}0.72     \\
 \hline
$\mtt$ &\cellcolor{Gray}$0.1158^{+0.0020}_{-0.0021}$ \
&\cellcolor{Gray}0.17&\cellcolor{Gray}$0.1157^{+0.0027}_{-0.0027}$ \
&\cellcolor{Gray}0.19&\cellcolor{Gray}$0.1173^{+0.0022}_{-0.0022}$ \
&\cellcolor{Gray}0.22     \\
 \hline
$\yt$ &$0.1149^{+0.0018}_{-0.0019}$ \
&0.91&\cellcolor{Gray}$0.1155^{+0.0018}_{-0.0018}$ \
&\cellcolor{Gray}0.22&&     \\
\hline
$\ytt$ &\cellcolor{Gray}$0.1162^{+0.0019}_{-0.0020}$ \
&\cellcolor{Gray}2.04&\cellcolor{Gray}$0.1154^{+0.0025}_{-0.0025}$ \
&\cellcolor{Gray}0.57&&    \\
\hhline{|=|==|==|==|}
Average &$0.1158^{+0.0014}_{-0.0015}$&&$0.1157^{+0.0013}_{-0.0013}$ &&$0.1172^{+0.0016}_{-0.0016}$&\\
\hline
\end{tabular} \\[10pt]

\begin{tabular}{ |c|c c|c c|c c| }
\hline
\multicolumn{1}{|c|}{} & \multicolumn{6}{|c|}{CMS}\\
\hline
& \multicolumn{2}{|c|}{CT14} & \multicolumn{2}{|c|}{NNPDF30} & \multicolumn{2}{|c|}{NNPDF31} \\
& $\as$ & $\chisqmin$ & $\as$ & $\chisqmin$ & $\as$ & $\chisqmin$ \\
\hline
$\pT$ &\cellcolor{Gray}$0.1164^{+0.0014}_{-0.0015}$ \
&\cellcolor{Gray}5.01&$0.1121^{+0.0026}_{-0.0026}$ \
&3.62&\cellcolor{Gray}$0.1178^{+0.0019}_{-0.0019}$ \
&\cellcolor{Gray}2.56     \\
 \hline
$\mtt$ &$0.1147^{+0.0014}_{-0.0014}$ \
&9.68&$0.1100^{+0.0014}_{-0.0043}$ \
&6.97&\cellcolor{Gray}$0.1161^{+0.0018}_{-0.0018}$ \
&\cellcolor{Gray}6.01     \\
 \hline
$\yt$ &\cellcolor{Gray}$0.1165^{+0.0019}_{-0.0020}$ \
&\cellcolor{Gray}2.36&\cellcolor{Gray}$0.1199^{+0.0020}_{-0.0020}$ \
&\cellcolor{Gray}2.99&&     \\
\hline
$\ytt$ &\cellcolor{Gray}$0.1154^{+0.0016}_{-0.0017}$ \
&\cellcolor{Gray}2.14&$0.1149^{+0.0017}_{-0.0017}$ \
&0.89&&     \\
\hline
\end{tabular}

\caption{
Tabulated values of best-fit $\as$ (with uncertainties) and associated $\chisqmin$ from extractions of $\as$
using ATLAS (upper table) and CMS (lower table) measurements of normalised distributions and 
the total cross section.  
Results are shown for three different PDF sets and $\mt$ has been set to the world average value of 173.3~GeV.
The cells highlighted in grey correspond to extractions that satisfy the quality conditions described in sec~\ref{sec:averaging}.
Where shown, the average values have been calculated from the shaded cells in the manner described in sec.~\ref{sec:averaging}. As explained in \sec{nnpdf31}, the rapidity distributions are not used for extractions with the PDF set NNPDF3.1.
}
\label{tab:as-extr-1d}
\end{table}

Looking at \tab{as-extr-1d}, we notice that for a given experiment and distribution, there can be non-trivial differences 
between extracted values of $\as$. 
For example for $\pT$ from ATLAS, there is a 0.0014 difference in the best-fit values when 
using CT14 and NNPDF3.1, although these values are consistent within uncertainties. 
Comparing the results obtained when using measurements from the two experiments, it is clear that there are 
significant differences, particularly in the values of $\chisqmin$ for $\pT$ and $\mtt$ using CT14 or NNPDF3.0. 
As detailed in ref.~\cite{Czakon:2016olj}, this indicates a certain tension between the two sets of measurements.

\begin{table}[h!]
\centering

\begin{tabular}{ |c|c c|c c|c c| }
\hline
\multicolumn{1}{|c|}{} & \multicolumn{6}{|c|}{ATLAS}\\
\hline
& \multicolumn{2}{|c|}{CT14} & \multicolumn{2}{|c|}{NNPDF30} & \multicolumn{2}{|c|}{NNPDF31} \\
& $\mt$ & $\chisqmin$ & $\mt$ & $\chisqmin$ & $\mt$ & $\chisqmin$ \\
\hline
$\pT$ &\cellcolor{Gray}$174.5^{+1.0}_{-1.0}$ \
&\cellcolor{Gray}0.39&\cellcolor{Gray}$174.6^{+1.0}_{-1.0}$ \
&\cellcolor{Gray}0.36&\cellcolor{Gray}$174.6^{+1.0}_{-1.0}$ \
&\cellcolor{Gray}0.54     \\
 \hline
$\mtt$ &\cellcolor{Gray}$173.4^{+0.6}_{-0.5}$ \
&\cellcolor{Gray}0.35&\cellcolor{Gray}$173.4^{+0.6}_{-0.5}$ \
&\cellcolor{Gray}0.28&\cellcolor{Gray}$173.5^{+0.5}_{-0.5}$ \
&\cellcolor{Gray}0.21     \\
 \hline
$\yt$ &\cellcolor{Gray}$175.6^{+1.2}_{-1.2}$ \
&\cellcolor{Gray}0.87&\cellcolor{Gray}$174.8^{+1.3}_{-1.2}$ \
&\cellcolor{Gray}0.30&&     \\
 \hline
$\ytt$ &\cellcolor{Gray}$175.0^{+1.3}_{-1.3}$ \
&\cellcolor{Gray}1.85&\cellcolor{Gray}$174.4^{+1.3}_{-1.3}$ \
&\cellcolor{Gray}0.62&&     \\
\hhline{|=|==|==|==|}
Average &$174.3^{+0.9}_{-0.8}$&&$174.3^{+0.7}_{-0.7}$ &&$173.9^{+0.7}_{-0.7}$&\\
\hline
\end{tabular} \\[10pt]

\begin{tabular}{ |c|c c|c c|c c| }
\hline
\multicolumn{1}{|c|}{} & \multicolumn{6}{|c|}{CMS}\\
 \hline
& \multicolumn{2}{|c|}{CT14} & \multicolumn{2}{|c|}{NNPDF30} & \multicolumn{2}{|c|}{NNPDF31} \\
& $\mt$ & $\chisqmin$ & $\mt$ & $\chisqmin$ & $\mt$ & $\chisqmin$ \\
 \hline
$\pT$ &\cellcolor{Gray}$170.6^{+0.6}_{-0.6}$ \
&\cellcolor{Gray}2.29&\cellcolor{Gray}$170.9^{+0.6}_{-0.6}$ \
&\cellcolor{Gray}1.82&\cellcolor{Gray}$171.4^{+0.6}_{-0.6}$ \
&\cellcolor{Gray}1.09     \\
 \hline
$\mtt$ &\cellcolor{Gray}$170.4^{+0.6}_{-0.7}$ \
&\cellcolor{Gray}7.31&\cellcolor{Gray}$170.5^{+0.7}_{-0.7}$ \
&\cellcolor{Gray}5.98&\cellcolor{Gray}$170.8^{+0.7}_{-0.7}$ \
&\cellcolor{Gray}3.87     \\
 \hline
$\yt$ &\cellcolor{Gray}$173.8^{+1.1}_{-1.1}$ \
&\cellcolor{Gray}2.40&\cellcolor{Gray}$173.0^{+1.1}_{-1.1}$ \
&\cellcolor{Gray}3.08&&    \\
 \hline
$\ytt$ &$176.2^{+1.3}_{-1.3}$ \
&1.87&\cellcolor{Gray}$174.8^{+1.3}_{-1.2}$ \
&\cellcolor{Gray}1.05&&     \\
 \hline
\end{tabular}

\caption{As in \tab{as-extr-1d} but for $\mt$ rather than $\as$, which has been set to the world average value of 0.118. Masses are quoted in GeV.}
\label{tab:mt-extr-1d}
\end{table}
 
Turning our attention to the top mass extractions shown in \tab{mt-extr-1d}, we observe similar behaviour to the $\as$ case: that is, we see large differences $>1$~GeV between the extracted values of the mass, particularly when using the data on $\mtt$ for ATLAS or $\yt$/$\ytt$ for CMS. We also note that the uncertainties when extracting $\mt$ from the $\pT$ distributions are roughly a factor of two
larger than those obtained when extracting using $\mtt$. As a consistency check we see that the pattern of top-mass sensitivity of \fig{mt-sensitivity} is largely
reflected in the uncertainties on the extracted values, i.e. that extractions based on 
measurements of $\mtt$ have smaller uncertainties than extractions based on $\yt$ and $\ytt$.

\subsubsection{Extraction from individual bins of distributions}

\begin{figure}[t]
\includegraphics[trim=0.4cm 0.2cm 1.8cm 1.5cm,clip,width=0.22\textwidth]{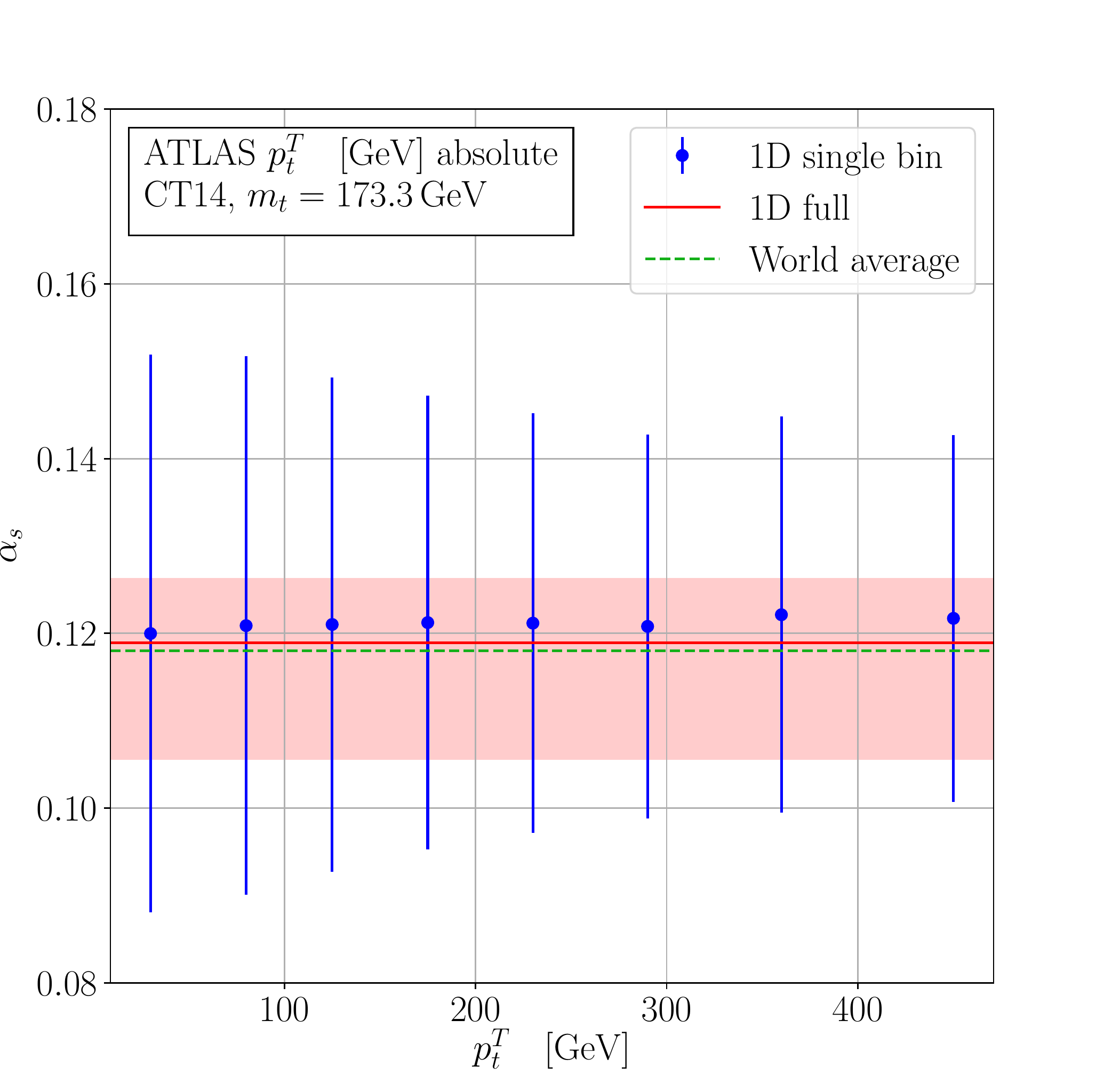} \hspace{0.4cm}
\includegraphics[trim=0.4cm 0.2cm 1.8cm 1.5cm,clip,width=0.22\textwidth]{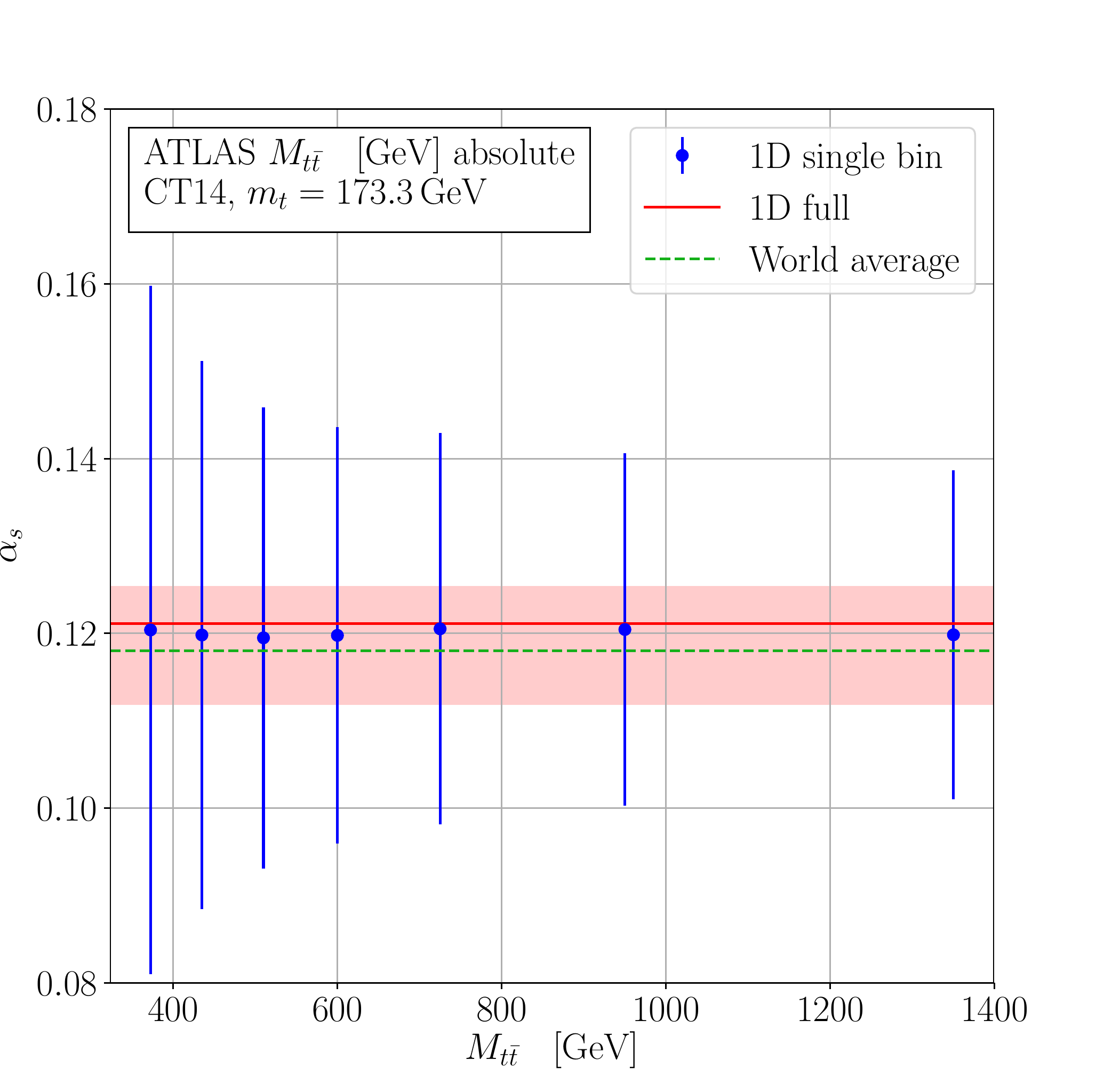} \hspace{0.4cm}
\includegraphics[trim=0.4cm 0.2cm 1.8cm 1.5cm,clip,width=0.22\textwidth]{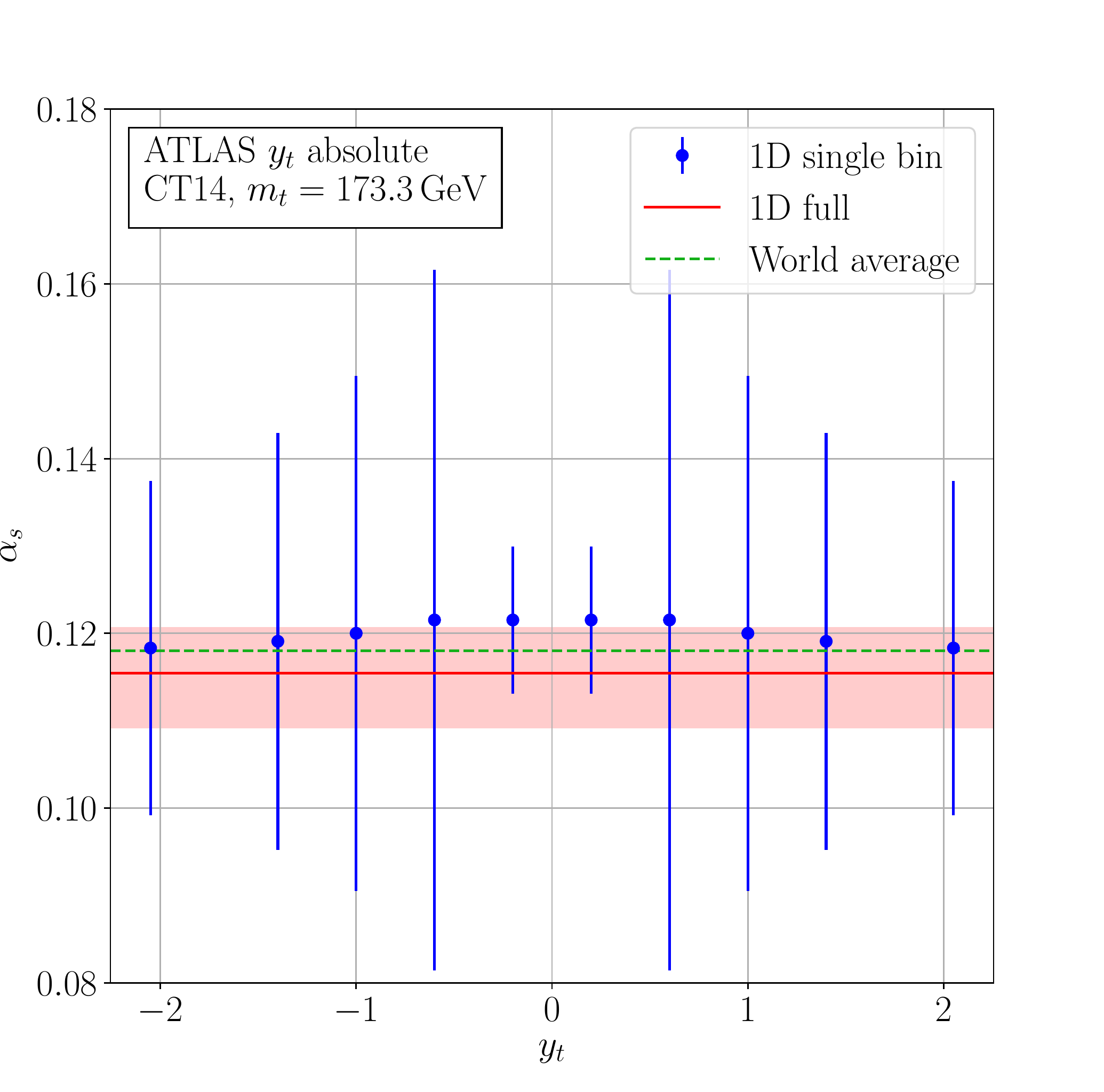} \hspace{0.4cm}
\includegraphics[trim=0.4cm 0.2cm 1.8cm 1.5cm,clip,width=0.22\textwidth]{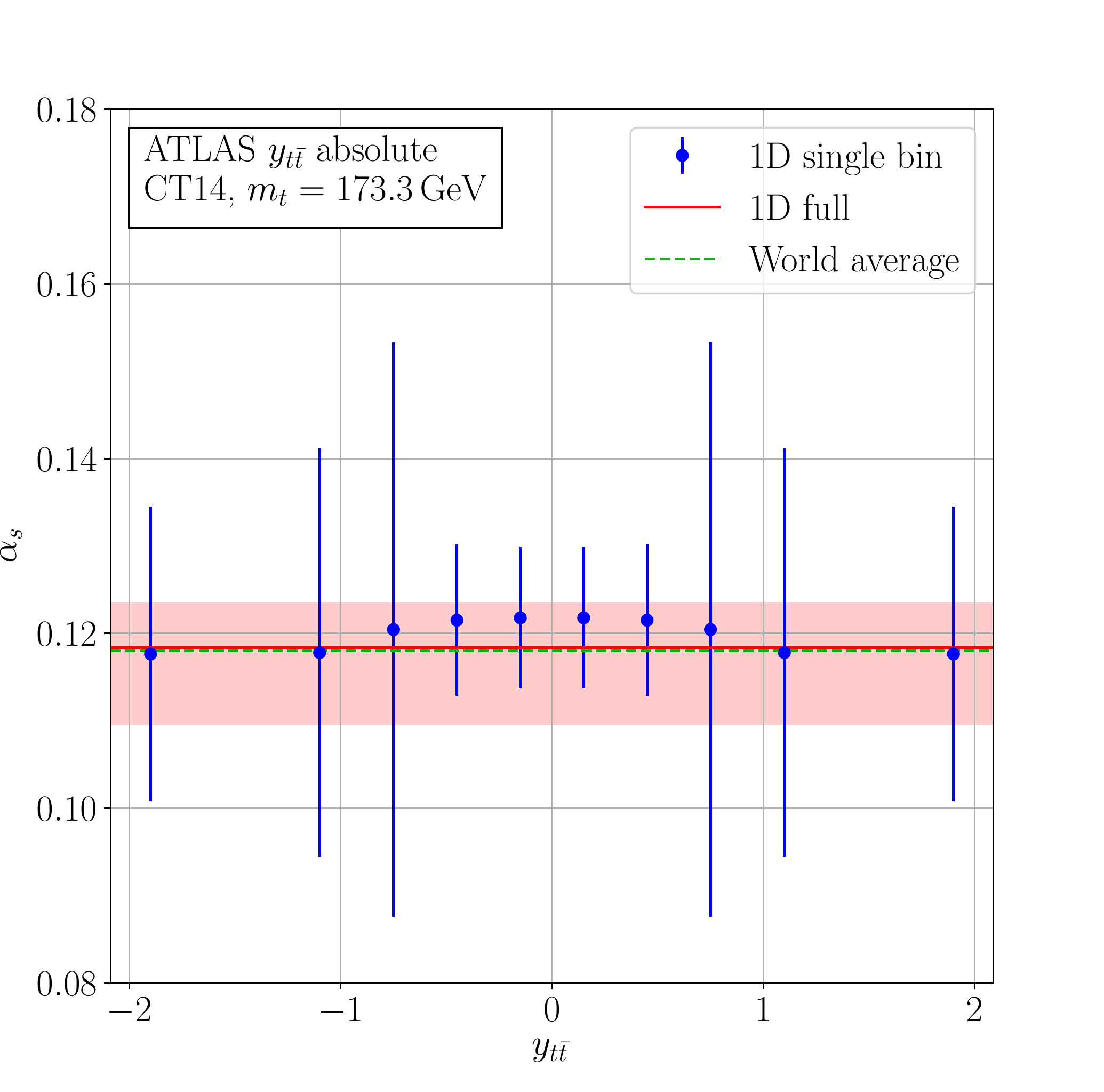} 
\includegraphics[trim=0.4cm 0.2cm 1.8cm 1.5cm,clip,width=0.22\textwidth]{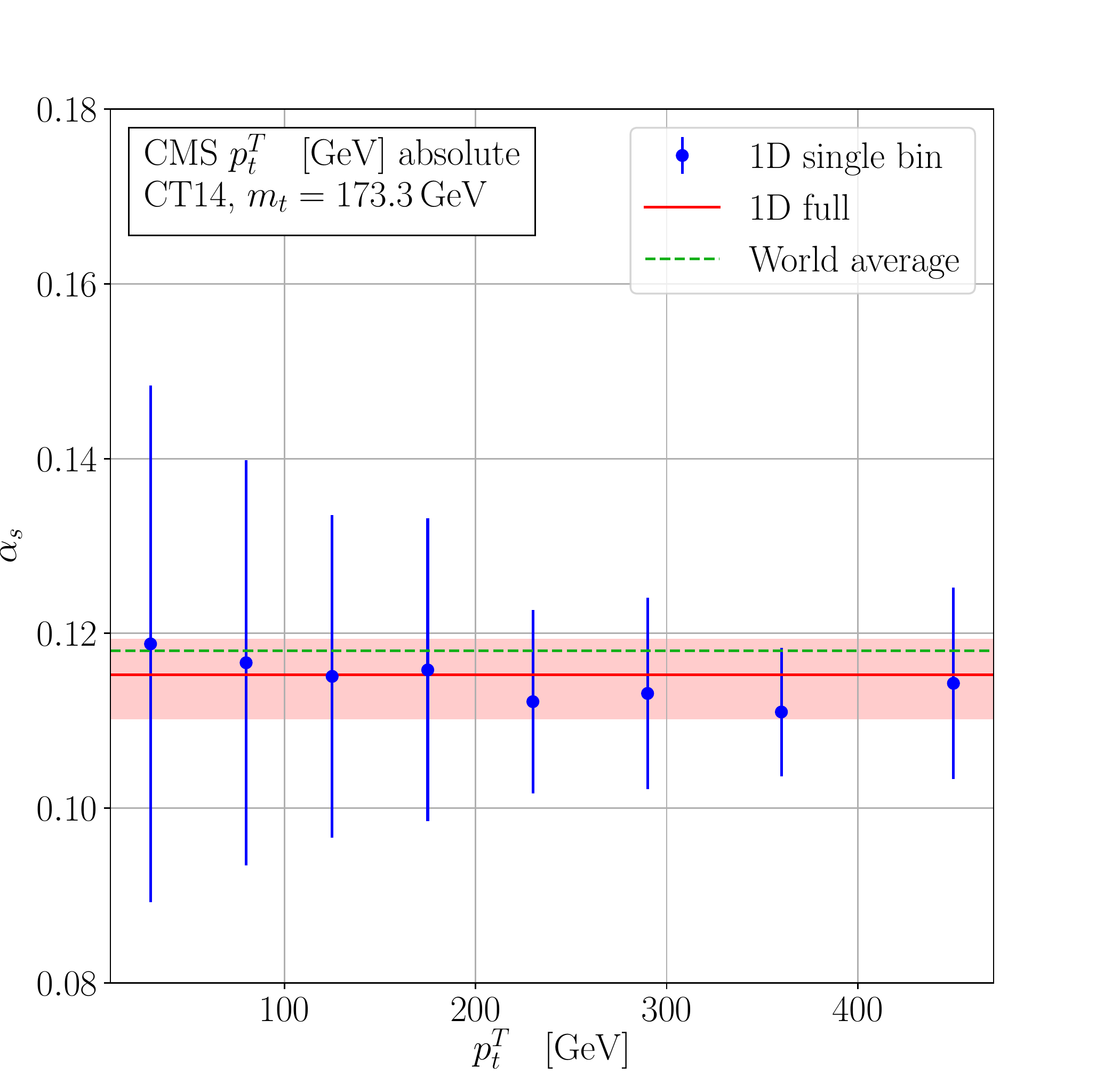} \hspace{0.4cm}
\includegraphics[trim=0.4cm 0.2cm 1.8cm 1.5cm,clip,width=0.22\textwidth]{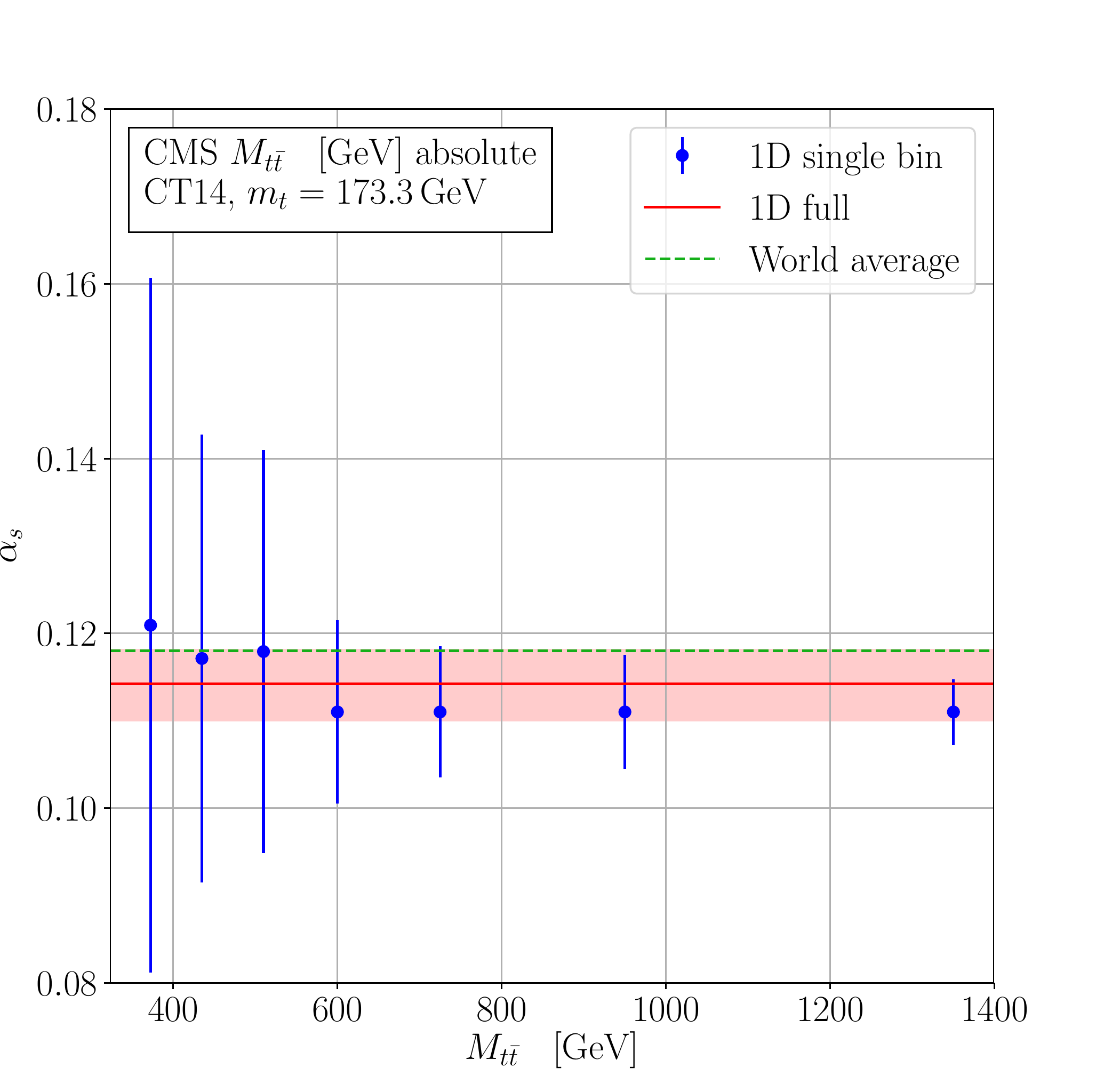} \hspace{0.4cm}
\includegraphics[trim=0.4cm 0.2cm 1.8cm 1.5cm,clip,width=0.22\textwidth]{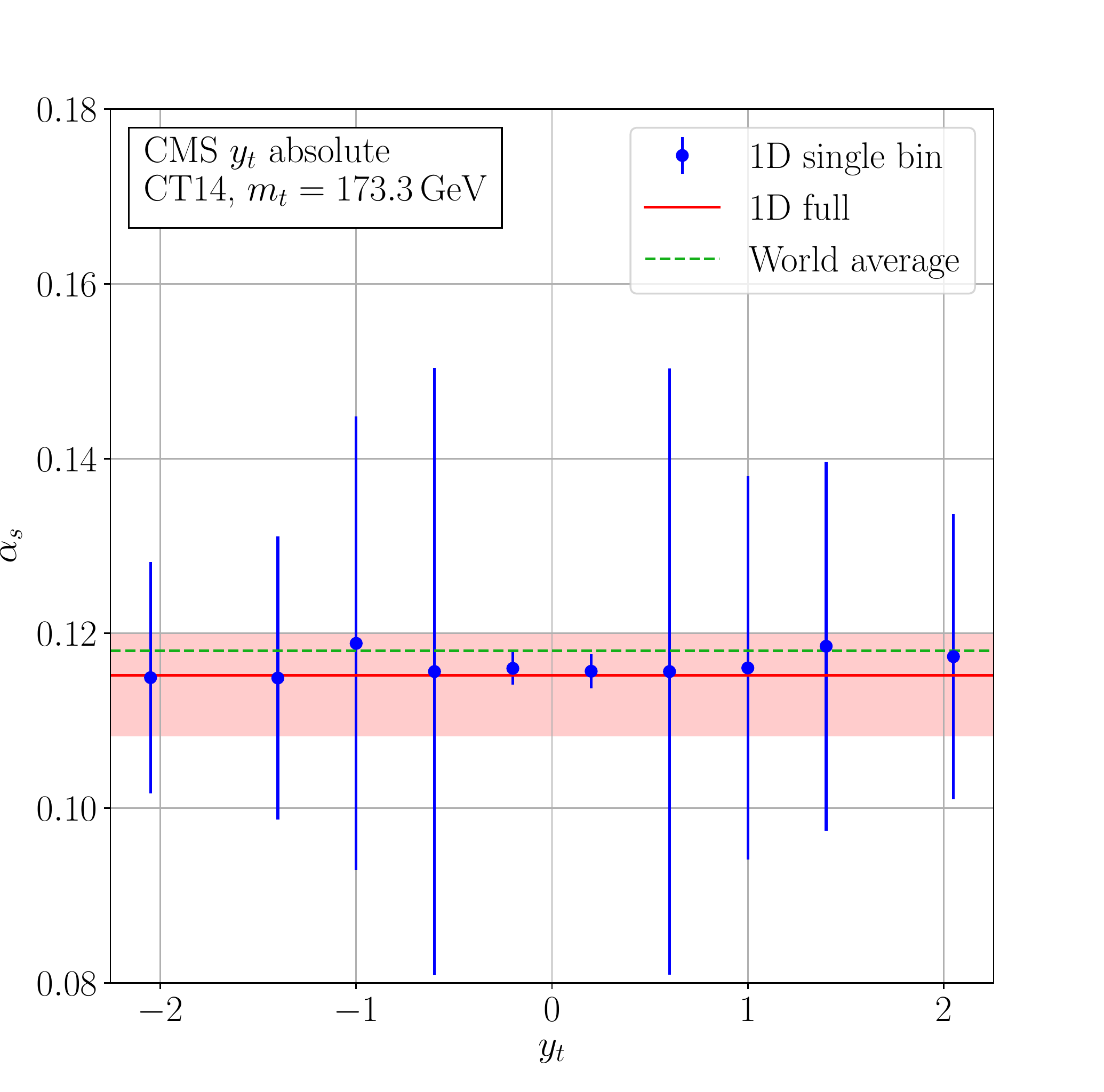}\hspace{0.4cm}
\includegraphics[trim=0.4cm 0.2cm 1.8cm 1.5cm,clip,width=0.22\textwidth]{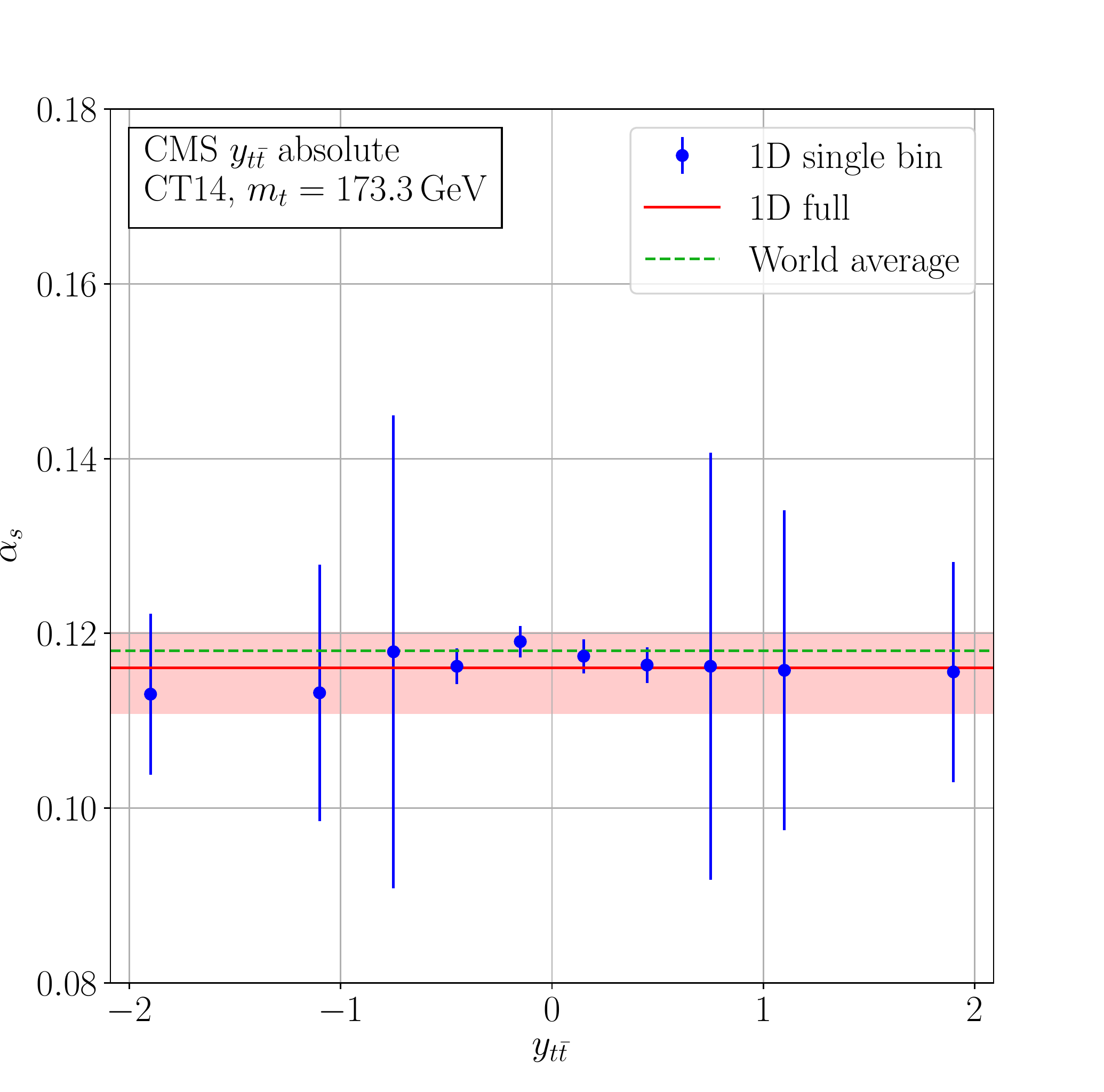}
\caption{Plots showing the extraction of $\as$ for $\mt=173.3$~GeV using CT14 for each individual bin of every distribution. 
Extractions based on ATLAS and CMS \emph{absolute} measurements are at the top and bottom respectively, and extractions
from measurements of $\pT$, $\mtt$, $\yt$ and $\ytt$ are shown in columns 1--4. The blue points show the values extracted from
the single bins with their associated errors, while the dashed green line shows the world average value. The equivalent result
from a full 1D extraction including correlations is shown as a red line with an associated error band.}
\label{fig:bin-by-bin-as}
\end{figure}

In order to scrutinise the patterns observed in the one-dimensional extractions we perform
bin-by-bin extractions of $\as$ and $\mt$ from bins of absolute distributions and compare
to the best-fit values obtained from an extraction from all bins. 
This exercise serves not only as a check on our overall extraction methodology, but also 
allows us to assess the importance of the effects of correlations across different bins. 
In this case, the form of the $\chi^2$ objective that is minimised is the same as that of \eq{chisq-abs}, 
with the sum over all bins replaced by just the $\chi^2$ function for a single bin. 
The results for the $\as$ and $\mt$ extractions are shown in \figs{bin-by-bin-as}{bin-by-bin-mt} respectively, 
where the blue points indicate the best-fit values using just the individual bin weight, whilst
the red horizontal lines indicate the overall best-fit values (see also \tabs{as-extr-1d-abs}{mt-extr-1d-abs} in appendix~\ref{app:B}).
For comparison, the world average values of $\as$ and $\mt$ are also shown in dashed green. 
The results have been obtained using theoretical predictions computed with CT14 PDFs, but 
very similar results are obtained using the other PDF sets we have considered. 

\begin{figure}[t]
\includegraphics[trim=0.4cm 0.2cm 1.8cm 1.5cm,clip,width=0.22\textwidth]{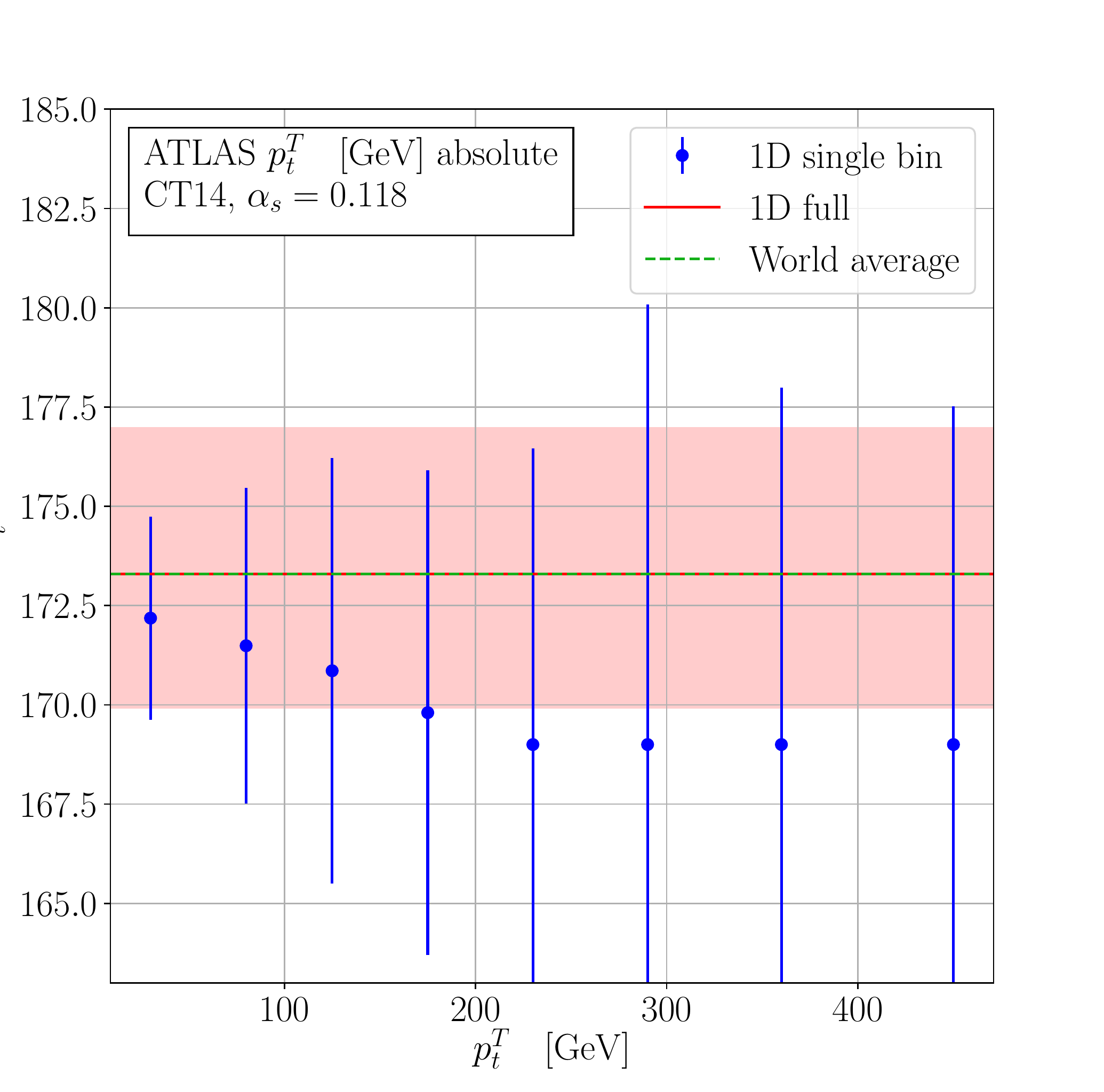} \hspace{0.4cm}
\includegraphics[trim=0.4cm 0.2cm 1.8cm 1.5cm,clip,width=0.22\textwidth]{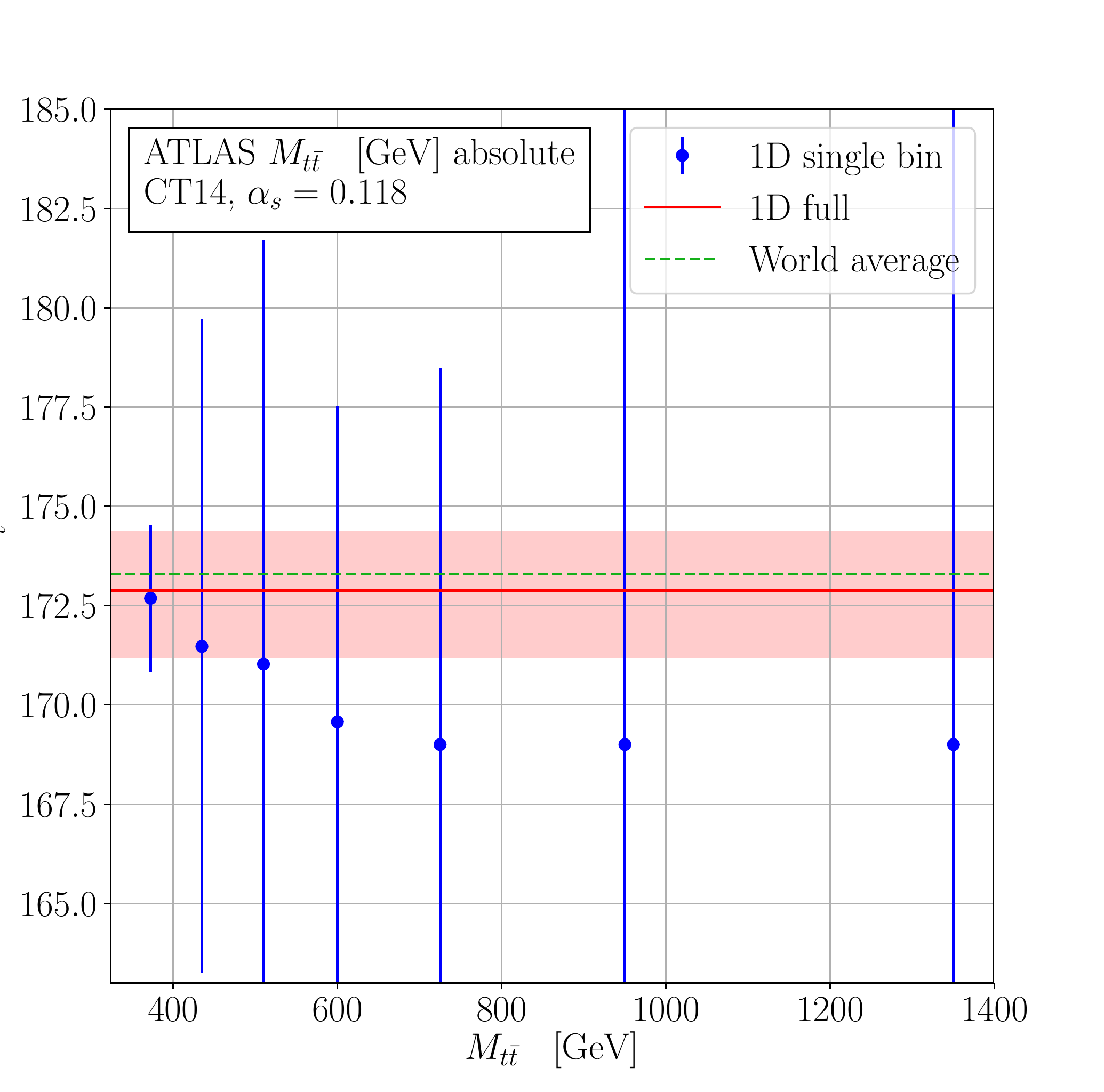} \hspace{0.4cm}
\includegraphics[trim=0.4cm 0.2cm 1.8cm 1.5cm,clip,width=0.22\textwidth]{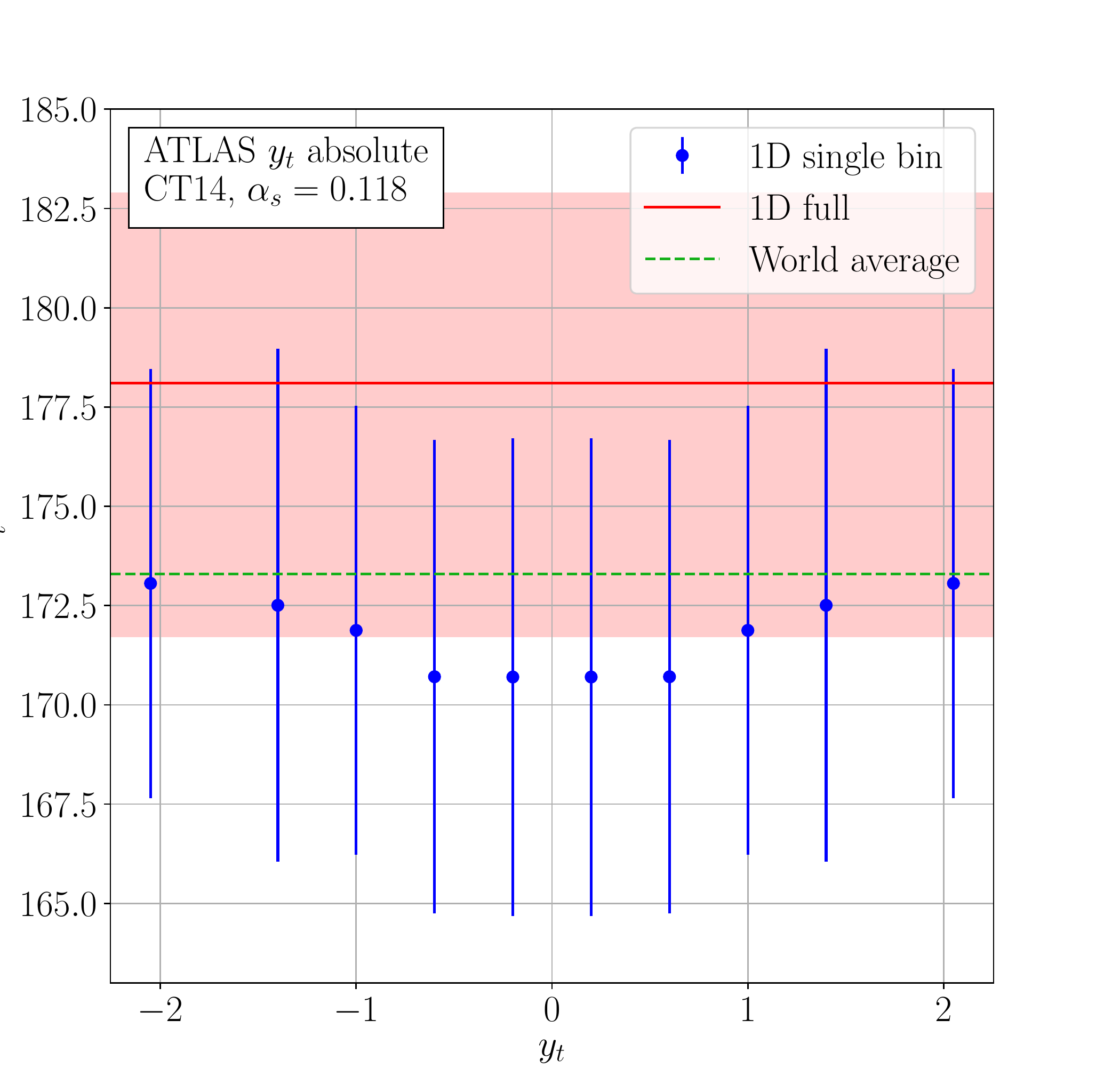} \hspace{0.4cm}
\includegraphics[trim=0.4cm 0.2cm 1.8cm 1.5cm,clip,width=0.22\textwidth]{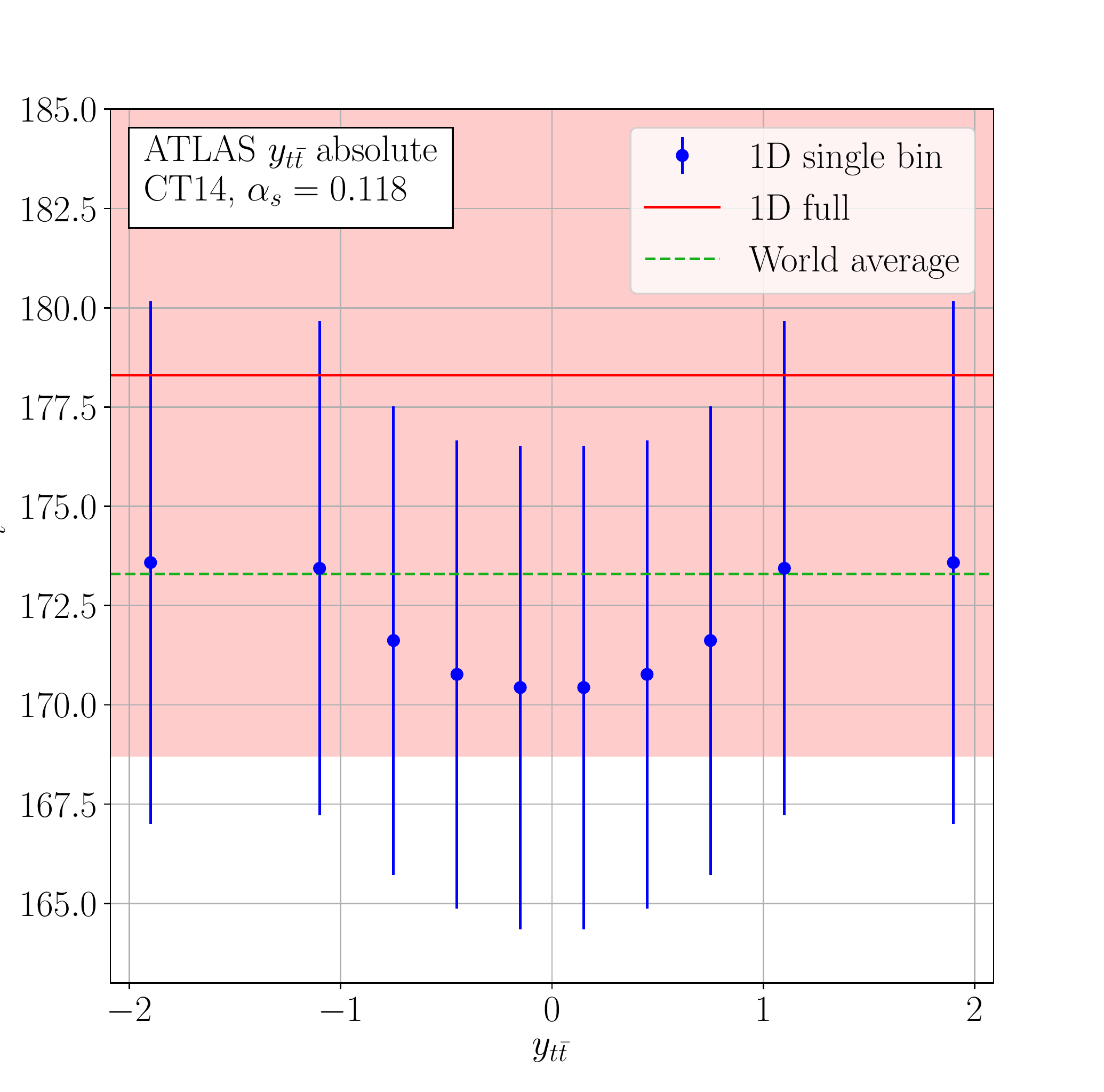}
\includegraphics[trim=0.4cm 0.2cm 1.8cm 1.5cm,clip,width=0.22\textwidth]{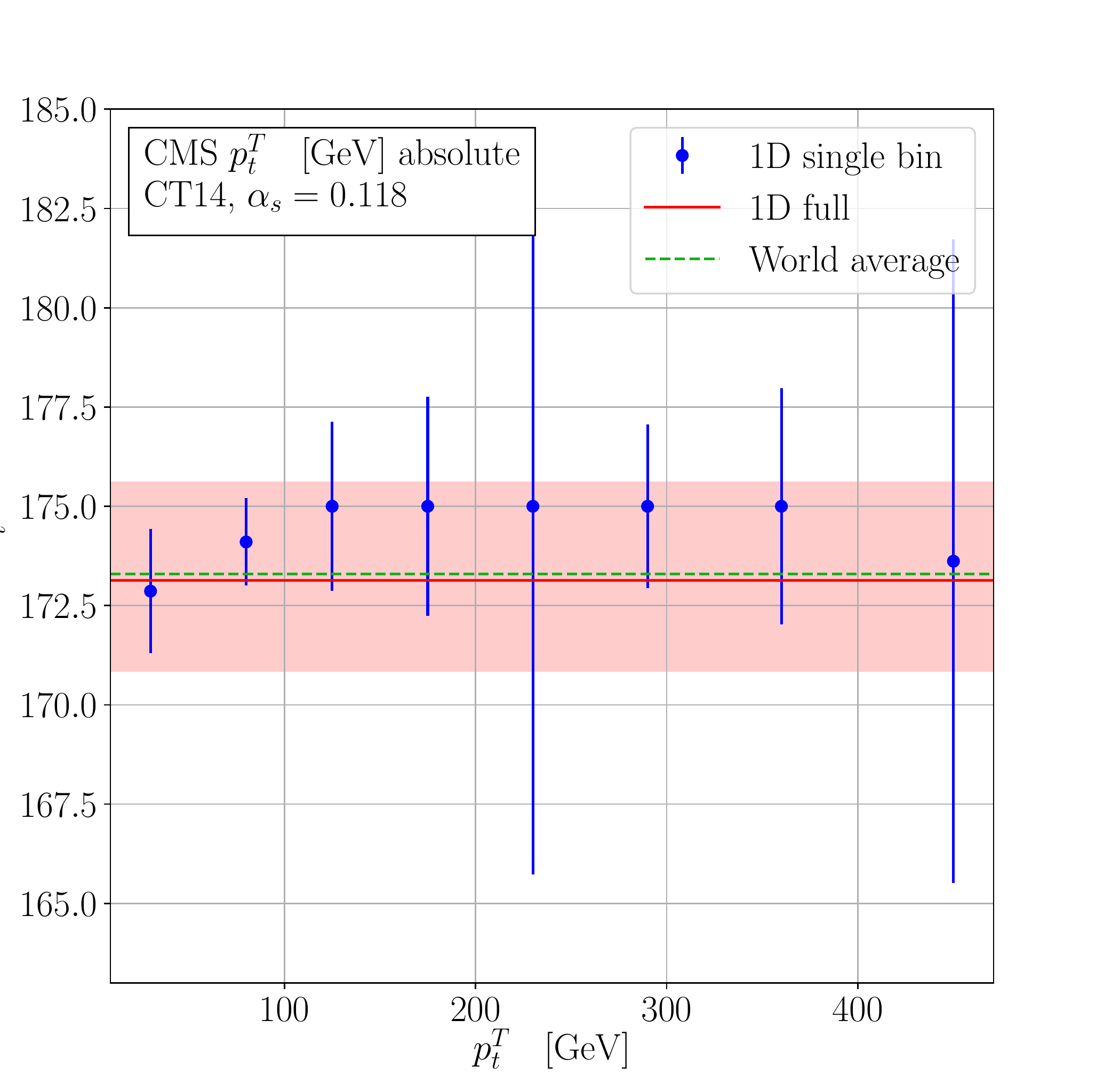}\hspace{0.4cm}
\includegraphics[trim=0.4cm 0.2cm 1.8cm 1.5cm,clip,width=0.22\textwidth]{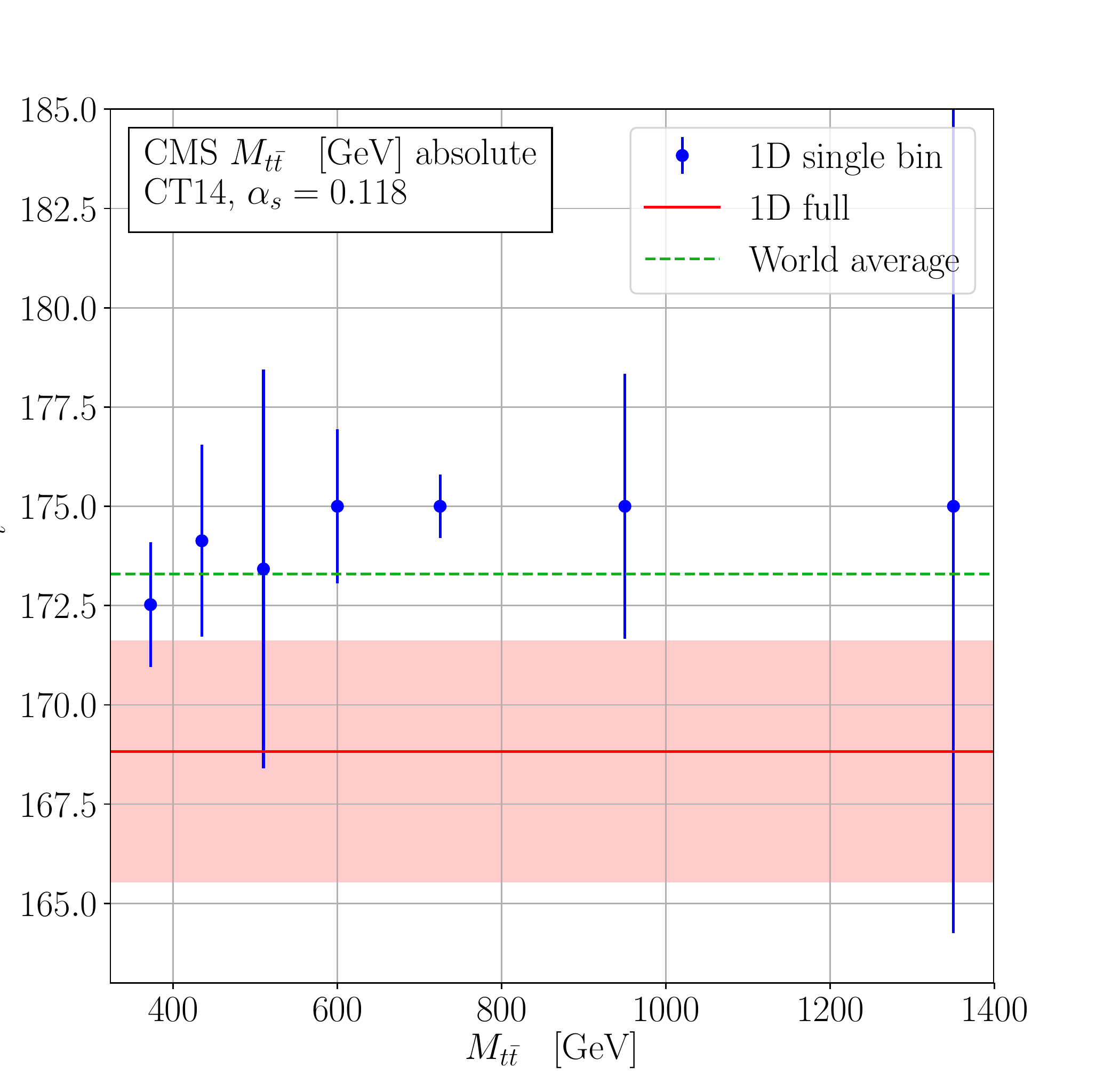} \hspace{0.4cm}
\includegraphics[trim=0.4cm 0.2cm 1.8cm 1.5cm,clip,width=0.22\textwidth]{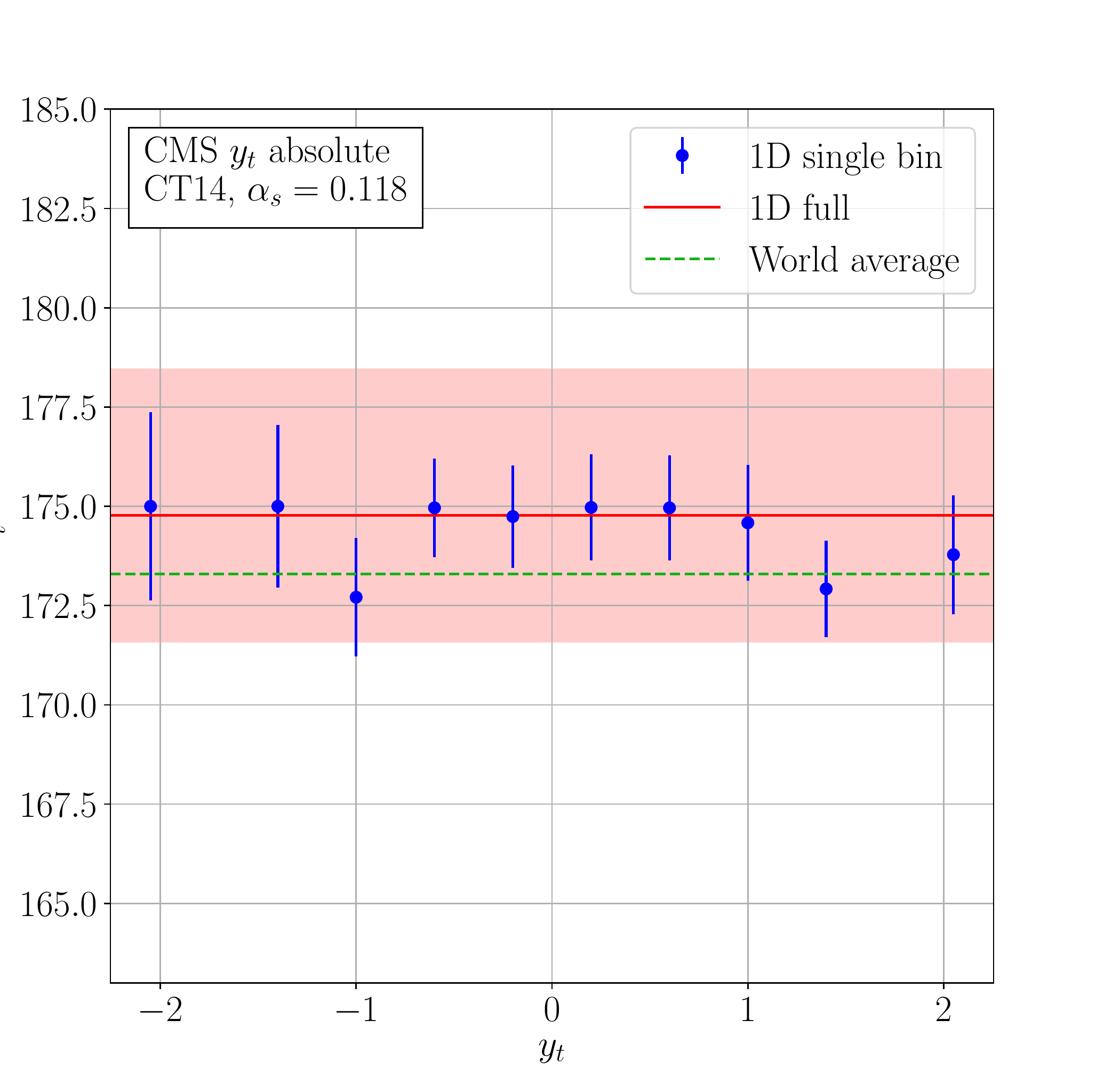} \hspace{0.4cm}
\includegraphics[trim=0.4cm 0.2cm 1.8cm 1.5cm,clip,width=0.22\textwidth]{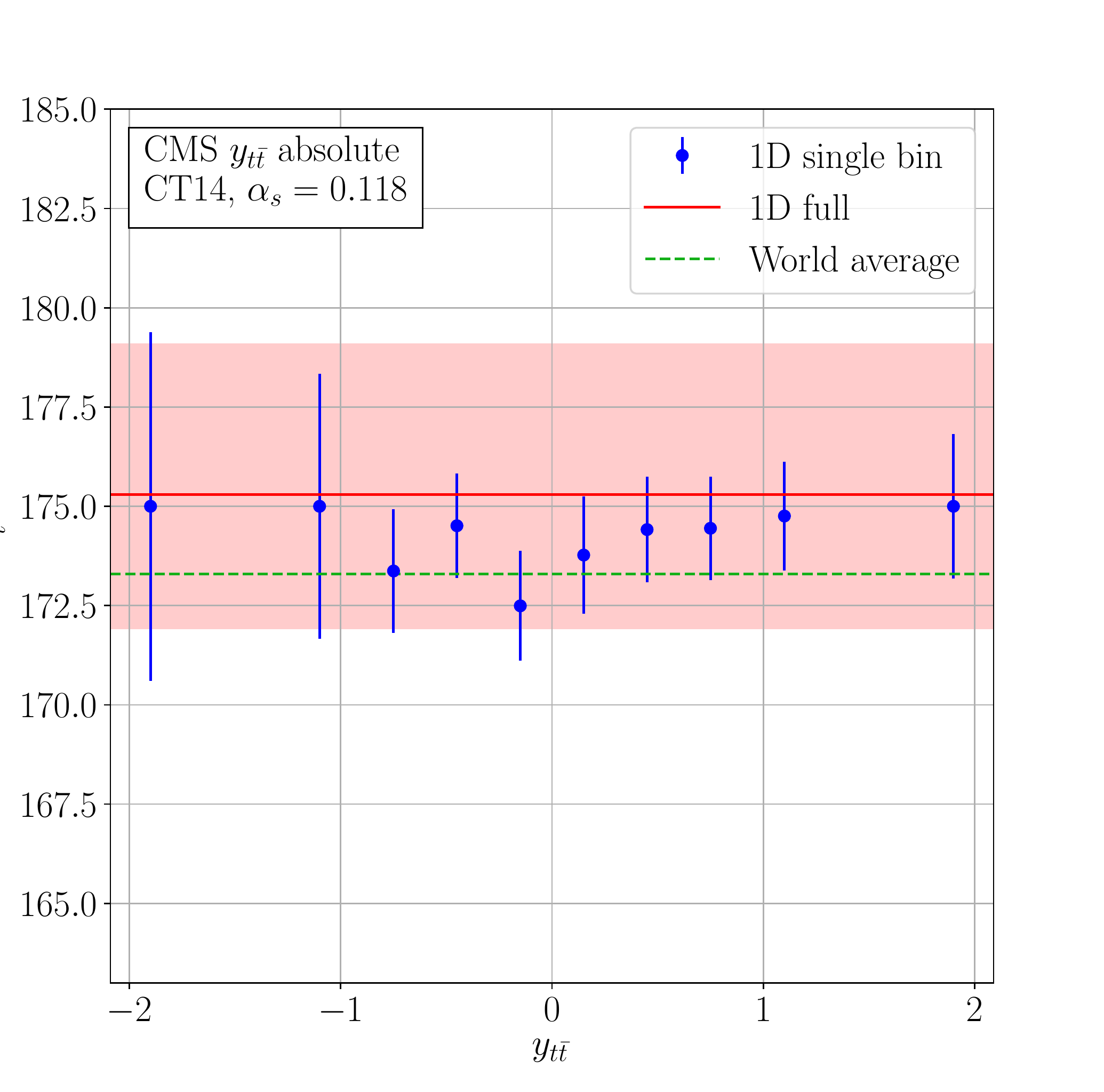}
\caption{As in fig.~\ref{fig:bin-by-bin-as} but for the extraction of $\mt$ for $\as=0.118$.}
\label{fig:bin-by-bin-mt}
\end{figure}

There are a few important features to identify in these plots.
Firstly, we note that, as expected, the uncertainties on the extracted values of
$\as$ and $\mt$ on the whole tend to be significantly larger for the individual-bin
extractions compared to the overall extractions. Since the sensitivity to $\as$ is largest in the tails of distributions (see \fig{as-sensitivity}),
if experimental uncertainties were negligible we might expect that uncertainties
on the extracted values of $\as$ would decrease if we use the weights of bins in the tails.
The fact that this is not more generally observed indicates that
experimental uncertainties are not negligible, and compete directly with the $\as$-sensitivity
to determine the overall size of uncertainties on the best-fit values.
In contrast, the extractions of $\mt$ from $\pT$ and $\mtt$ do roughly follow expectations
and the bin-by-bin extractions show increased uncertainties for bins with low $\mt$-sensitivities
(those in the tails of distributions).
This is due to the fact that the largest $\mt$ sensitivity for the absolute $\pT$ and $\mtt$
distributions lies in the bins with largest cross section and therefore smallest relative
experimental uncertainties.

Another interesting feature of these plots is the difference between
the values of $\as$ and $\mt$ extracted from individual bin weights, and the best-fit
values that result from an extraction using all bins.
If there were no correlations between different bins in a particular distribution,
then the overall extracted value would lie in the range of values from the individual-bin
extractions.
However, we observe that for some distributions, for example $\mt$ from the CMS measurements of $\mtt$, the overall extracted values
lie outside the range of individual-bin extractions.
This is a clear indication that correlations between bins (via the off-diagonal terms
in the experimental covariance matrices) can have a large effect and in some cases
shift the overall extracted value significantly. This may also reflect a d'Agostini bias
in the construction of the covariance matrices~\cite{DAgostini:1993arp}.

\subsection{Simultaneous (two-dimensional) extractions of $\as$ and $\mt$}

\subsubsection{Extraction from single distributions}

We now turn our attention to the simultaneous extraction of $\as$ and $\mt$ from differential measurements,
which uses the two-dimensional form of our $\chi^2$ objective in \eq{chisq-norm}.
The results of the joint $\as$ and $\mt$ extractions are shown in \tab{as-mt-extr-2d}.

\begin{table}[h!]
\centering

\footnotesize
\begingroup
\setlength{\tabcolsep}{3pt}
\begin{tabular}{ |c|c c c|c c c|c c c| }
\hline
\multicolumn{1}{|c|}{} & \multicolumn{9}{|c|}{ATLAS}\\
 \hline
& \multicolumn{3}{|c|}{CT14} & \multicolumn{3}{|c|}{NNPDF30} & \multicolumn{3}{|c|}{NNPDF31} \\
&   $\as$ & $\mt$ & $\chisqmin$   &   $\as$ & $\mt$ & $\chisqmin$   &   $\as$ & $\mt$ & $\chisqmin$ \\
\hline
$\pT$ &\cellcolor{Gray}$0.1171^{+0.0020}_{-0.0021}$ \
&\cellcolor{Gray}$ 174.2^{+1.0}_{-1.0}$ &\cellcolor{Gray}$0.38$ \
&\cellcolor{Gray}$0.1188^{+0.0028}_{-0.0028} $ \
&\cellcolor{Gray}$174.7^{+1.0}_{-1.0}$ &\cellcolor{Gray}$0.35$ \
&\cellcolor{Gray}$0.1210^{+0.0024}_{-0.0023}$ \
&\cellcolor{Gray}$175.6^{+1.0}_{-1.0}$ &\cellcolor{Gray}$0.43$     \\
 \hline
$\mtt$ &\cellcolor{Gray}$0.1155^{+0.0020}_{-0.0022}$ \
&\cellcolor{Gray}$ 173.1^{+0.6}_{-0.5}$ &\cellcolor{Gray}$0.16$ \
&\cellcolor{Gray}$0.1157^{+0.0027}_{-0.0027} $ \
&\cellcolor{Gray}$173.2^{+0.6}_{-0.5}$ &\cellcolor{Gray}$0.19$ \
&\cellcolor{Gray}$0.1176^{+0.0022}_{-0.0022}$ \
&\cellcolor{Gray}$173.5^{+0.5}_{-0.5}$ &\cellcolor{Gray}$0.21$     \\
\hline
$\yt$ &\cellcolor{Gray}$0.1171^{+0.0016}_{-0.0018}$ \
&\cellcolor{Gray}$ 175.0^{+1.3}_{-1.3}$ &\cellcolor{Gray}$0.86$ \
&\cellcolor{Gray}$0.1156^{+0.0018}_{-0.0018} $ \
&\cellcolor{Gray}$173.4^{+1.3}_{-1.3}$ &\cellcolor{Gray}$0.22$ \
&&&    \\
\hline
$\ytt$ &$0.1205^{+0.0017}_{-0.0019}$ \
&$176.6^{+1.3}_{-1.2}$ &$1.76$ \
&\cellcolor{Gray}$0.1150^{+0.0025}_{-0.0024}$ \
&\cellcolor{Gray}$173.0^{+1.3}_{-1.3}$ &\cellcolor{Gray}$0.57$ \
&&&     \\
\hhline{|=|===|===|===|}
Average &$0.1163^{+0.0015}_{-0.0016}$&$173.8^{+0.8}_{-0.8}$&&$0.1164^{+0.0019}_{-0.0019}$ &$173.6^{+0.8}_{-0.8}$&&$0.1192^{+0.0023}_{-0.0023}$&$174.5^{+1.2}_{-1.2}$&\\
\hline

\end{tabular} \\[10pt]

\begin{tabular}{ |c|c c c|c c c|c c c| }
\hline
\multicolumn{1}{|c|}{} & \multicolumn{9}{|c|}{CMS} \\
 \hline
& \multicolumn{3}{|c|}{CT14} & \multicolumn{3}{|c|}{NNPDF30} & \multicolumn{3}{|c|}{NNPDF31} \\
&   $\as$ & $\mt$ & $\chisqmin$   &   $\as$ & $\mt$ & $\chisqmin$   &   $\as$ & $\mt$ & $\chisqmin$ \\
\hline
$\pT$ &$0.1096^{+0.0017}_{-0.0015}$ &$ 169.0^{+0.6}_{-0.6}$ &$0.68$ \
&$0.1109^{+0.0023}_{-0.0022} $ &$170.5^{+0.6}_{-0.6}$ &$0.67$ \
&$0.1136^{+0.0020}_{-0.0021}$ &$170.7^{+0.6}_{-0.6}$ &$0.64$     \\
 \hline
$\mtt$ &$0.1108^{+0.0013}_{-0.0012}$ &$ 168.5^{+0.8}_{-0.8}$ &$4.43$ \
&$0.1055^{+0.0021}_{-0.0020} $ &$168.8^{+0.9}_{-0.9}$ &$2.02$ \
&$0.1100^{+0.0022}_{-0.0030}$ &$169.1^{+0.8}_{-0.9}$ &$2.41$     \\
\hline
$\yt$ &$0.1100^{+0.0021}_{-0.0018}$ &$ 169.7^{+1.2}_{-1.1}$ &$2.20$ \
&$0.1233^{+0.0022}_{-0.0021} $ &$175.3^{+1.1}_{-1.0}$ &$2.89$ \
&&&     \\
\hline
$\ytt$ &$0.1191^{+0.0013}_{-0.0015}$ &$ 177.0^{+1.3}_{-1.3}$ &$1.85$ \
&$0.1132^{+0.0016}_{-0.0016} $ &$171.8^{+1.2}_{-1.2}$ &$0.85$ \
&&&     \\
 \hline
\end{tabular}
\normalsize
\endgroup

\caption{
As \tab{as-extr-1d} but for the simultaneous extractions of $\mt$ and $\as$ using ATLAS (upper table) and CMS (lower table) differential distributions.
}
\label{tab:as-mt-extr-2d}
\end{table}

The first feature to emphasise is that, compared to the separate extractions in \tabs{as-extr-1d}{mt-extr-1d},
for virtually every combination of kinematic distribution and PDF set, the $\chisqmin$ value is reduced.
This indicates an improvement in the agreement between theory and data.
None of the combinations show an increase in $\chisqmin$.

\begin{figure}[t]
  \begin{centering}
  \includegraphics[trim=0.0cm 0.0cm 0.0cm 0.0cm,clip,width=0.24\textwidth]{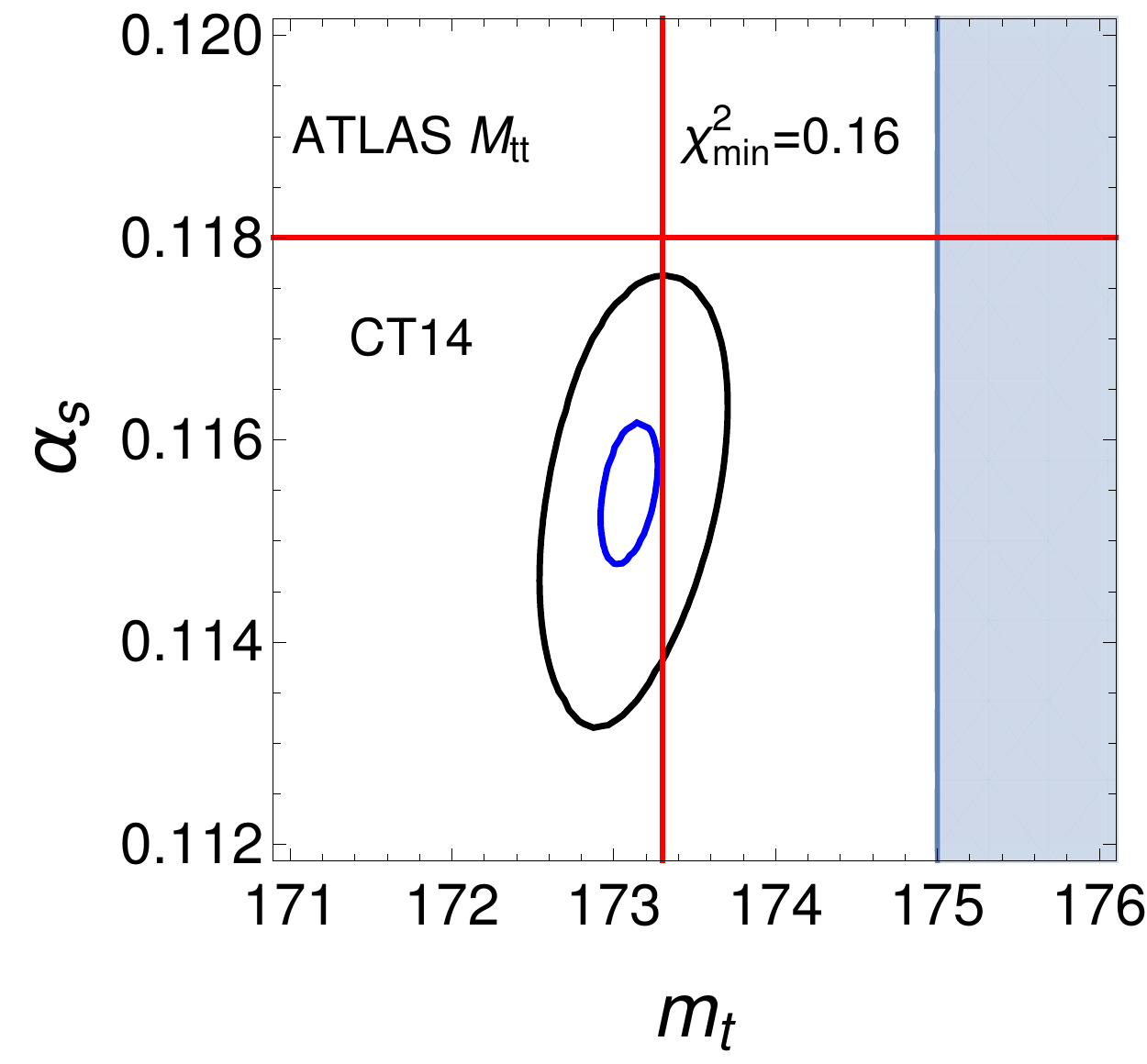}
  \includegraphics[trim=0.0cm 0.0cm 0.0cm 0.0cm,clip,width=0.24\textwidth]{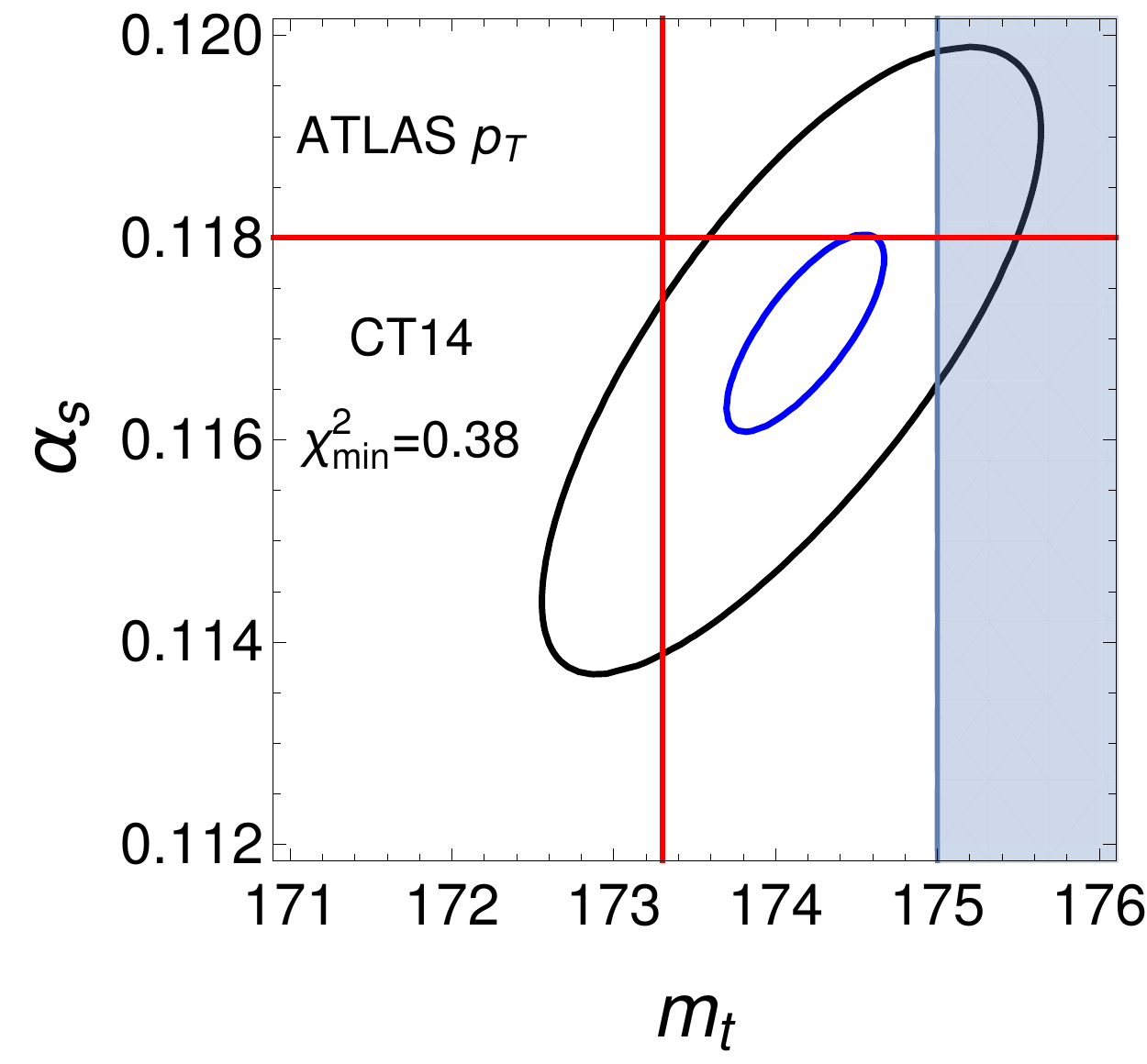}
  \includegraphics[trim=0.0cm 0.0cm 0.0cm 0.0cm,clip,width=0.24\textwidth]{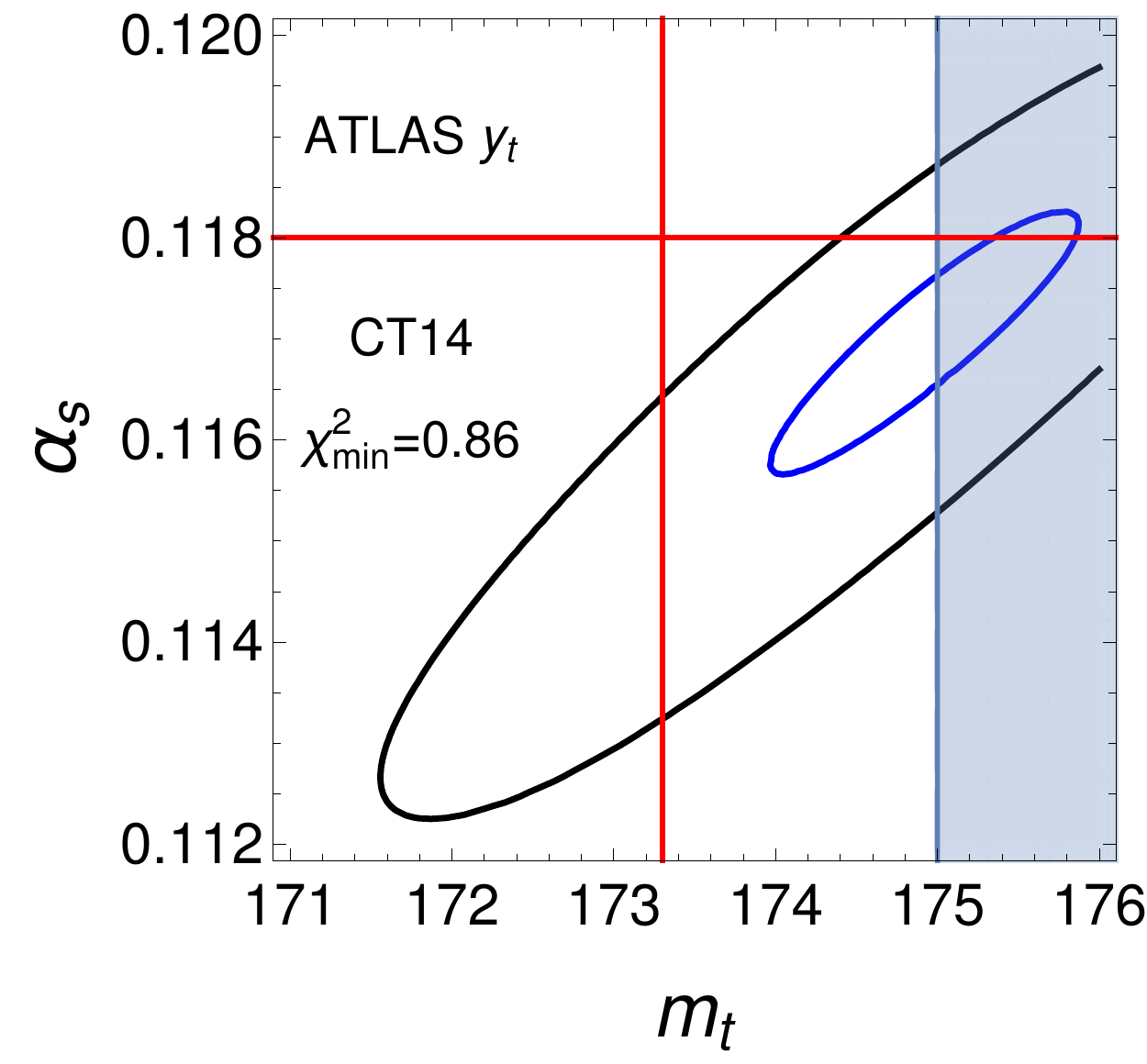}\\
\includegraphics[trim=0.0cm 0.0cm 0.0cm 0.0cm,clip,width=0.24\textwidth]{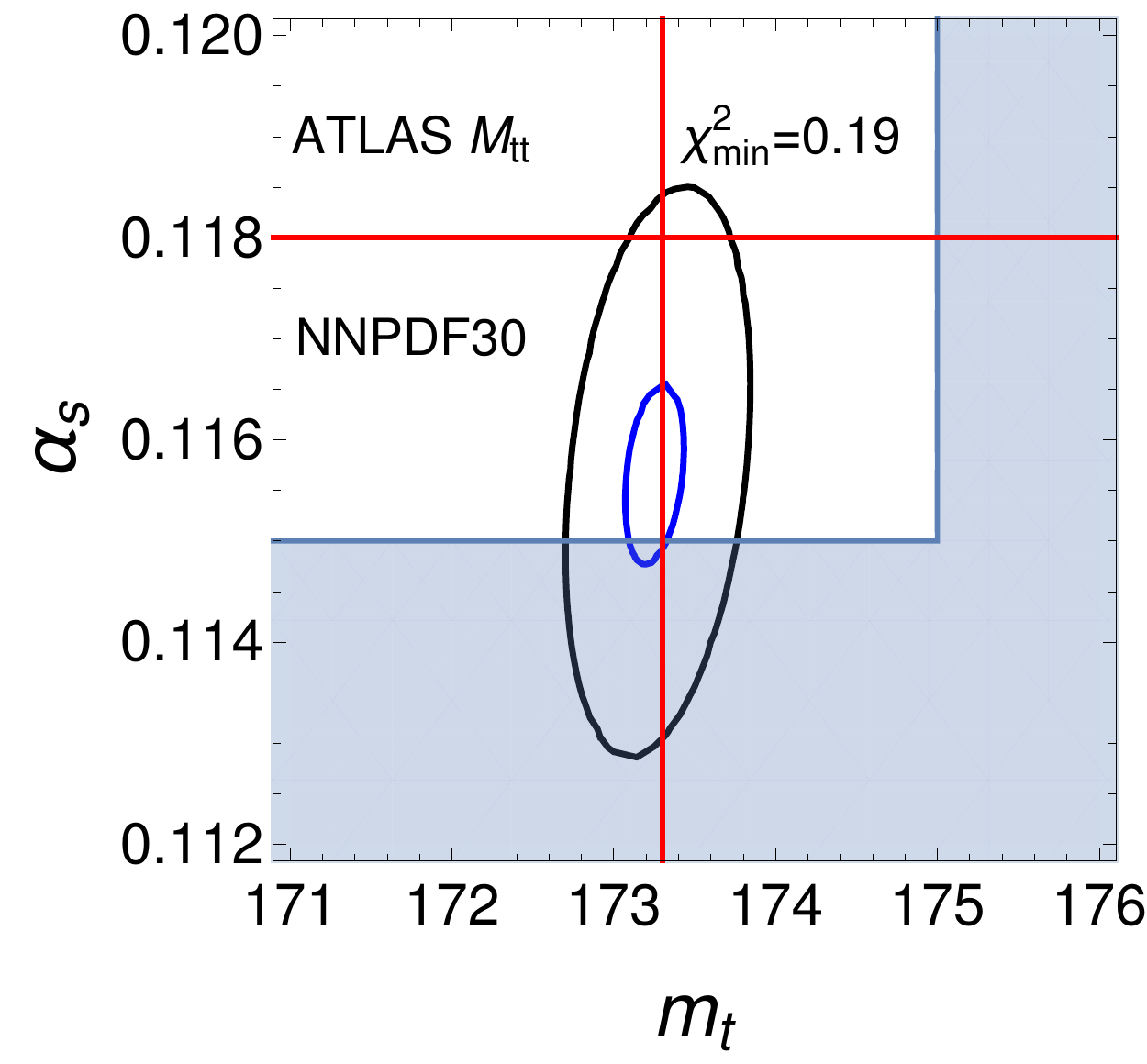}
\includegraphics[trim=0.0cm 0.0cm 0.0cm 0.0cm,clip,width=0.24\textwidth]{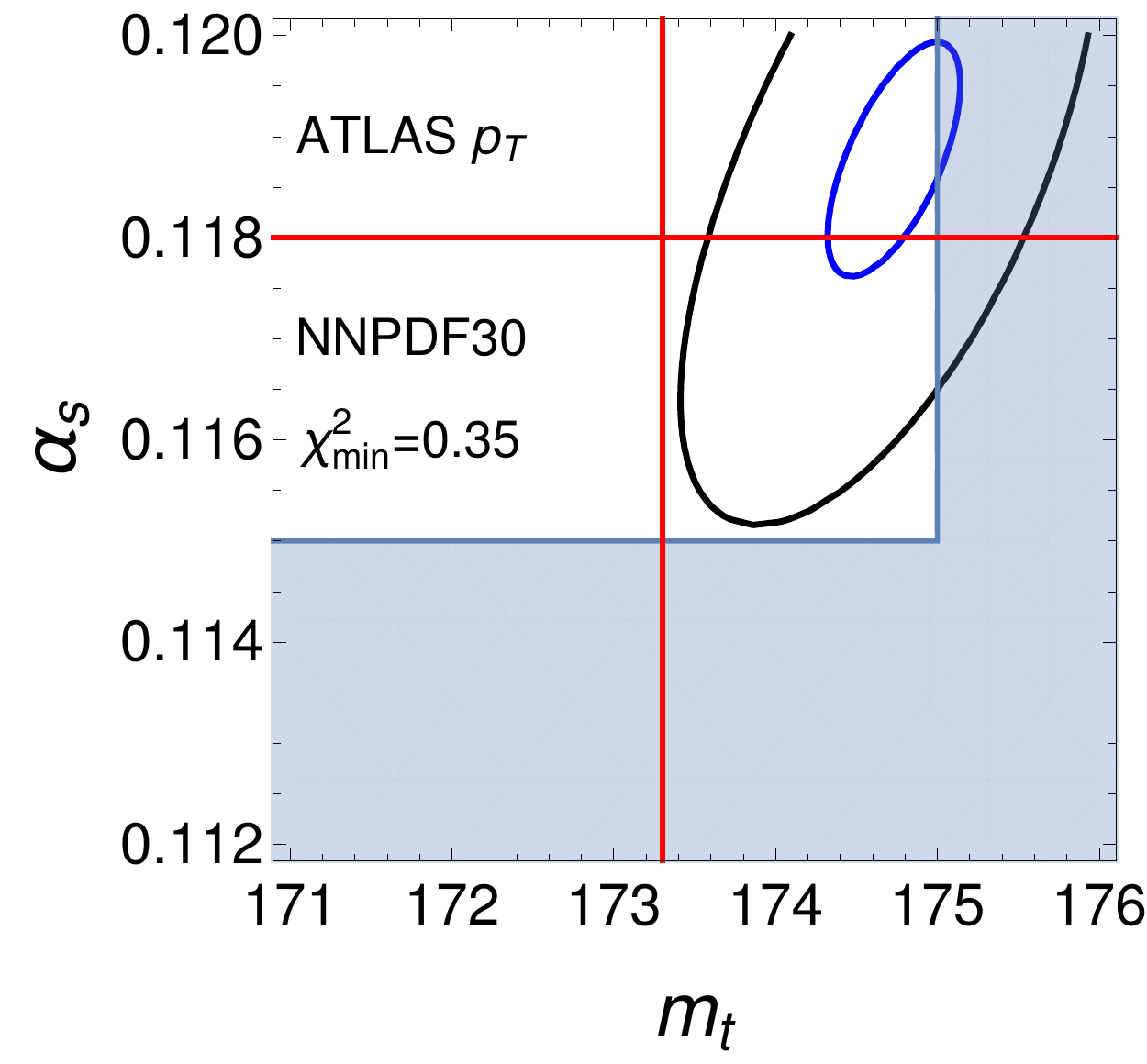}
\includegraphics[trim=0.0cm 0.0cm 0.0cm 0.0cm,clip,width=0.24\textwidth]{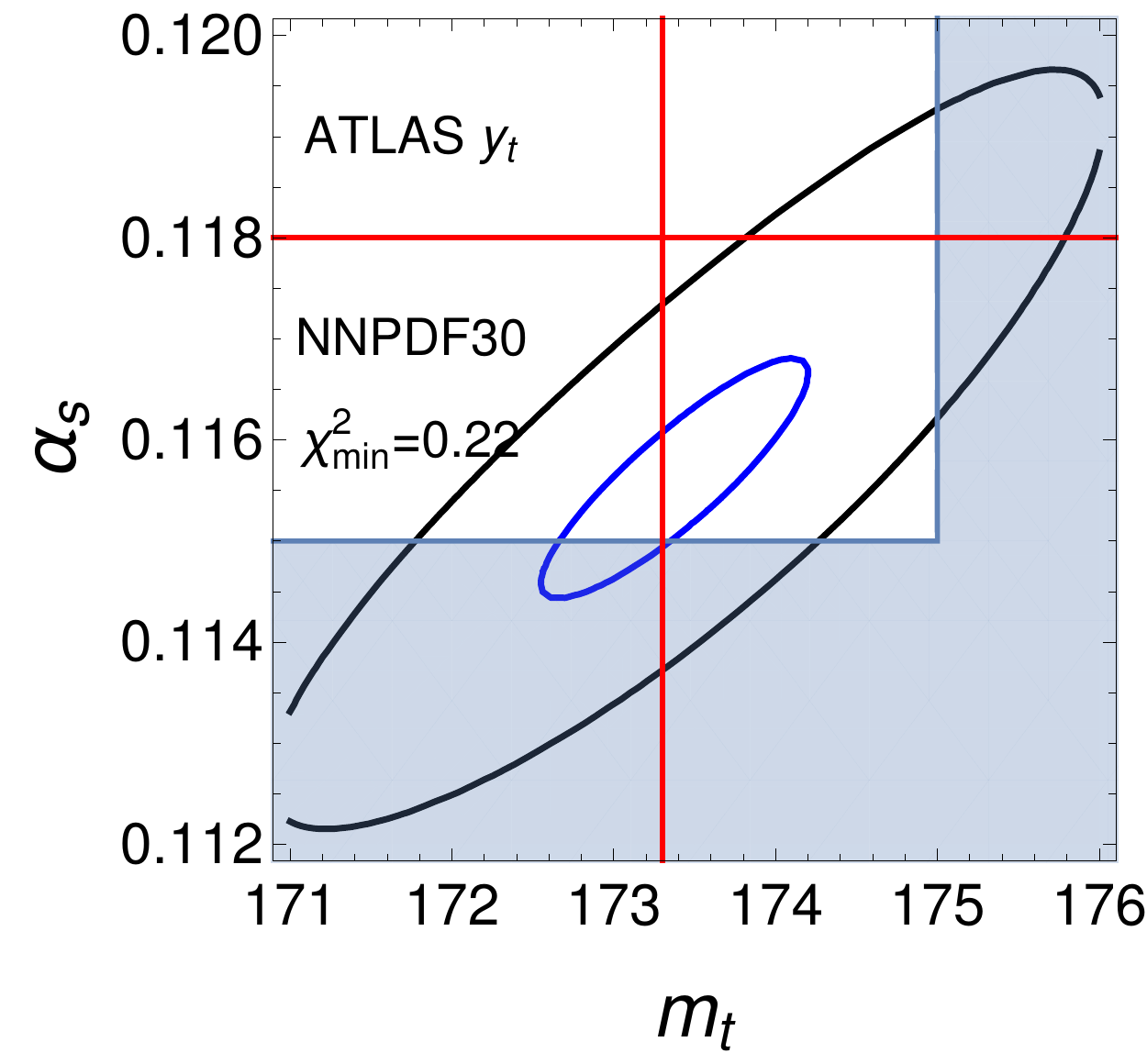} \\
\end{centering}
\caption{$\Delta\chi^2$ contour plots from a two-dimensional extraction of $\as$ and $\mt$ using ATLAS $\mtt$ (first column), ATLAS $\pT$ (second column) and ATLAS $\yt$ distributions (third column). The blue and black contours correspond to a $\Delta\chi^2$ of 0.1 and 1.0. CT14 and NNPDF3.0 PDFs have been used in the upper and lower plots respectively. The red lines denote the world average values for $\as$ and $\mt$. The white area shows the parameter space where interpolation is possible while the blue-grey area denotes the parameter space where the fits are extrapolated.}
\label{fig:2d-extraction-single-exp}
\end{figure}

In \fig{2d-extraction-single-exp} we show contour plots of $\Delta\chi^2$
corresponding to our extractions using measurements of $\pT$, $\mtt$ and $\yt$ by ATLAS.
The red vertical and horizontal lines indicate the world-average values of $\mt$ and $\as$.
We have indicated the regions in the $(\as,\mt)$-plane inside which we have `exact' NNLO theory predictions
(through interpolations of the discrete points computed) as the white area.
The outer blue-grey regions indicate the parameter space where extrapolation of the theory predictions is required.

These plots highlight that, by including the distribution shape information together with the measurements of the cross section,
it is possible to obtain joint constraints on $\as$ and $\mt$.
The particular set of contour plots shown in \fig{2d-extraction-single-exp} correspond to extractions
where the experimental data is particularly well described by the theoretical predictions for both CT14 and NNPDF3.0
PDFs.
The shapes of the $\Delta\chi^2$ contours for the extraction using the ATLAS $\yt$ data (third column) indicate
a large amount of compensation between $\as$ and $\mt$, similar to the pattern observed for the inclusive
cross section.
This is an indication that for these distributions the $\chi^2$ objective is, to a large extent, driven
by the measurements of the cross section, and that the normalised distributions provide relatively
weak constraining power.

\subsubsection{Extraction from a combination of two ATLAS distributions}
\label{sec:twoatlas}
In this section, we make use of the full information about correlations between bins made available by
the ATLAS experiment in order to extract values of the parameters from combinations of distributions. We form
covariance matrices which are block diagonal in systematic uncertainties but which contain off-diagonal entries
encoding information about the correlation of statistical uncertainties between bins of different distributions.
Information on systematic correlations between the distributions is available, but several studies have established that some decorrelation of the largest systematic uncertainties is needed to obtain acceptable PDF fits~\cite{ATLAS:2018owm, Amat:2019upj, Bailey:2019yze}. This implies that no systematic correlation between the distributions is a better approximation than full correlation. It is beyond the scope of this paper to investigate ATLAS systematic correlations further.

\begin{figure}[t]
\includegraphics[trim=0.0cm 0.0cm 0.0cm 0.0cm,clip,width=0.24\textwidth]{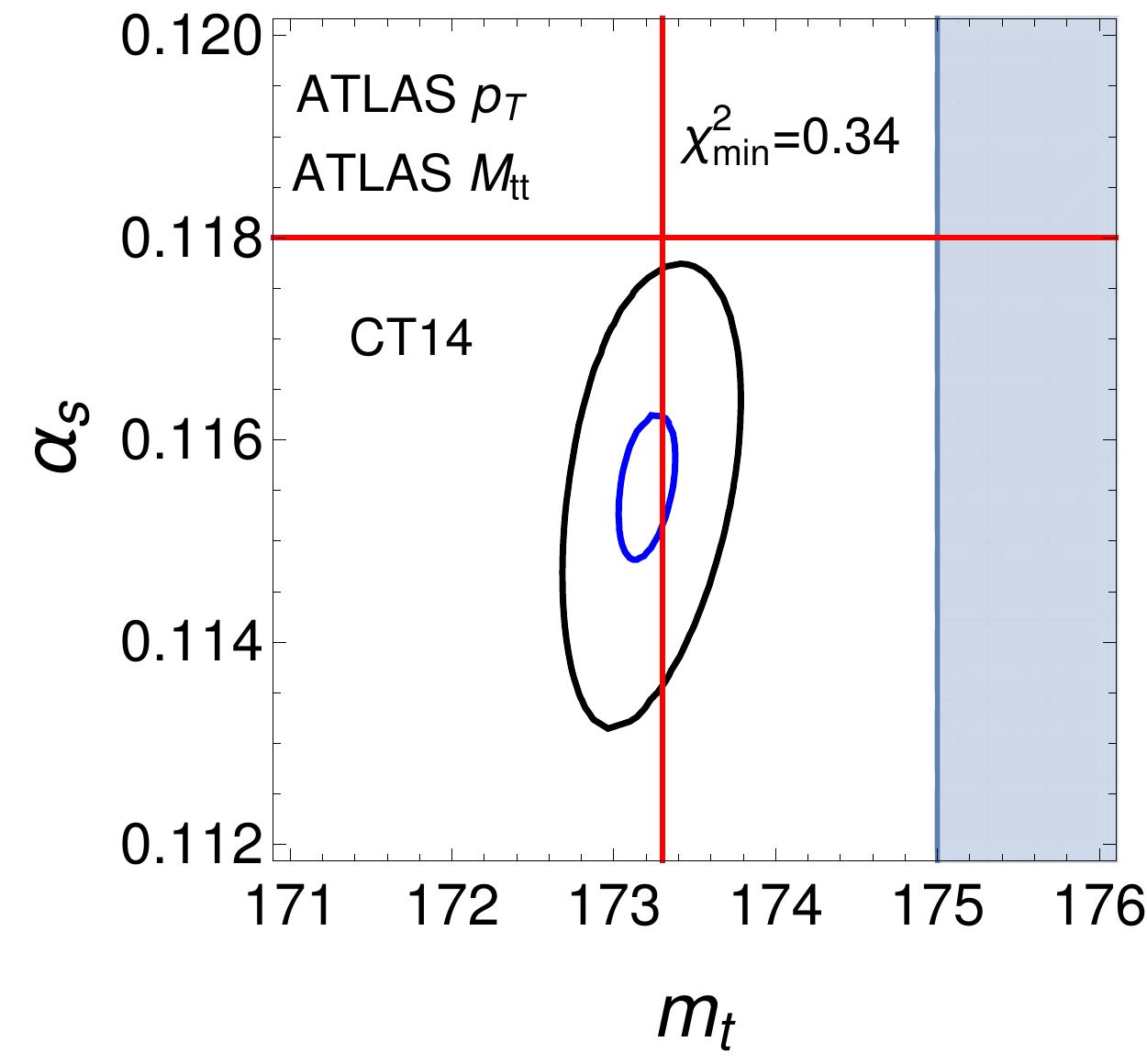}
\includegraphics[trim=0.0cm 0.0cm 0.0cm 0.0cm,clip,width=0.24\textwidth]{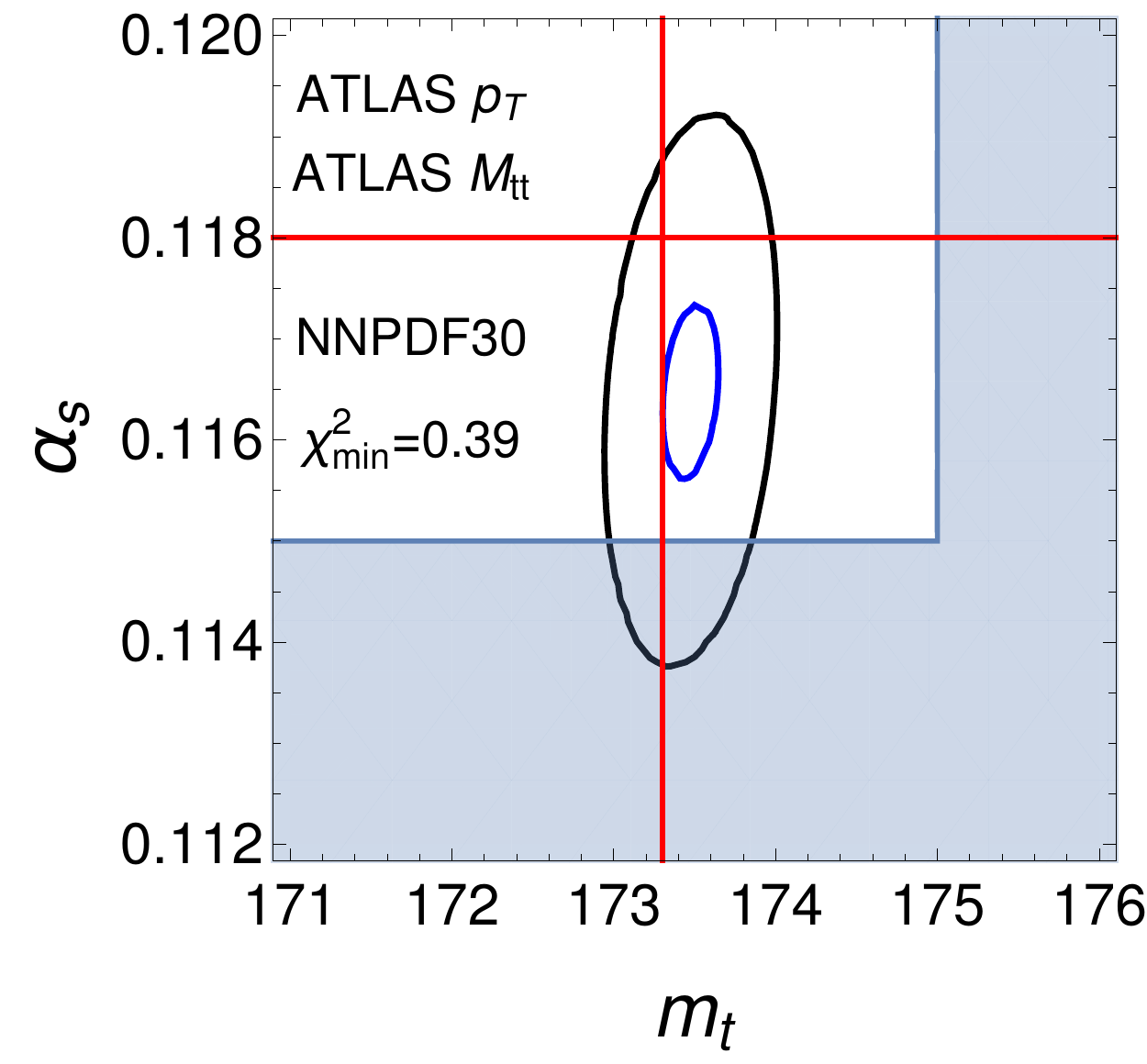}
\includegraphics[trim=0.0cm 0.0cm 0.0cm 0.0cm,clip,width=0.24\textwidth]{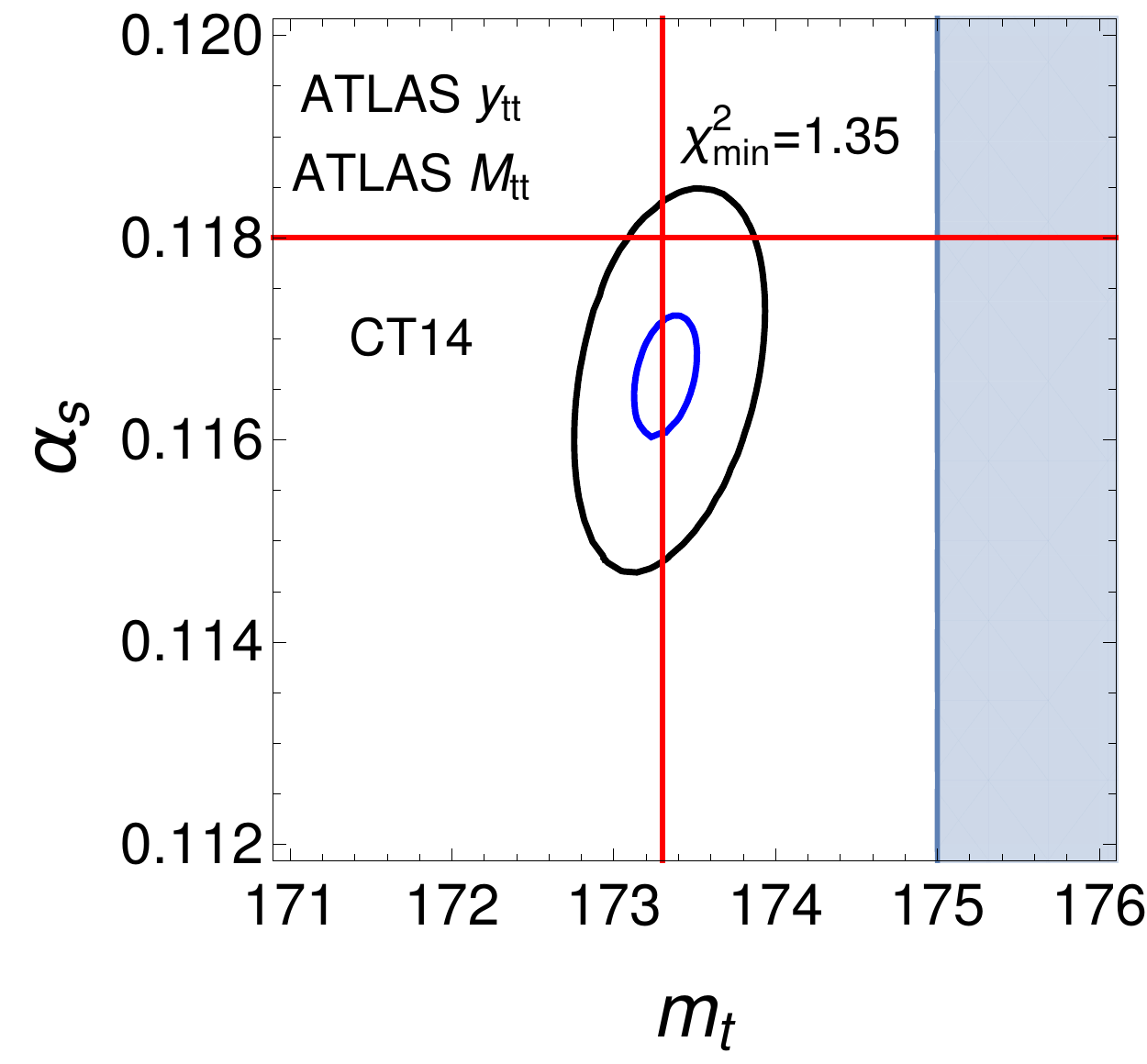}
\includegraphics[trim=0.0cm 0.0cm 0.0cm 0.0cm,clip,width=0.24\textwidth]{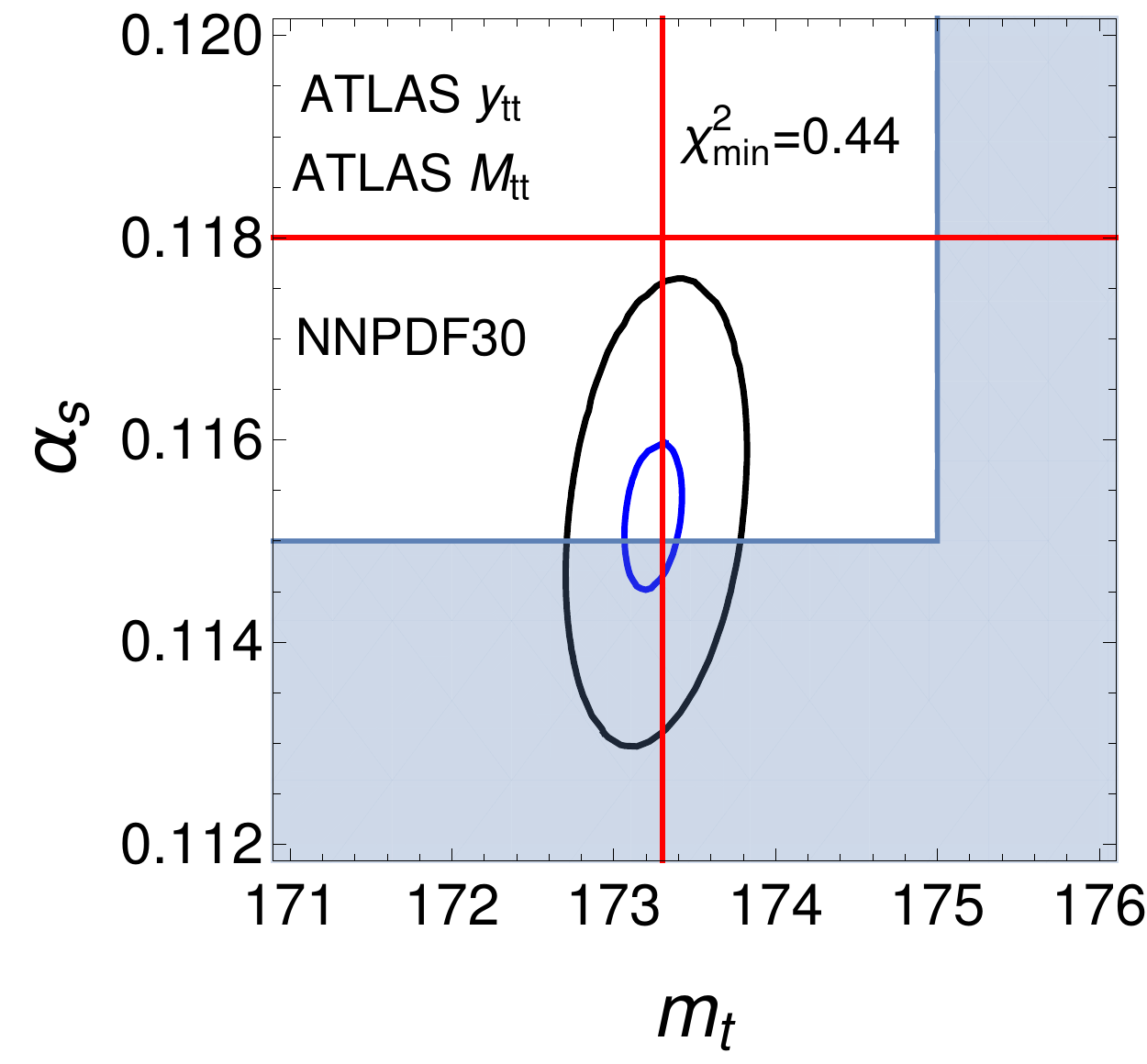} \\[-25pt]
\caption{As in fig.~\ref{fig:2d-extraction-single-exp} but for pairs of distributions from ATLAS: \{$\pT$, $\mtt$\} (left two) and \{$\mtt$, $\ytt$\} (right two). CT14 and NNPDF3.0 PDF results are shown alternately.}
\label{fig:2d-extraction-1exp2dists}
\end{figure}

We consider pairwise combinations of distributions and present in \tab{as-mt-extr-2d-1exp2dists} the
values obtained from a simultaneous extraction. Corresponding contour plots are shown in \fig{2d-extraction-1exp2dists}.

\begin{table}[t]
\centering
\footnotesize

\begingroup
\setlength{\tabcolsep}{3pt}
\begin{tabular}{ |c|c|c c c|c c c|c c c| }
\hline
ATLAS & ATLAS & \multicolumn{3}{|c|}{CT14} & \multicolumn{3}{|c|}{NNPDF30} & \multicolumn{3}{|c|}{NNPDF31} \\
1 & 2 & $\as$ & $\mt$ & $\chisqmin$   &   $\as$ & $\mt$ & $\chisqmin$   &   $\as$ & $\mt$ & $\chisqmin$ \\
\hline
$\pT$ & $\mtt$ &\cellcolor{Gray}$0.1155^{+0.0020}_{-0.0022} $ &\cellcolor{Gray}$ 173.2^{+0.5}_{-0.5}$ &\cellcolor{Gray}$ 0.34$ &\cellcolor{Gray}$0.1165^{+0.0027}_{-0.0027}$ &\cellcolor{Gray}$173.5^{+0.5}_{-0.5}$ &\cellcolor{Gray}$ 0.39$ &\cellcolor{Gray}$0.1180^{+0.0022}_{-0.0022}$ &\cellcolor{Gray}$173.8^{+0.5}_{-0.5}$ &\cellcolor{Gray}$ 0.49$  \\
\hline
$\pT$ & $y_t$ &\cellcolor{Gray}$0.1159^{+0.0016}_{-0.0017} $ &\cellcolor{Gray}$ 174.6^{+0.9}_{-1.0}$ &\cellcolor{Gray}$ 0.66$ &\cellcolor{Gray}$0.1168^{+0.0015}_{-0.0015}$ &\cellcolor{Gray}$174.5^{+0.9}_{-0.9}$ &\cellcolor{Gray}$ 0.31$ &&&  \\
\hline
$\pT$ & $\ytt$ &\cellcolor{Gray}$0.1187^{+0.0017}_{-0.0018} $ &\cellcolor{Gray}$175.1^{+0.9}_{-1.0}$ &\cellcolor{Gray}$ 1.28$ &\cellcolor{Gray}$0.1170^{+0.0024}_{-0.0023}$ &\cellcolor{Gray}$174.4^{+1.0}_{-1.0}$ &\cellcolor{Gray}$ 0.53$ &&&  \\
\hline
$\yt$ & $\mtt$ &  $0.1142^{+0.0017}_{-0.0017} $ &  $173.2^{+0.6}_{-0.5}$ &  $ 0.63$ &\cellcolor{Gray}$0.1155^{+0.0015}_{-0.0014}$ &\cellcolor{Gray}$173.3^{+0.6}_{-0.5}$ &\cellcolor{Gray}$ 0.22$ &&&  \\
\hline
$\ytt$ & $\mtt$ &\cellcolor{Gray}$0.1166^{+0.0017}_{-0.0018} $ &\cellcolor{Gray}$ 173.3^{+0.6}_{-0.5}$ &\cellcolor{Gray}$ 1.35$ &\cellcolor{Gray}$0.1152^{+0.0023}_{-0.0022}$ &\cellcolor{Gray}$173.2^{+0.6}_{-0.5}$ &\cellcolor{Gray}$ 0.44$ &&&  \\
\hline
$\yt$ & $\ytt$ &  $0.1182^{+0.0013}_{-0.0014} $ &  $176.4^{+1.0}_{-1.0}$ &  $ 1.32$ &\cellcolor{Gray}$0.1156^{+0.0014}_{-0.0014}$ &\cellcolor{Gray}$173.5^{+1.1}_{-1.2}$ &\cellcolor{Gray}$ 0.40$ &&&  \\
\hhline{|==|===|===|===|}
\multicolumn{2}{|c|}{Average}&$0.1159^{+0.0013}_{-0.0014}$&$173.8^{+0.8}_{-0.8}$&&$0.1161^{+0.0011}_{-0.0010}$ &$173.7^{+0.6}_{-0.6}$&&$0.1180^{+0.0022}_{-0.0022}$&$173.8^{+0.5}_{-0.5}$&\\
\hline

\end{tabular}

\endgroup

\normalsize
\caption{
As in \tab{as-extr-1d} but for the simultaneous extractions of $\mt$ and $\as$ from two ATLAS distributions. Measured observables has been combined with appropriate correlations included.
}
\label{tab:as-mt-extr-2d-1exp2dists}
\end{table}

We notice again that the inclusion of $\mtt$ data seems to lower the extracted value of $\mt$ by about 1 GeV compared to combinations where it is not included and does so consistently across PDF sets.

\subsubsection{Extraction from a combination of one ATLAS and one CMS distribution}

The final extraction we discuss in this section is one that is based on using measurements
of two normalised distributions, one from ATLAS and one from CMS. 
As mentioned in \sec{methodology}, we have performed these extractions assuming that 
the correlations between measurements of the two experiments are negligible. 
While this exercise may not be fully complete in terms of including 
uncertainties such as luminosity systematics, it illustrates the potential of exploiting 
measurements from \emph{both} experiments. Unlike the other ATLAS extractions presented thus far,
the covariance matrices used in this section are diagonal in statistical uncertainties
in order to match those used for the CMS extractions. 
For completeness we present results for all 16 possible combinations of pairs of kinematic 
distributions from the two experiments -- however, we mainly focus our attention and draw insights 
from the combinations where \emph{different} distributions are combined (avoiding potential 
shared systematics that may enter in measuring a given distribution). 
The extractions we present are two-dimensional extractions based on the definition of 
$\chi^2$ given in \eq{chi2-norm-2exps}.
The results of these extractions are shown in \tab{as-mt-extr-2d-2exp}.

\begin{table}[t]
\centering
\footnotesize

\begingroup
\setlength{\tabcolsep}{3pt}
\begin{tabular}{ |c|c|c c c|c c c|c c c| }
\hline
ATLAS & CMS & \multicolumn{3}{|c|}{CT14} & \multicolumn{3}{|c|}{NNPDF30} & \multicolumn{3}{|c|}{NNPDF31} \\
& & $\as$ & $\mt$ & $\chisqmin$   &   $\as$ & $\mt$ & $\chisqmin$   &   $\as$ & $\mt$ & $\chisqmin$ \\
\hline
$\pT$ & $\pT$ &$0.1142^{+0.0014}_{-0.0014} $ &$ 172.1^{+0.4}_{-0.4}$&$ 2.64$ &$0.1148^{+0.0017}_{-0.0017}$ &$ 172.7^{+0.4}_{-0.4}$ &$2.60$ &\cellcolor{Gray}$0.1176^{+0.0015}_{-0.0015}$&\cellcolor{Gray}$ 173.3^{+0.4}_{-0.4}$ &\cellcolor{Gray}$ 2.57$  \\
\hline
$\pT$ & $\mtt$ &$0.1135^{+0.0013}_{-0.0013} $ &$ 172.3^{+0.3}_{-0.4}$&$ 4.61$ &$0.1129^{+0.0017}_{-0.0017}$ &$ 172.6^{+0.3}_{-0.3}$ &$4.28$ &\cellcolor{Gray}$0.1158^{+0.0015}_{-0.0015}$&\cellcolor{Gray}$ 172.9^{+0.4}_{-0.3}$ &\cellcolor{Gray}$ 4.14$  \\
\hline
$\pT$ & $\yt$ &\cellcolor{Gray}$0.1178^{+0.0013}_{-0.0014} $&\cellcolor{Gray}$ 174.4^{+0.5}_{-0.5}$ &\cellcolor{Gray}$ 1.62$&\cellcolor{Gray}$0.1209^{+0.0016}_{-0.0016}$ &\cellcolor{Gray}$174.8^{+0.5}_{-0.5}$ &\cellcolor{Gray}$ 1.93$&&&  \\
\hline
$\pT$ & $\ytt$ &\cellcolor{Gray}$0.1172^{+0.0012}_{-0.0013} $&\cellcolor{Gray}$ 174.7^{+0.5}_{-0.5}$ &\cellcolor{Gray}$ 1.33$&\cellcolor{Gray}$0.1173^{+0.0015}_{-0.0015}$ &\cellcolor{Gray}$174.7^{+0.5}_{-0.5}$ &\cellcolor{Gray}$ 0.78$&&&  \\
\hline
$\mtt$ & $\pT$ &\cellcolor{Gray}$0.1162^{+0.0012}_{-0.0013} $&\cellcolor{Gray}$ 172.9^{+0.2}_{-0.2}$ &\cellcolor{Gray}$ 3.06$&\cellcolor{Gray}$0.1158^{+0.0017}_{-0.0017}$ &\cellcolor{Gray}$173.0^{+0.2}_{-0.2}$ &\cellcolor{Gray}$ 2.87$&\cellcolor{Gray}$0.1186^{+0.0015}_{-0.0015}$ &\cellcolor{Gray}$173.2^{+0.2}_{-0.2}$ &\cellcolor{Gray}$ 2.93$  \\
\hline
$\mtt$ & $\mtt$ &$0.1148^{+0.0013}_{-0.0013} $ &$172.9^{+0.2}_{-0.2}$ &$ 5.25$ &$0.1140^{+0.0017}_{-0.0017}$ &$172.9^{+0.2}_{-0.2}$ &$ 4.79$&\cellcolor{Gray}$0.1172^{+0.0015}_{-0.0015}$ &\cellcolor{Gray}$173.2^{+0.2}_{-0.2}$ &\cellcolor{Gray}$ 4.75$  \\
\hline
$\mtt$ & $\yt$ &\cellcolor{Gray}$0.1168^{+0.0014}_{-0.0015} $&\cellcolor{Gray}$ 173.2^{+0.2}_{-0.2}$ &\cellcolor{Gray}$ 1.94$&\cellcolor{Gray}$0.1198^{+0.0015}_{-0.0015}$ &\cellcolor{Gray}$173.3^{+0.2}_{-0.2}$ &\cellcolor{Gray}$ 2.50$&&&  \\
\hline
$\mtt$ & $\ytt$ &\cellcolor{Gray}$0.1161^{+0.0013}_{-0.0014} $&\cellcolor{Gray}$ 173.3^{+0.2}_{-0.2}$ &\cellcolor{Gray}$ 1.80$&\cellcolor{Gray}$0.1166^{+0.0014}_{-0.0014}$ &\cellcolor{Gray}$173.3^{+0.2}_{-0.2}$ &\cellcolor{Gray}$ 1.36$&&&  \\
\hline
$\yt$ & $\pT$ &$0.1128^{+0.0009}_{-0.0008} $ &$ 170.8^{+0.5}_{-0.5}$&$ 5.83$ &$0.1109^{+0.0010}_{-0.0010}$ &$ 170.5^{+0.5}_{-0.5}$ &$1.11$ &&&  \\
\hline
$\yt$ & $\mtt$ &$0.1127^{+0.0008}_{-0.0008} $ &$ 170.9^{+0.6}_{-0.6}$&$ 8.71$ &$0.1091^{+0.0009}_{-0.0009}$ &$ 169.3^{+0.7}_{-0.7}$ &$2.34$ &&&  \\
\hline
$\yt$ & $\yt$ &$0.1175^{+0.0010}_{-0.0011} $ &$ 176.4^{+0.8}_{-0.8}$&$ 4.79$ &$0.1142^{+0.0010}_{-0.0010}$ &$ 172.4^{+0.8}_{-0.8}$ &$3.24$ &&& \\
\hline
$\yt$ & $\ytt$ &$0.1186^{+0.0008}_{-0.0009} $ &$ 178.6^{+0.8}_{-0.8}$&$ 3.71$ &$0.1120^{+0.0009}_{-0.0009}$ &$ 171.4^{+0.9}_{-0.9}$ &$1.19$ &&&  \\
\hline
$\ytt$ & $\pT$ &$0.1140^{+0.0008}_{-0.0008} $ &$ 171.7^{+0.5}_{-0.5}$&$ 11.00$ &$0.1115^{+0.0008}_{-0.0008}$ &$ 170.7^{+0.5}_{-0.5}$ &$1.76$ &&& \\
\hline
$\ytt$ & $\mtt$ &$0.1139^{+0.0008}_{-0.0008} $ &$172.1^{+0.5}_{-0.7}$ &$ 14.27$ &$0.1105^{+0.0008}_{-0.0008}$ &$169.9^{+0.6}_{-0.6}$ &$ 3.29$ &&&  \\
\hline
$\ytt$ & $\yt$ &$0.1176^{+0.0009}_{-0.0009} $ &$ 176.7^{+0.7}_{-0.7}$&$ 8.11$ &$0.1134^{+0.0008}_{-0.0008}$ &$ 172.0^{+0.8}_{-0.8}$ &$3.74$ &&&  \\
\hline
$\ytt$ & $\ytt$ &$0.1179^{+0.0008}_{-0.0009} $ &$178.0^{+0.7}_{-0.7}$ &$ 7.03$ &$0.1125^{+0.0008}_{-0.0008}$ &$172.0^{+0.8}_{-0.8}$ &$ 1.66$&&&  \\
\hhline{|==|===|===|===|}
\multicolumn{2}{|c|}{Average}&$0.1172^{+0.0009}_{-0.0010}$&$174.5^{+0.5}_{-0.5}$&&$0.1172^{+0.0013}_{-0.0013}$ &$174.5^{+0.5}_{-0.5}$&&$0.1178^{+0.0013}_{-0.0013}$&$173.3^{+0.3}_{-0.3}$&\\
\hline
\end{tabular}
\endgroup

\normalsize
\caption{
As in \tab{as-extr-1d} but for the simultaneous extractions of $\mt$ and $\as$ from one ATLAS and one CMS differential distribution. Data from the two experiments has been combined assuming no correlations between distributions.}
\label{tab:as-mt-extr-2d-2exp}
\end{table}
\begin{figure}[t]
\includegraphics[trim=0.0cm 0.0cm 0.0cm 0.0cm,clip,width=0.24\textwidth]{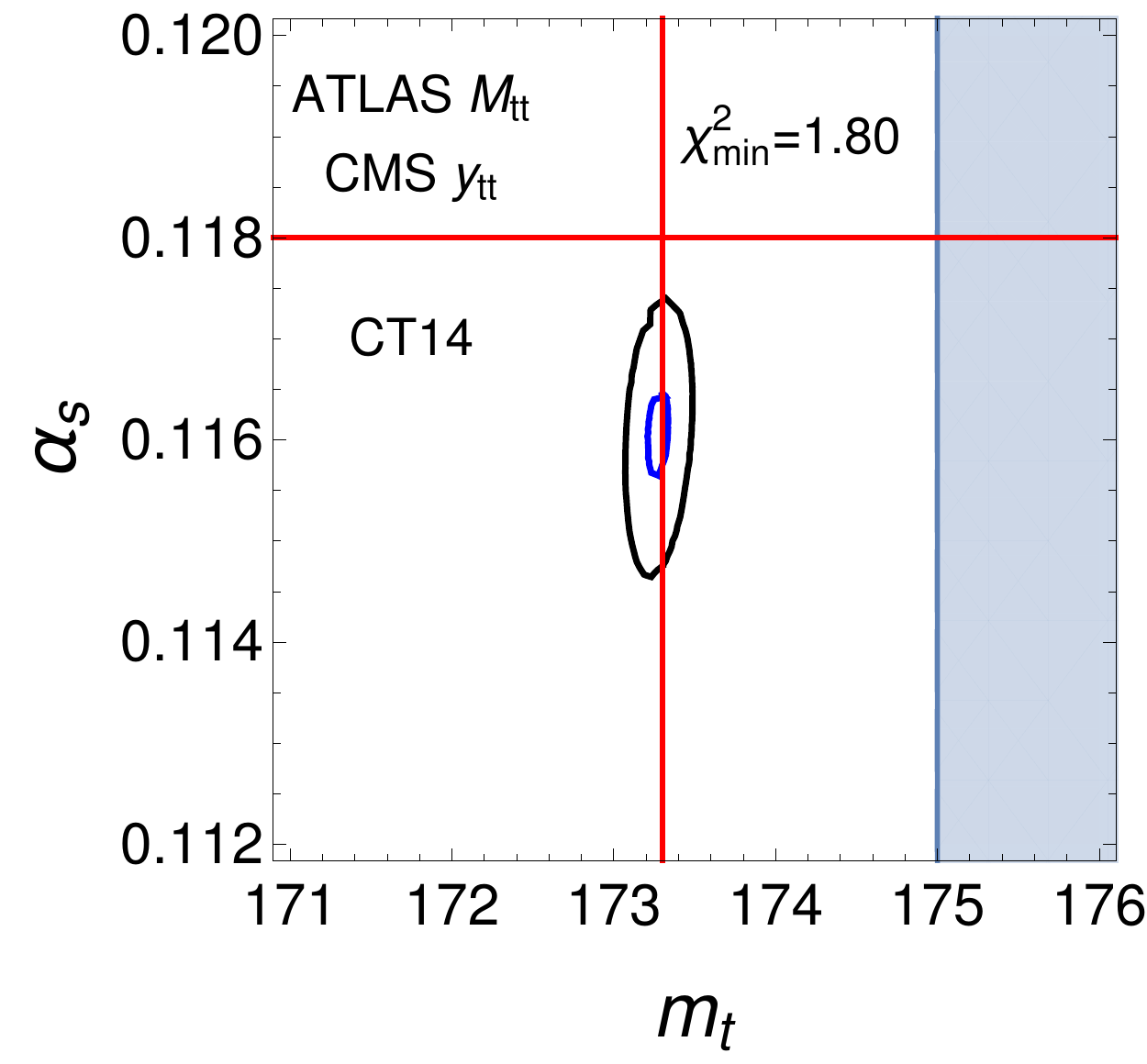}
\includegraphics[trim=0.0cm 0.0cm 0.0cm 0.0cm,clip,width=0.24\textwidth]{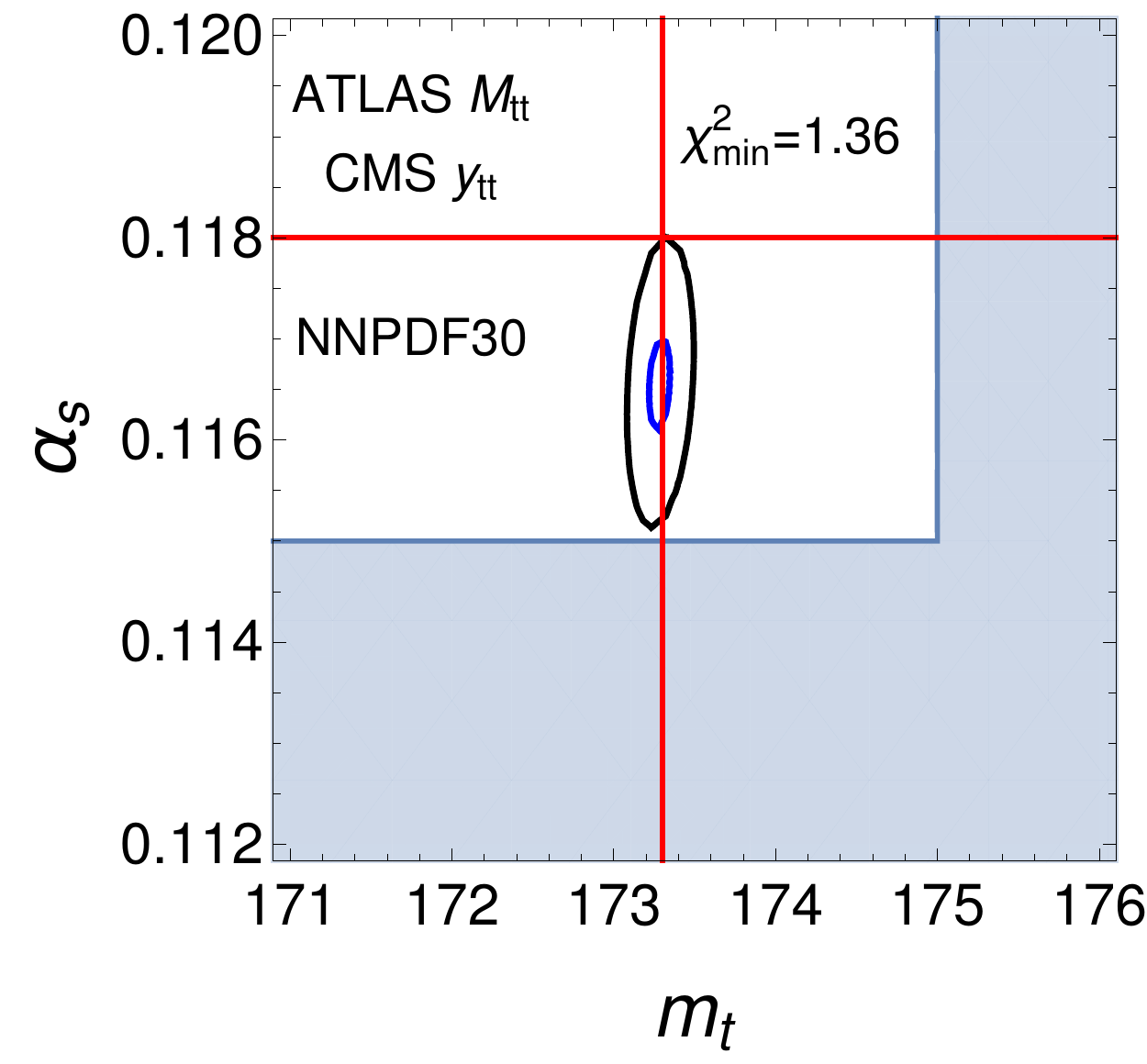}\
\includegraphics[trim=0.0cm 0.0cm 0.0cm 0.0cm,clip,width=0.24\textwidth]{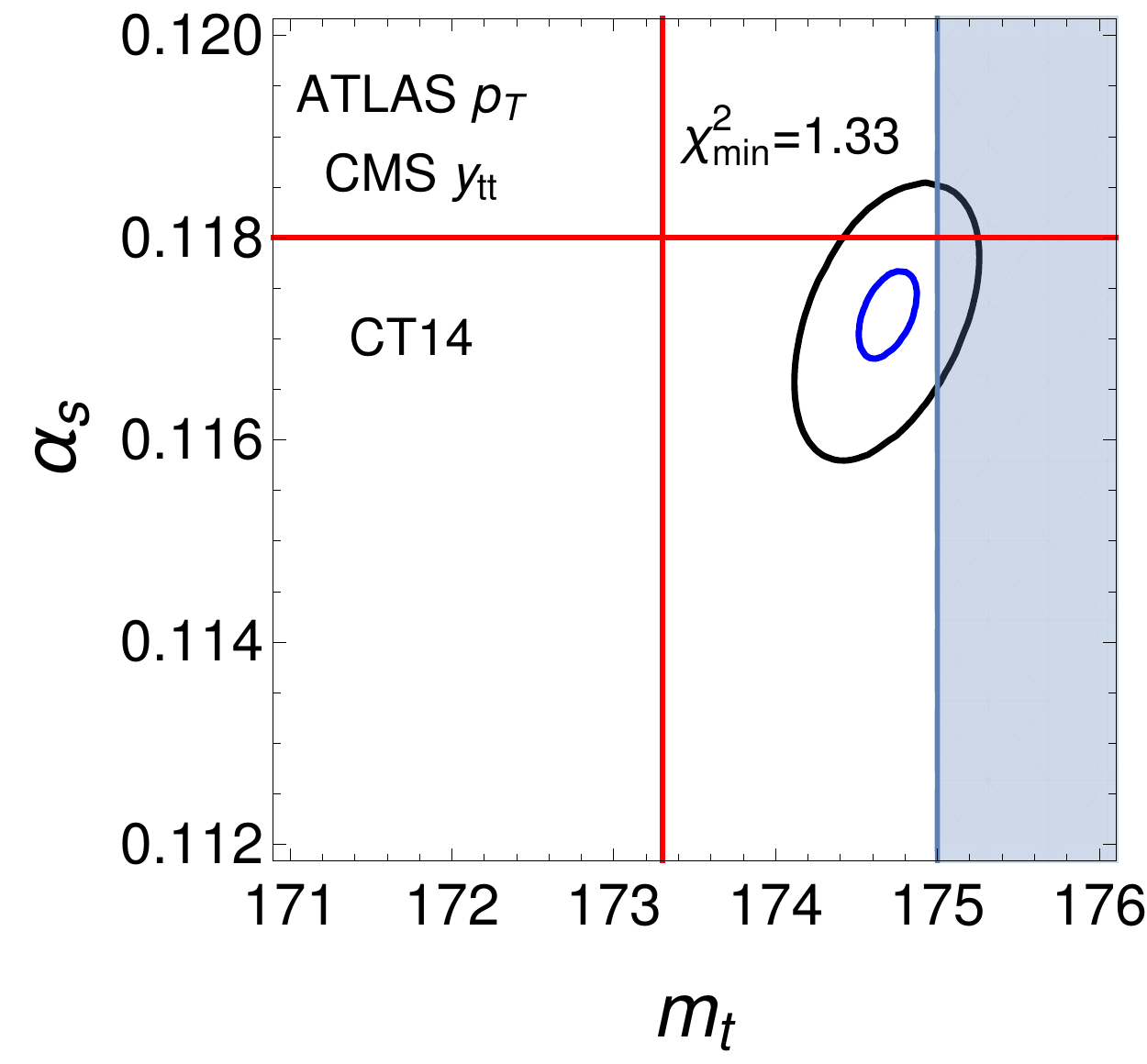}
\includegraphics[trim=0.0cm 0.0cm 0.0cm 0.0cm,clip,width=0.24\textwidth]{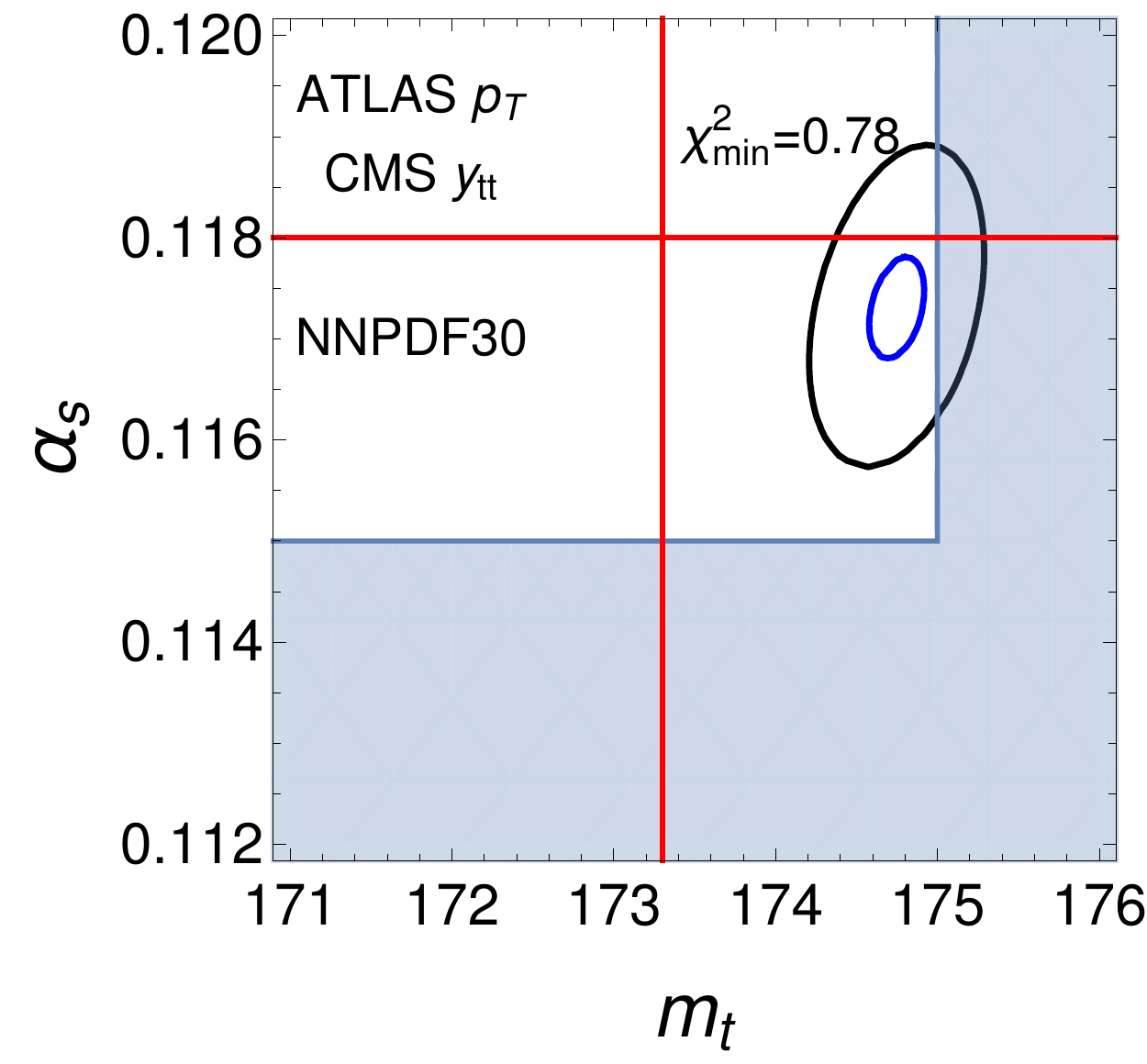} \\[-25pt]
\caption{As in fig.~\ref{fig:2d-extraction-single-exp} but for pairs of distributions from \{ATLAS, CMS\}: \{$\mtt$, $\ytt$\} (left two) and \{$\pT$, $\ytt$\} (right two). CT14 and NNPDF3.0 PDF results are shown alternately.}
\label{fig:2d-extraction-both-exp}
\end{figure}

Examining \tab{as-mt-extr-2d-2exp}, we note that compared to the other extractions presented so far,
the $\chisqmin$ values returned by the combinations of ATLAS and CMS data are rather large.
Indeed, taking for example the ATLAS $\ytt$ and CMS $\mtt$ distributions together, with the CT14 PDF set one obtains the highest value of $\chisqmin$
encountered thus far, at 14.27. These large values clearly indicate a poor description of the data by the theory,
both across PDF sets and combinations of distributions.

Bearing this in mind, we note that the
combination of distributions yielding the best results for the CT14 and NNPDF3.0 PDF sets are
$\pT$ or $\mtt$ from ATLAS combined with $\ytt$ from CMS. This observation is consistent with the results of the simultaneous extractions of $\as$ and $\mt$ from each individual experiment (discussed in the previous section and tabulated in \tab{as-mt-extr-2d}).
These two combinations give best-fit values of $\as$ and $\mt$ that differ by about 1\% -- in the case of
the $\as$ values, these remain fully consistent within their uncertainties but for $\mt$ the values lie
slightly outside the $1\sigma$ interval.

In \fig{2d-extraction-both-exp} we show the $\Delta\chi^2$ contours for the extractions
based on the combinations of distributions \{$\pT$, $\ytt$\} and \{$\mtt$, $\ytt$\}.
Compared to the contour plots arising from other extractions, we do see reduced uncertainties despite the
slightly higher $\chisqmin$ values, particularly in the $\mt$ direction.

We comment further on the combined CMS and ATLAS data extractions in \sec{summary}.

\section{Discussion}\label{sec:discussion}

\subsection{Averaging of $\as$ and $\mt$ extractions}\label{sec:averaging}

A large number of $\as$ and/or $\mt$ extractions were presented in \sec{results} and appendices \ref{app:A},\ref{app:B},\ref{app:C}. Upon reviewing them it quickly becomes apparent that some extractions can be of low quality (indicated by a high value of $\chisqmin$) and some can be non-sensible (the extracted parameters differ greatly from current world average values). This inability of theory to simultaneously describe all data sets used in this work has already been noted in ref.~\cite{Czakon:2016olj}. To proceed with our extraction of $\as$ and $\mt$ we follow the following strategy. We produce a weighted average for the values of the extracted parameters over as many data sets as possible in order to reduce the role of biases from theory and modelling (as already discussed in \sec{methodology}). In practical terms, for each type of extraction discussed in \sec{results} we produce one value of $\as$ and $\mt$ for each PDF set used. We do not combine extractions based on different PDF sets because they contain different systematic uncertainties, but we check if the extractions based on different PDF sets are compatible with each other. We exclude from this approach the extractions based on the NNPDF3.1 PDF sets since they already contain fits to the top-quark data used in this work. We return to make further comments on the use of this PDF set in \sec{nnpdf31} below.

In implementing our averaging procedure, described in detail in \app{averaging}, we have borne the following considerations in mind. We have decided to exclude from the averaging procedure extractions which return values of the parameters significantly different from previous results, restricting ourselves to those which yield best-fit values of $\as$ and $\mt$ that lie roughly within $\pm 3\sigma$ of the current world average values. Specifically, we require that:
\begin{align} \label{eq:rangerestriction}
0.115 \leq \as \leq 0.121 \;\;\; \text{and} \;\;\; 170.0 \; \text{GeV} \leq \mt \leq 176.0 \; \text{GeV} \,.
\end{align}
To aid the reader, in all tables in this work we have highlighted in grey the extractions that satisfy the criteria in \eq{rangerestriction}. Our motivation for excluding such extractions is that they indicate the presence of some serious difference between theory and data as opposed to a legitimate statistical fluctuation which one may hope to remove with the help of a weighted average.

In principle we do not neglect extractions which return a high value of $\chisqmin$. Our averaging procedure is designed in such a way that it weights values based on their inverse $\chisqmin$, i.e. extractions with large $\chisqmin$ which are less likely to be reliable receive smaller weight in the determination of both the average value as well as in the uncertainty estimate on the average. This procedure fails, of course, in cases where all $\chisqmin$ are large. In such cases we have not performed averaging over the values of a given data set. Specifically, we do not average the CMS entries in \tabs{as-extr-1d}{mt-extr-1d} or the CMS entries in \tab{as-mt-extr-2d}. In addition, we exclude the entries in all of the tables appearing in \app{B} from the averaging procedure due to the larger uncertainties we obtain relative to the normalised cases.

To account for the likely systematic spread between the various extractions we have introduced a systematic component to the uncertainty estimate. It is also weighted by the inverse $\chisqmin$ in order to reduce the effect of outliers (see \app{averaging} for details). This error is designed such that for a large number $n$ of measurements to be averaged, all with similar values of $\chisqmin$, the error tends to a constant value and, unlike the estimate of the statistical uncertainty, does not decrease as $1/\sqrt{n}$.

Finally, we would like to mention that when averaging over $\as$ or $\mt$ values we do not introduce any correlations between them. The reasons for this are: first, all correlations available to us have already been accounted for in the individual extractions and, second, improper modelling of correlations can have a significant detrimental effect on the final uncertainty estimate.

\subsection{Summary of findings}\label{sec:summary}

In this section, we summarise the $\as$ and $\mt$ values obtained with the help of the averaging procedure outlined in \sec{averaging} above. We only use extractions derived from normalised differential distributions supplemented with measurements of the total cross section, since they have a higher precision than those based on absolute distributions.

\underline{For the one-dimensional extractions from single distributions we obtain}:
\begin{itemize}
\item Averaging the ATLAS results for $\as$ (see \tab{as-extr-1d}) for $\mt=173.3$~GeV:
\begin{align}
  \as&=0.1158^{+0.0014}_{-0.0015}\qquad \mathrm{CT14}\\
  \as&=0.1157^{+0.0013}_{-0.0013}\qquad \mathrm{NNPDF3.0}
  \label{eq:bestassingle}
\end{align}
\item Averaging the ATLAS results for $\mt$ (see \tab{mt-extr-1d}) for $\as=0.118$:
\begin{align}
  \mt=&174.3^{+0.9}_{-0.8} ~\mathrm{GeV}\qquad \mathrm{CT14}\\
  \mt=&174.3^{+0.7}_{-0.7} ~\mathrm{GeV}\qquad \mathrm{NNPDF3.0}
  \label{eq:bestmtsingle}
\end{align}
\end{itemize}

\underline{For the simultaneous (two-dimensional) extraction of $\as$ and $\mt$ we obtain}:
\begin{itemize}
\item Extraction from a single (ATLAS) distribution (see \tab{as-mt-extr-2d})
\begin{align}
  \as&=0.1163^{+0.0015}_{-0.0016},\quad\mt=173.8^{+0.8}_{-0.8} ~\mathrm{GeV}\qquad \mathrm{CT14}\\
  \as&=0.1164^{+0.0019}_{-0.0019},\quad\mt=173.6^{+0.8}_{-0.8} ~\mathrm{GeV}\qquad \mathrm{NNPDF3.0}
\end{align}

\item Extraction from multiple ATLAS distributions (see \tab{as-mt-extr-2d-1exp2dists})
\begin{align}
  \as&=0.1159^{+0.0013}_{-0.0014},\quad\mt=173.8^{+0.8}_{-0.8} ~\mathrm{GeV}\qquad \mathrm{CT14}\\
  \as&=0.1161^{+0.0011}_{-0.0010},\quad\mt=173.7^{+0.6}_{-0.6} ~\mathrm{GeV}\qquad \mathrm{NNPDF3.0}
\end{align}
\end{itemize}

The above results have a number of interesting features and show an impressive mutual consistency. In particular, they appear to be almost identical for the two PDF sets we consider, CT14 and NNPDF3.0, both in terms of central values and uncertainties. This is not neccessarily the case for the individual extractions, as can be noted by inspection of the tables appearing in \sec{results}.

The independent extractions of $\as$ and $\mt$ agree well with the simultaneous extractions, both from one differential distribution and from a pair of differential distributions. The size of the uncertainties in all cases is comparable. Perhaps the only (mild) exception we observe is the uncertainty on the $\as$ extraction with NNPDF3.0, which increases by about 50\% from the independent extraction to the simultaneous extraction from a single distribution but then decreases for the simultaneous extraction from a pair of distributions. In general, as expected, the uncertainty on the extractions from a pair of distributions is smallest, both for $\as$ and $\mt$ and for each PDF set. It is interesting to compare the above numbers to the extraction from the total cross section only (see appendix~\ref{app:A}). Regarding $\as$, while all central values are fairly close and compatible within uncertainties, the extractions from the differential distributions have notably smaller uncertainties, in some cases by as much as a factor of two. This is consistent with the expectation that the inclusion of differential data and correlations should improve the extraction of $\as$. In the case of $\mt$, we find very good consistency between the extraction from the total cross section and that from the independent (one-dimensional) $\mt$ fit. The difference in central values is larger when compared to two-dimensional extractions and can reach up to 1 GeV (although this difference is within the uncertainties). In all cases, as expected, the uncertainties for the differential extractions are smaller, sometimes by a factor of two.

Finally, it is interesting to consider the extraction from the simultaneous fit to one ATLAS and one CMS distribution given in \tab{as-mt-extr-2d-2exp} and to compare it to the above extractions. As emphasised in \sec{results}, the fits produced are of low quality as indicated by the large $\chisqmin$ values which they return. Since one might \textit{a priori} consider this particular extraction to be the most reliable due to the overlapping data sets from different experiments, we have listed in \tab{as-mt-extr-2d-2exp} the average extracted values. We observe that the extractions with both PDF sets, CT14 and NNPDF3.0, lead to almost identical central values and uncertainty estimates. The central values for $\as$ are closer to the world average than those discussed above, while the central value for the average $\mt$ extraction is further away. Notably, the estimated uncertainties are rather small and are, in fact, significantly lower than in the cases discussed above. On the one hand these results show the power of averaging (reasonably independent) results. On the other hand, one might infer the low quality of the fits from the fact that the central value for $\mt$ returned is 2.4$\sigma$ away from the world average.

\subsection{Extraction from PDF sets which have already fitted top-quark data}\label{sec:nnpdf31}

In this section we briefly consider the question of consistency of parameter extraction with a PDF set that has itself been fit to the same data. Of the three PDF sets used in this work, only one, NNPDF3.1, has fit top-quark pair differential distributions. The general expectation regarding $\mt$ is that one should obtain from the fits using this set a value closer to $\mt=173.3$ GeV, which was used to fit the PDF initially. It is therefore interesting to examine whether this expectation is borne out by the fits. Regarding $\as$, the expectation is less clear than in the case of $\mt$ since the strong coupling is not a parameter relevant only to the process of top-quark pair production and in fact features in the theoretical predictions for the majority of processes which enter into the PDF fit.

A second point one might hope to assess is whether any impact on the parameter extraction appears in fits from all differential distributions, or only in fits from distributions directly exploited by the PDF collaboration. Of the distributions which we consider in this work, the NNPDF3.1 sets contain information only on the $\yt$ data from ATLAS and the $\ytt$ data from CMS. In the following we shed some light on this issue, however, we caution that we cannot offer a comprehensive answer to this question here, since that would require detailed information about the specifics of the NNPDF3.1 PDF fit that are not available to us. To that end one would need to examine, for example, the relative impacts of different data sets not related to top-quark data, any modification of fitting methodology etc. which distinguish NNPDF3.1 with respect to say NNPDF3.0 but are not directly related to the inclusion of top-quark data.

\begin{figure}[t]
\includegraphics[trim=0.0cm 0.0cm 0.0cm 0.0cm,clip,width=0.32\textwidth]{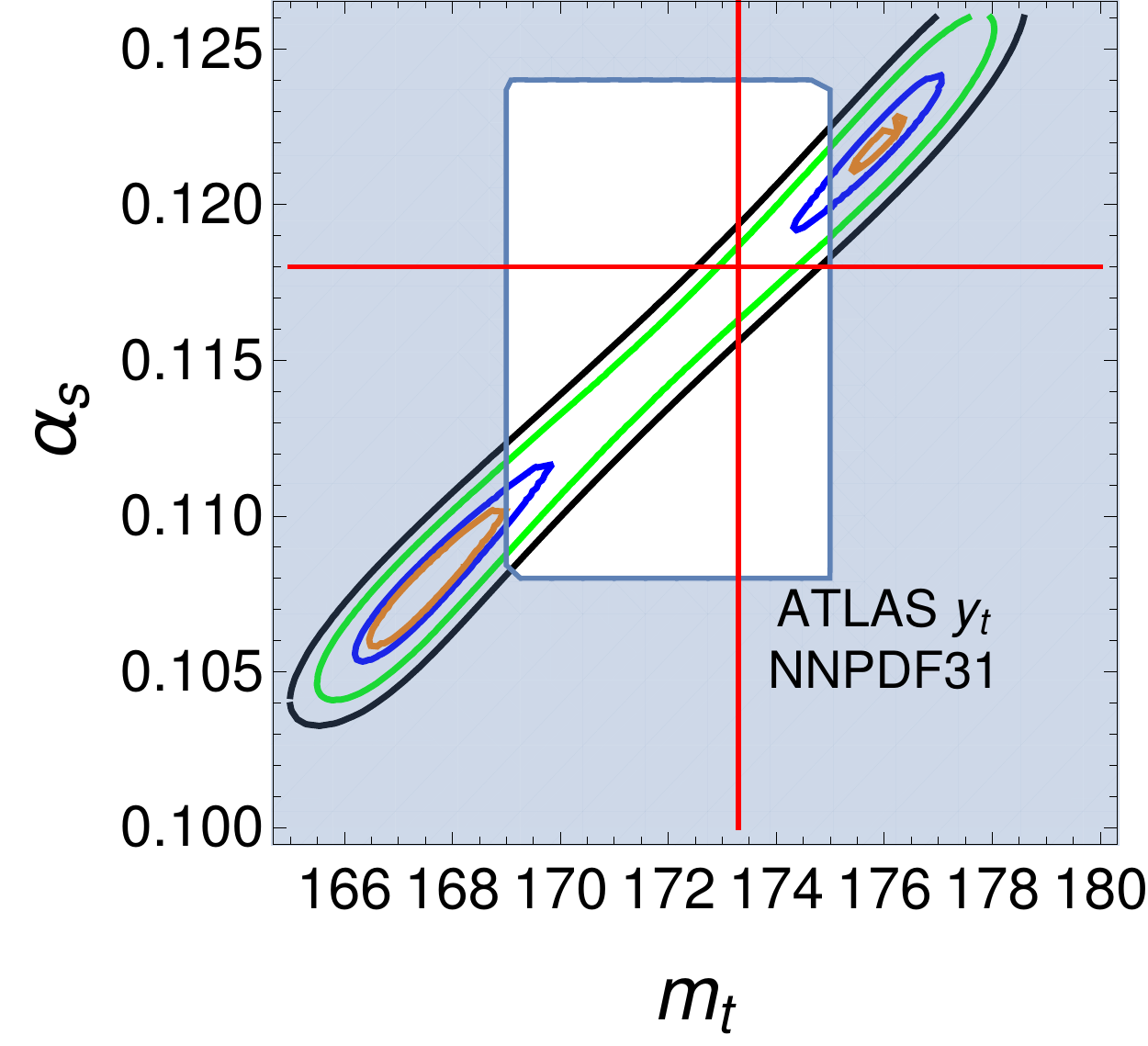}
\includegraphics[trim=0.0cm 0.0cm 0.0cm 0.0cm,clip,width=0.32\textwidth]{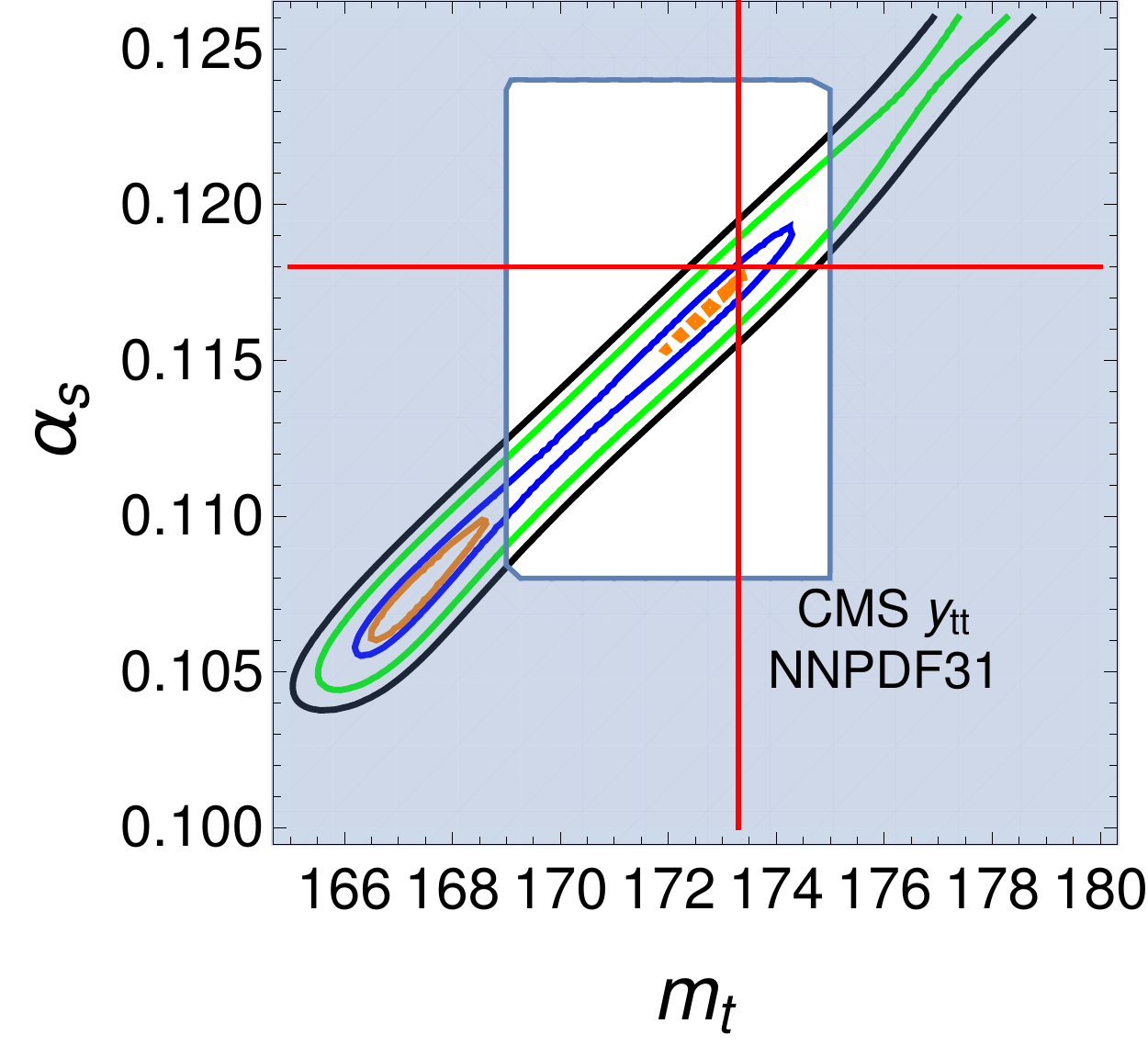}
\includegraphics[trim=0.0cm 0.0cm 0.0cm 0.0cm,clip,width=0.32\textwidth]{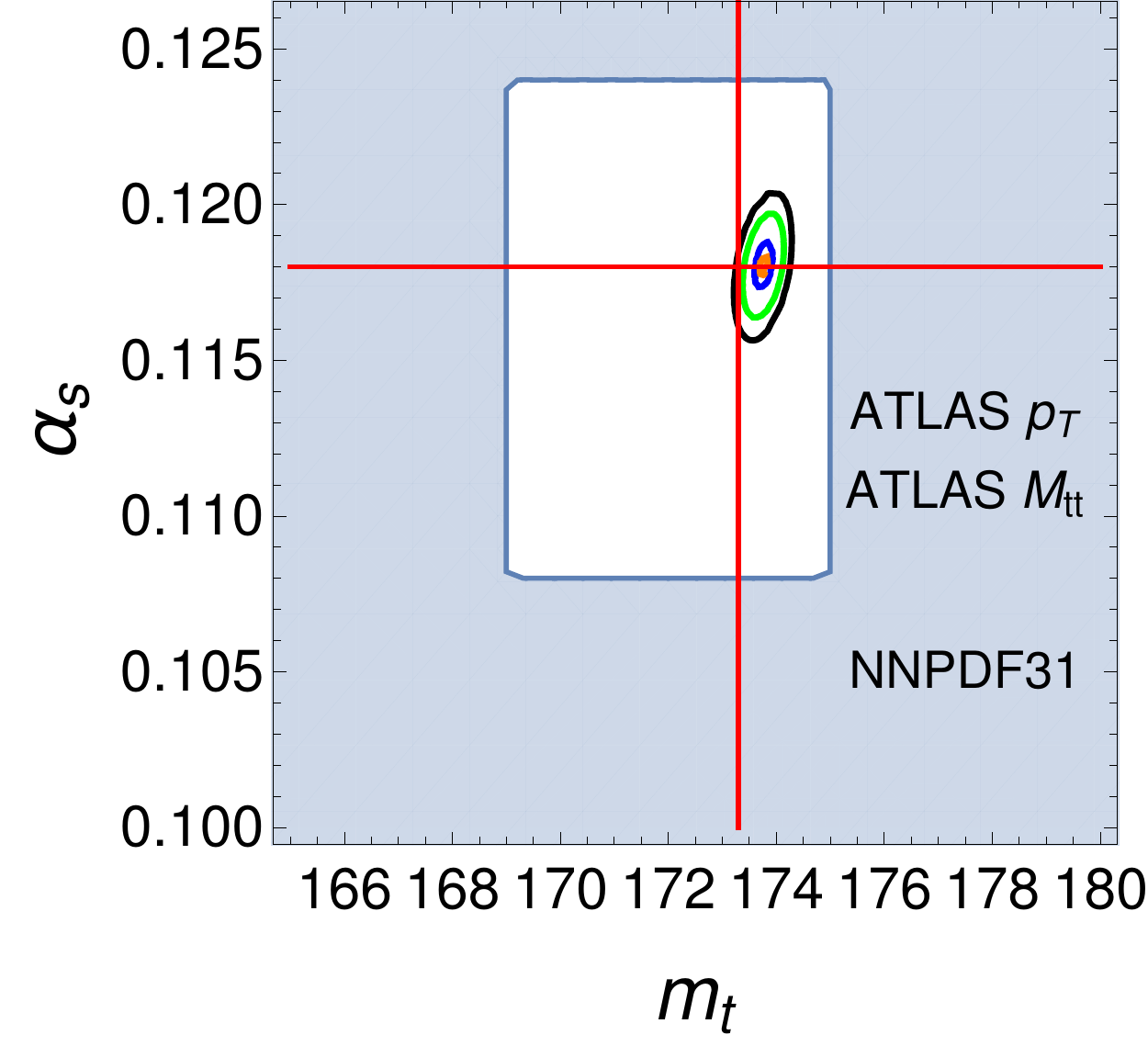}\\
\includegraphics[trim=0.0cm 0.0cm 0.0cm 0.0cm,clip,width=0.32\textwidth]{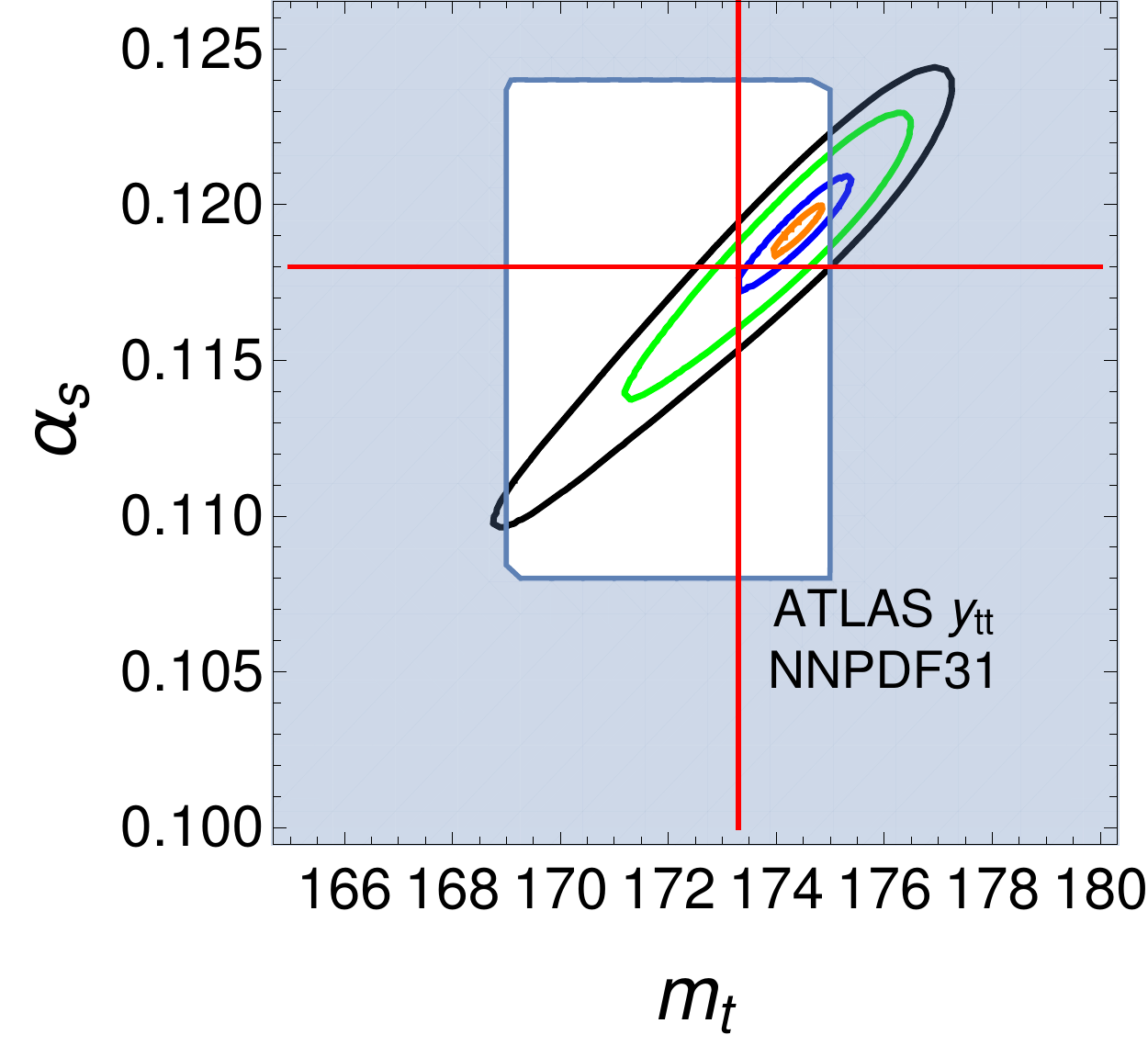}
\includegraphics[trim=0.0cm 0.0cm 0.0cm 0.0cm,clip,width=0.32\textwidth]{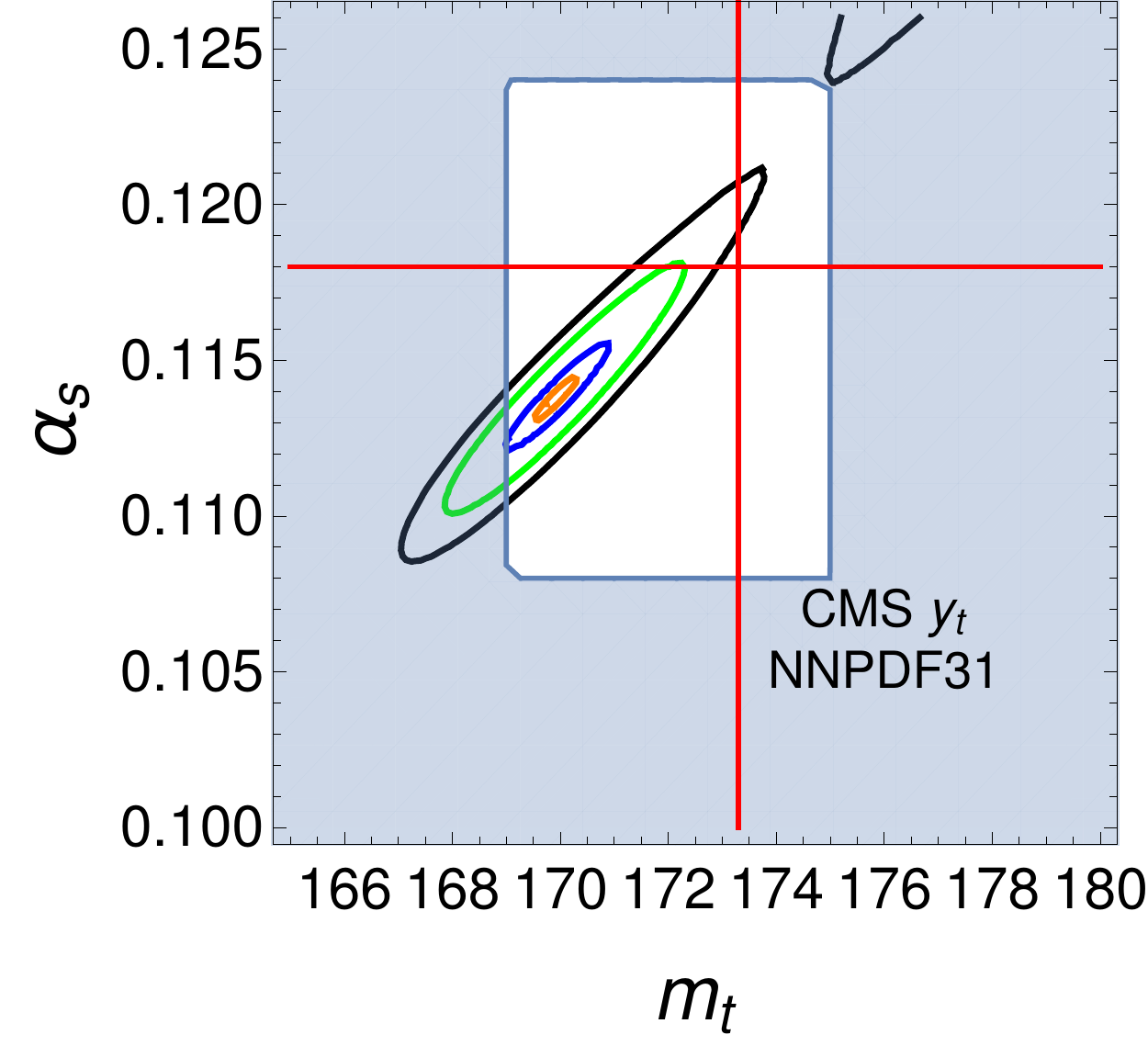}
\includegraphics[trim=0.0cm 0.0cm 0.0cm 0.0cm,clip,width=0.32\textwidth]{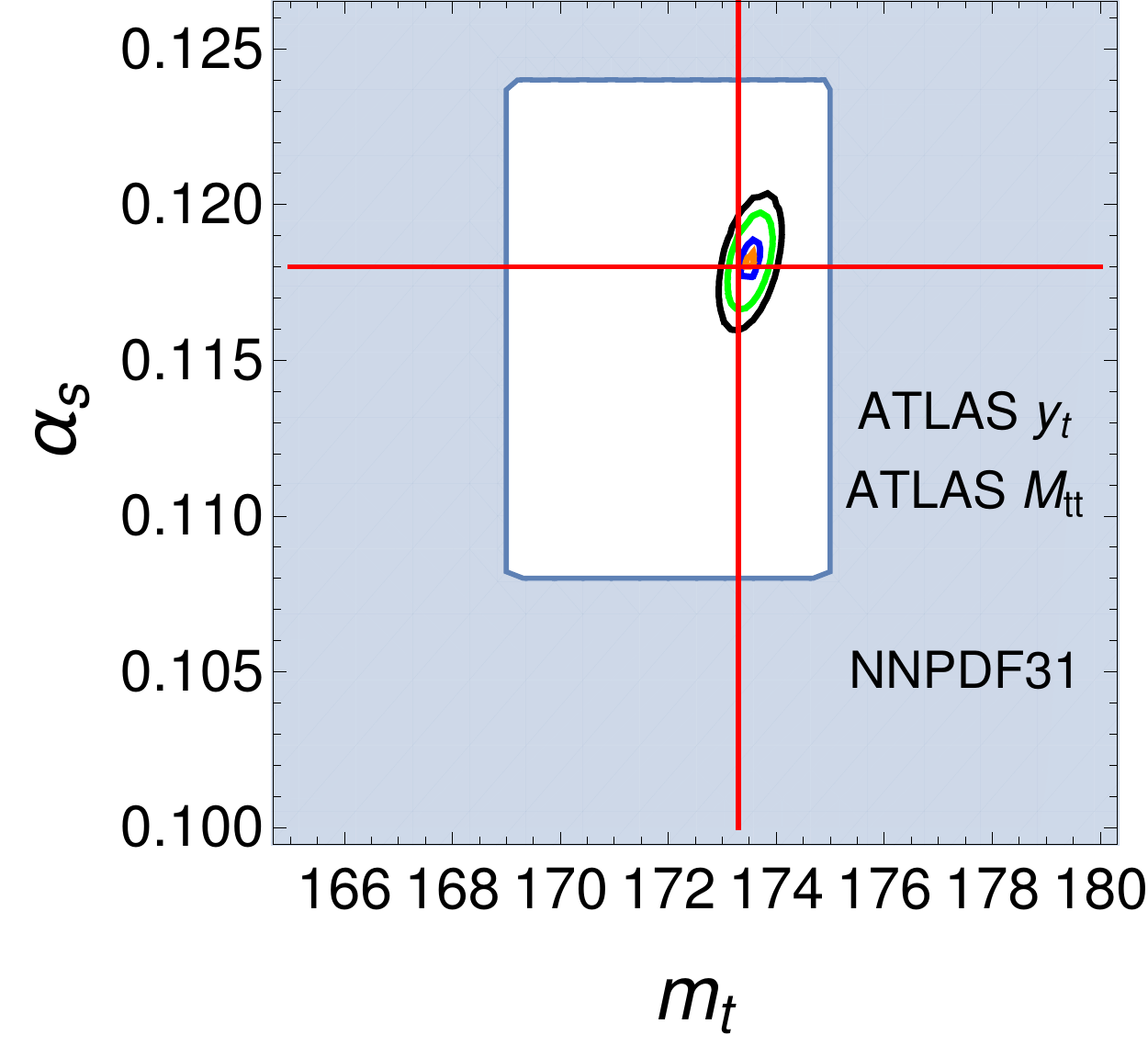}
\caption{
Examples of simultaneous $\as$ and $\mt$ extractions from differential distributions using the PDF set NNPDF3.1. Some of the plots show extractions from distributions that have been used in the PDF fit while others are from distributions that have not. See \sec{nnpdf31} for details.
}
\label{fig:nnpdf31}
\end{figure}

In \fig{nnpdf31} we show the results of several two-dimensional $\as$ and $\mt$ extractions from various differential distributions computed with the NNPDF3.1 PDF set. To emphasise the important features of these plots, we have increased the displayed area in parameter space and have shown several contours corresponding to the values $\Delta\chi^2=$ 1.0 (black), 0.5 (green), 0.1 (blue) and 0.02 (orange).  In the upper left and upper centre plots we see the extractions from the distributions directly fitted by the PDF set. In line with our general expectations above, we observe that the behaviour of the fits is strikingly different compared to the fits performed with CT14 or NNPDF3.1. In particular we observe strong, nearly perfect, correlation between the two fitted parameters as indicated by the $\chi^2$ contours that do not show a localised minimum in the $\as-\mt$ plane but rather a degenerate flat `valley' of minima. Clearly, the distributions used in the PDF fits are not suitable for extraction of top-quark-related parameters. In the lower left and lower centre plots we show the fits to the same two rapidity distributions but now fitted from the opposite experiment: $\ytt$ from ATLAS and $\yt$ from CMS. We again observe strong correlation between $\as$ and $\mt$, although the degeneracy is not quite as marked. Inferring that the two rapidity distributions are rather correlated, we conclude that we cannot use either for $\as$ and $\mt$ fits. For this reason we have not shown any results in the tables in \sec{results} for NNPDF3.1 with $\yt$ or $\ytt$.

It is next interesting to consider the extraction of $\as$ and $\mt$ from distributions that have not been directly used in the PDF fits. For the case at hand these would be the $\mtt$ and $\pT$ distributions. The resulting $\chi^2$ fit can be seen in \fig{nnpdf31} (upper right). From that figure we conclude that the fit based on these two differential distributions does not appear to have any obvious degeneracy between the two parameters as was observed in the cases of the rapidity distributions. This may be an indication that distributions which have not been directly included in PDF fits can be considered to be unaffected. This is especially relevant for the $\mtt$ distribution, which is very important to $\mt$ determination and searches for BSM physics. We caution that this finding should not be considered as definitive proof of a lack of correlation but rather as an evidence of absence of obvious correlation. Lastly, it is interesting to consider an intermediate case where $\as$ and $\mt$ are fit to two distributions, one of which has been included in the PDF fit and one of which has not. In \fig{nnpdf31} (lower right) we show the $\Delta\chi^2$ contours for the two-dimensional extraction from the combined ATLAS data on $\yt$ and $\mtt$. We see only a small change relative to the $\mtt$ and $\pT$ case and observe $\Delta\chi^2$ contours which do not show obvious signs of a degeneracy. Nevertheless, in order to be conservative, we have consistently omitted extractions using a combination of either $\yt$ or $\ytt$ and any other distribution from the NNPDF3.1 column of the tables in \sec{results}.

\section{Conclusions}\label{sec:conclusions}

In this work we have presented the first joint extraction using NNLO theory predictions of $\as$ and the top-quark pole mass $\mt$ from measurements
of both the top-quark pair production cross section as well as differential distributions. We have taken experimental measurements of the kinematic variables $\pT, \mtt, \yt$ and $\ytt$ in $t\bar{t}$ events provided by the ATLAS and CMS experiments at the 8 TeV LHC. These have been compared with theoretical predictions including NNLO QCD corrections for different values of the top-quark mass and using different PDF sets. The \texttt{fastNLO} tables used to produce the results in this paper, for values of the top-quark pole mass $m_t=\{169.0,171.0,172.5,173.3,175.0\}$~GeV, are available for download from ref.~\cite{website}. Functional forms of the bin weights of each predicted distribution have been determined by smooth interpolation. A least-squares method has been used to perform individual and simultaneous extractions of the SM parameters $\mt$ and $\as$---we have also considered combining distributions from different experiments in the absence of correlations.

Our best result is thus a two-dimensional extraction, taking into account $\as$ and $\mt$ correlations and using two distributions to maximise the differential information included. We select the case of two different ATLAS distributions since the experimental correlations between the distributions are properly included and the overall quality of the fits is good. Within this, we choose the more conservative estimate using CT14 (which contains no top-quark data). Our final values are therefore
\begin{equation}
  \as=0.1159^{+0.0013}_{-0.0014},\quad\mt=173.8^{+0.8}_{-0.8} ~\mathrm{GeV}\,.
\end{equation}

These values are compatible with those obtained from extractions using the total cross section for top-quark pair production. The $\as$ extraction is about 1.5$\sigma$ below the world average of $0.118$; it is of high precision which is competitive with measurements from other processes (for example, the CMS Drell-Yan measurement in ref.~\cite{dEnterria:2019aat}). Subtleties related to extracting $\as$ from PDFs with fixed $\as$ value aside \cite{Forte:2020pyp}, this indicates that the determination of the strong coupling constant from $t\bar t$ production has significant potential. The extracted $\mt$ value is in perfect agreement with the world average of $173.3$ GeV and has impressively small uncertainty.

In performing our analysis, we have noticed several features of the data which hinder a clean extraction. We observe certain tensions between
ATLAS and CMS measurements of the same distribution (in keeping with previous observations~\cite{Czakon:2016olj}), which may point to different treatments by the two experiments. Furthermore, in some cases we also observe considerable differences between PDF sets, which may indicate that a simultaneous fit of $\as$, $\mt$ and the PDFs must be considered. In any case, despite these difficulties we believe that the approach presented in this work is promising and is likely to bear greater fruit when provided with the significantly larger data sets offered by newer LHC runs.

There are a few ways in which our treatment could be further refined in a future study. Firstly, as we have highlighted in \sec{methodology}, the uncertainties on the extracted values of 
$\as$ and $\mt$ which we have presented only include experimental effects, both systematic and statistical.
In a more complete treatment, theory uncertainties due to the choice of renormalisation
and factorisation scales as well as PDF uncertainties should be included in the definition
of the $\chi^2$ objective function.
Since the scale uncertainty is greatly reduced at NNLO when considering normalised (compared
to absolute) distributions, we expect that including these will not increase the uncertainties
on the extracted parameters dramatically.
On the other hand, in one-dimensional extractions of $\as$, the scale uncertainty on the
cross section is known to contribute a significant fraction of the overall uncertainty on the
best fit $\as$~\cite{Klijnsma:2017eqp}, and we would therefore expect this to also play a r\^{o}le
for our study here.
Additionally, in the case of our extractions from combined measurements of ATLAS and CMS,
a complete treatment would also include systematic uncertainties correlated between the
two experiments, such as that of luminosity.

Finally, given the constraining power that differential measurements of top-pair production
have on PDFs~\cite{Czakon:2016olj,Sirunyan:2017azo,Bailey:2019yze,Czakon:2019yrx,Kadir:2020yml}, a joint fit at NNLO in QCD of $\as$ and $\mt$ together
with PDFs using differential data would simultaneously provide useful constraints on all
these three fundamental ingredients of QCD.

\begin{acknowledgments}
M.A.L. is grateful to Emma Slade for useful discussions. A.M. thanks the Department of Physics at Princeton University for hospitality during the completion of this work.  A.M.C.S. acknowledges the support of the Leverhulme Trust. The work of M.C. was supported in part by a grant of the BMBF and by the Deutsche Forschungsgemeinschaft under grant 396021762 -- TRR 257. The work of M.A.L., A.M. and A.S.P. has received funding from the European Research Council (ERC) under the European Union's Horizon 2020 research and innovation programme (grant agreement No 683211); it was also supported by the UK STFC grants ST/L002760/1 and ST/K004883/1. M.A.L. acknowledges funding from Fondazione Cariplo and Regione Lombardia, grant 2017-2070 as well as from the Cambridge Philosophical Society. A.P. is a cross-disciplinary postdoctoral fellow supported by funding from the University of Edinburgh and Medical Research Council (core grant to the MRC Institute of Genetics and Molecular Medicine). 
\end{acknowledgments}

\appendix
\section{Extractions using measurements of the total cross section} \label{app:A}

In this appendix, see \tab{as-mt-extr-xs}, we present the results for the extraction of $\mt$ and $\as$ using the total cross section alone. Results for the central value of the $\as$ extraction can be compared with Table 6 of \mycite{Klijnsma:2017eqp}, where full agreement is found.

\begin{table}[h!]
\centering

\begin{tabular}{|c|c c| c c | c c |}
\hline
& \multicolumn{2}{|c|}{CT14} & \multicolumn{2}{|c|}{NNPDF30} & \multicolumn{2}{|c|}{NNPDF31} \\
 & $\as$ & $\mt$ & $\as$ & $\mt$ & $\as$ & $\mt$ \\
\hline
ATLAS & $0.1158^{+0.0021}_{-0.0020}$ & $174.8^{+1.4}_{-1.3}$ &
        $0.1156^{+0.0027}_{-0.0027}$ & $174.4^{+1.4}_{-1.3}$ &
        $0.1172^{+0.0021}_{-0.0022}$ & $173.8^{+1.4}_{-1.3}$ \\
\hline
CMS   & $0.1163^{+0.0020}_{-0.0019}$ & $174.4^{+1.3}_{-1.2}$ &
        $0.1164^{+0.0026}_{-0.0026}$ & $174.1^{+1.3}_{-1.2}$ &
        $0.1178^{+0.0021}_{-0.0021}$ & $173.4^{+1.3}_{-1.2}$ \\
\hline
\end{tabular}

\caption{Best-fit $\as$ and $\mt$ from the total cross section: extracted $\as$ or $\mt$ (with uncertainties) from ATLAS and CMS measurements at 8 TeV. The extractions have been performed after fixing $\mt=173.3$~GeV and $\as=0.118$ respectively. We show results obtained from three different PDF sets. Masses are quoted in GeV.}
\label{tab:as-mt-extr-xs}
\end{table}

\section{Extractions using measurements of absolute distributions} \label{app:B}
In this appendix, we present results for the extraction of $\mt$ and $\as$ using measurements of absolute distributions by ATLAS and CMS. We show the results of one-dimensional extractions for $\as$ in \tab{as-extr-1d-abs} and for $\mt$ in \tab{mt-extr-1d-abs}. Results from two-dimensional extractions from a single distribution appear in \tab{as-mt-extr-2d-abs}, while those from combinations of distributions appear in \tab{as-mt-extr-2d-1exp2dists-abs} (two ATLAS) and \tab{as-mt-extr-2d-2exp-abs} (ATLAS and CMS).

\begin{table}[t]
\centering

\begin{tabular}{ |c|c c|c c|c c| }
\hline
\multicolumn{1}{|c|}{} & \multicolumn{6}{|c|}{ATLAS}\\
\hline
& \multicolumn{2}{|c|}{CT14} & \multicolumn{2}{|c|}{NNPDF30} & \multicolumn{2}{|c|}{NNPDF31} \\
& $\as$ & $\chisqmin$ & $\as$ & $\chisqmin$ & $\as$ & $\chisqmin$ \\
\hline
$\pT$ &\cellcolor{Gray}$0.1189^{+0.0035}_{-0.0041}$ &\cellcolor{Gray}0.32 &  $\cellcolor{Gray}0.1203^{+0.0055}_{-0.0055}$ &\cellcolor{Gray}0.32 &\cellcolor{Gray}$0.1194^{+0.0043}_{-0.0046}$ & \cellcolor{Gray}0.49     \\
\hline
$\mtt$ &  $0.1211^{+0.0020}_{-0.0026}$ &  0.20 & $0.1247^{+0.0047}_{-0.0047}$ &  0.21 &  $0.1251^{+0.0040}_{-0.0038}$&  0.25     \\
\hline
$\mttu$ &  $0.1211^{+0.0020}_{-0.0026}$ &  0.17 & $0.1247^{+0.0047}_{-0.0047}$ &  0.18 &  $0.1251^{+0.0041}_{-0.0038}$ &  0.22   \\
\hline
$\yt$ &\cellcolor{Gray}$0.1154^{+0.0029}_{-0.0033}$ &\cellcolor{Gray}2.55 & $0.1135^{+0.0033}_{-0.0033}$ &  0.72 &&     \\
\hline
$\ytt$ &\cellcolor{Gray}$0.1184^{+0.0029}_{-0.0042}$ &\cellcolor{Gray}2.48 & $0.1127^{+0.0054}_{-0.0054}$ &  0.66 &&     \\
\hline
\end{tabular} \\[10pt]

\begin{tabular}{ |c|c c|c c|c c| }
\hline
\multicolumn{1}{|c|}{} & \multicolumn{6}{|c|}{CMS}\\
\hline
& \multicolumn{2}{|c|}{CT14} & \multicolumn{2}{|c|}{NNPDF30} & \multicolumn{2}{|c|}{NNPDF31} \\
& $\as$ & $\chisqmin$ & $\as$ & $\chisqmin$ & $\as$ & $\chisqmin$ \\
\hline
$\pT$ &\cellcolor{Gray}$0.1153^{+0.0016}_{-0.0017}$ &\cellcolor{Gray}1.63 &  $0.1138^{+0.0023}_{-0.0023}$ &  1.30 &\cellcolor{Gray}$0.1164^{+0.0018}_{-0.0019}$ &\cellcolor{Gray}1.01     \\
\hline
$\mtt$ &  $0.1142^{+0.0016}_{-0.0017}$ &  6.20 & $0.1100^{+0.0020}_{-0.0034}$ &  4.38 &\cellcolor{Gray}$0.1157^{+0.0021}_{-0.0022}$ &\cellcolor{Gray}3.91    \\
\hline
$\mttu$ &  $0.1142^{+0.0016}_{-0.0017}$ &  6.18 & $0.1100^{+0.0020}_{-0.0034}$ &  4.36 &\cellcolor{Gray}$0.1157^{+0.0021}_{-0.0022}$ &\cellcolor{Gray}3.89   \\
\hline
$\yt$ &\cellcolor{Gray}$0.1152^{+0.0017}_{-0.0019}$ &\cellcolor{Gray}1.82 &\cellcolor{Gray}$0.1177^{+0.0021}_{-0.0021}$ &\cellcolor{Gray}2.80 &&     \\
\hline
$\ytt$ &\cellcolor{Gray}$0.1161^{+0.0015}_{-0.0016}$ &\cellcolor{Gray}1.57 &\cellcolor{Gray}$0.1154^{+0.0018}_{-0.0018}$ &\cellcolor{Gray}0.74 &&     \\
\hline
\end{tabular}

\caption{
Tabulated values of best-fit $\as$ (with uncertainties) and associated $\chisqmin$ from extractions of $\as$ using ATLAS (upper table) and CMS (lower table) measurements of absolute distributions. Results are shown for three different PDF sets and $\mt$ has been set to the world average value of 173.3~GeV. The cells highlighted in grey correspond to extractions that satisfy the quality conditions described in sec~\ref{sec:averaging}. As explained in \sec{nnpdf31}, the rapidity distributions are not used for extractions with the PDF set NNPDF3.1. The quantity $\mttu$ is defined in sec.~\ref{sec:lower-edge-mtt}.}
\label{tab:as-extr-1d-abs}
\end{table}

\begin{table}[h!]
\centering

\begin{tabular}{ |c|c c|c c|c c| }
\hline
\multicolumn{1}{|c|}{} & \multicolumn{6}{|c|}{ATLAS}\\
\hline
& \multicolumn{2}{|c|}{CT14} & \multicolumn{2}{|c|}{NNPDF30} & \multicolumn{2}{|c|}{NNPDF31} \\
& $\mt$ & $\chisqmin$ & $\mt$ & $\chisqmin$ & $\mt$ & $\chisqmin$ \\
\hline
$\pT$ &\cellcolor{Gray}$173.3^{+1.3}_{-1.2}$ &\cellcolor{Gray}0.32 &\cellcolor{Gray}$173.5^{+1.3}_{-1.3}$ & \cellcolor{Gray}0.34 &\cellcolor{Gray}$173.7^{+1.3}_{-1.3}$ & \cellcolor{Gray}0.49     \\
\hline
$\mtt$ &\cellcolor{Gray}$172.9^{+0.5}_{-0.5}$ & \cellcolor{Gray}0.30 &\cellcolor{Gray}$173.0^{+0.5}_{-0.5}$ & \cellcolor{Gray}0.45 &\cellcolor{Gray}$173.1^{+0.6}_{-0.5}$ & \cellcolor{Gray}0.71     \\
\hline
$\mttu$ &\cellcolor{Gray}$172.9^{+0.5}_{-0.5}$ & \cellcolor{Gray}0.28 &\cellcolor{Gray}$173.0^{+0.5}_{-0.5}$ & \cellcolor{Gray}0.41 &\cellcolor{Gray}$173.1^{+0.5}_{-0.5}$ & \cellcolor{Gray}0.67     \\
\hline
$\yt$ &  $178.2^{+2.3}_{-2.6}$ &  2.07 &  $176.2^{+2.8}_{-3.7}$ & 0.96 &&     \\
\hline
$\ytt$ &  $178.3^{+3.5}_{-4.0}$ &  2.18 &\cellcolor{Gray}$173.7^{+4.5}_{-4.8}$ & \cellcolor{Gray}0.85 &&     \\
\hline
\end{tabular} \\[10pt]

\begin{tabular}{ |c|c c|c c|c c| }
\hline
\multicolumn{1}{|c|}{} & \multicolumn{6}{|c|}{CMS}\\
\hline
& \multicolumn{2}{|c|}{CT14} & \multicolumn{2}{|c|}{NNPDF30} & \multicolumn{2}{|c|}{NNPDF31} \\
& $\mt$ & $\chisqmin$ & $\mt$ & $\chisqmin$ & $\mt$ & $\chisqmin$ \\
\hline
$\pT$ &\cellcolor{Gray}$173.1^{+0.9}_{-0.8}$ &\cellcolor{Gray}2.04 &\cellcolor{Gray}$173.0^{+0.9}_{-0.8}$ & \cellcolor{Gray}1.69 &\cellcolor{Gray}$173.0^{+0.9}_{-0.8}$ & \cellcolor{Gray}1.10     \\
\hline
$\mtt$ &  $168.8^{+1.1}_{-1.2}$ &  4.36 &  $169.1^{+1.1}_{-1.2}$ & 3.59 &  $169.9^{+1.0}_{-1.1}$ &  2.47     \\
\hline
$\mttu$ &\cellcolor{Gray}$170.1^{+0.7}_{-0.7}$ &\cellcolor{Gray}4.57 &\cellcolor{Gray}$170.3^{+0.7}_{-0.7}$ &\cellcolor{Gray}3.77 &\cellcolor{Gray}$170.7^{+0.8}_{-0.7}$ &\cellcolor{Gray}2.58     \\
\hline
$\yt$ &\cellcolor{Gray}$174.8^{+1.1}_{-1.1}$ &\cellcolor{Gray}1.91 &\cellcolor{Gray}$174.7^{+1.1}_{-1.1}$ &\cellcolor{Gray}2.62 &&     \\
\hline
$\ytt$ &\cellcolor{Gray}$175.3^{+1.2}_{-1.1}$ &\cellcolor{Gray}1.43 &\cellcolor{Gray}$174.5^{+1.1}_{-1.1}$ & \cellcolor{Gray}0.81 &&     \\
\hline
\end{tabular}
\caption{As in \tab{as-extr-1d-abs} but for $\mt$ rather than $\as$, which has been set to the world average value of 0.118. Masses are quoted in GeV.}
\label{tab:mt-extr-1d-abs}
\end{table}

\begin{table}[h!]
\centering

\footnotesize
\begingroup
\setlength{\tabcolsep}{3pt}
\begin{tabular}{ |c|c c c|c c c|c c c| }
\hline
\multicolumn{1}{|c|}{} & \multicolumn{9}{|c|}{ATLAS}\\
 \hline
& \multicolumn{3}{|c|}{CT14} & \multicolumn{3}{|c|}{NNPDF30} & \multicolumn{3}{|c|}{NNPDF31} \\
&   $\as$ & $\mt$ & $\chisqmin$   &   $\as$ & $\mt$ & $\chisqmin$   &   $\as$ & $\mt$ & $\chisqmin$ \\
\hline
$\pT$ &\cellcolor{Gray}$0.1199^{+0.0031}_{-0.0039}$ &\cellcolor{Gray}$173.8^{+1.2}_{-1.2}$ &\cellcolor{Gray}$0.31$ & $0.1229^{+0.0057}_{-0.0057} $ &  $174.2^{+1.2}_{-1.2}$ &  $0.28$ & $0.1240^{+0.0052}_{-0.0041}$ &  $175.0^{+1.2}_{-1.2}$ &  $0.37$ \\
\hline
$\mtt$ &  $0.1211^{+0.0021}_{-0.0028}$ &  $173.0^{+0.5}_{-0.5}$ &  $0.14$ &  $0.1244^{+0.0046}_{-0.0046} $ & $173.1^{+0.5}_{-0.5}$ &  $0.19$ &  $0.1256^{+0.0042}_{-0.0039}$ & $173.5^{+0.5}_{-0.5}$ &  $0.24$     \\
\hline
$\mttu$ &  $0.1211^{+0.0021}_{-0.0028}$ &  $173.0^{+0.5}_{-0.5}$ &  $0.12$ &  $0.1243^{+0.0046}_{-0.0046} $ & $173.1^{+0.5}_{-0.4}$ &  $0.16$ &  $0.1254^{+0.0042}_{-0.0038}$ & $173.4^{+0.5}_{-0.4}$ &  $0.22$     \\
\hline
$\yt$ &  $0.1183^{+0.0024}_{-0.0036}$ &  $178.3^{+2.3}_{-2.5}$ & $2.07$ &  $0.1116^{+0.0027}_{-0.0027} $ &  $168.0^{+3.4}_{-5.7}$ & $0.47$ &&&     \\
\hline
$\ytt$ &  $0.1195^{+0.0028}_{-0.0042}$ &  $178.4^{+3.3}_{-3.7}$ &  $2.15$ &  $0.1121^{+0.0048}_{-0.0048} $ & $170.8^{+5.8}_{-4.3}$ &  $0.62$ &&&     \\
\hline
\end{tabular} \\[10pt]

\begin{tabular}{ |c|c c c|c c c|c c c| }
\hline
\multicolumn{1}{|c|}{} & \multicolumn{9}{|c|}{CMS}\\
 \hline
& \multicolumn{3}{|c|}{CT14} & \multicolumn{3}{|c|}{NNPDF30} & \multicolumn{3}{|c|}{NNPDF31} \\
&   $\as$ & $\mt$ & $\chisqmin$   &   $\as$ & $\mt$ & $\chisqmin$   &   $\as$ & $\mt$ & $\chisqmin$ \\
\hline
$\pT$ &  $0.1083^{+0.0018}_{-0.0017}$ &  $169.1^{+0.8}_{-0.8}$ & $0.72$ &  $0.1088^{+0.0022}_{-0.0022}$ &  $170.5^{+0.9}_{-0.8}$ & $0.66$ &  $0.1115^{+0.0018}_{-0.0018}$ &  $170.6^{+0.8}_{-0.8}$ & $0.62$     \\
\hline
$\mtt$ &  $0.1114^{+0.0018}_{-0.0018}$ & $166.6^{+1.5}_{-1.6}$ &  $2.81$ &  $0.1089^{+0.0026}_{-0.0026}$ & $168.3^{+1.3}_{-1.4}$ &  $1.88$ &  $0.1134^{+0.0022}_{-0.0024}$ & $169.0^{+1.2}_{-1.3}$ &  $1.94$     \\
\hline
$\mttu$ &  $0.1134^{+0.0018}_{-0.0018}$ & $169.7^{+0.8}_{-0.7}$ &  $3.56$ &  $0.1098^{+0.0026}_{-0.0026}$ & $170.1^{+0.8}_{-0.7}$ &  $2.38$ &  $0.1147^{+0.0022}_{-0.0023}$ & $170.5^{+0.8}_{-0.7}$ &  $2.26$     \\
\hline
$\yt$ &  $0.1086^{+0.0017}_{-0.0014}$ &  $169.6^{+1.0}_{-1.0}$ & $1.60$ &  $0.1241^{+0.0024}_{-0.0024}$ &  $177.0^{+1.2}_{-1.0}$ & $2.36$ &&&     \\
\hline
$\ytt$ &  $0.1195^{+0.0012}_{-0.0013}$ & $176.4^{+1.2}_{-1.1}$ &  $1.39$ &\cellcolor{Gray}$0.1152^{+0.0018}_{-0.0018}$ & \cellcolor{Gray}$173.2^{+1.1}_{-1.1}$ &\cellcolor{Gray}$0.74$ &&&    \\
\hline
\end{tabular}

\normalsize
\endgroup

\caption{As in \tab{as-extr-1d-abs} but for the simultaneous extractions of $\as$ and $\mt$.}
\label{tab:as-mt-extr-2d-abs}
\end{table}

\begin{table}[h!]
\centering
\footnotesize

\begingroup
\setlength{\tabcolsep}{3pt}
\begin{tabular}{ |c|c|c c c|c c c|c c c| }
\hline
ATLAS & ATLAS & \multicolumn{3}{|c|}{CT14} & \multicolumn{3}{|c|}{NNPDF30} & \multicolumn{3}{|c|}{NNPDF31} \\
 obs. 1 & obs. 2 & $\as$ & $\mt$ & $\chisqmin$   &   $\as$ & $\mt$ & $\chisqmin$   &   $\as$ & $\mt$ & $\chisqmin$ \\
\hline
$\pT$ & $\mtt$ &\cellcolor{Gray}$0.1202^{+0.0020}_{-0.0024} $ &\cellcolor{Gray}$173.1^{+0.5}_{-0.4}$ &\cellcolor{Gray}$ 0.28$ &  $0.1227^{+0.0036}_{-0.0036}$ &  $173.2^{+0.5}_{-0.5}$ &  $ 0.31$ &  $0.1233^{+0.0028}_{-0.0029}$ &  $173.6^{+0.5}_{-0.5}$ &  $ 0.43$  \\
\hline
$\pT$ & $\mttu$ &\cellcolor{Gray}$0.1202^{+0.0020}_{-0.0024} $ &\cellcolor{Gray}$173.1^{+0.4}_{-0.4}$ &\cellcolor{Gray}$ 0.26$ &  $0.1226^{+0.0036}_{-0.0036}$ &  $173.2^{+0.4}_{-0.4}$ &  $ 0.29$ &  $0.1230^{+0.0028}_{-0.0029}$ &  $173.5^{+0.5}_{-0.4}$ &  $ 0.43$  \\
\hline
$\pT$ & $y_t$ &\cellcolor{Gray}$0.1174^{+0.0016}_{-0.0019} $ &\cellcolor{Gray}$ 174.6^{+1.2}_{-1.2}$ &\cellcolor{Gray}$ 1.55$ & $0.1140^{+0.0021}_{-0.0021}$ & $ 172.7^{+1.3}_{-1.3}$ & $ 0.61$ &&&  \\
\hline
$\pT$ & $\ytt$ &\cellcolor{Gray}$0.1194^{+0.0017}_{-0.0021} $ &\cellcolor{Gray}$ 174.4^{+1.2}_{-1.2}$ &\cellcolor{Gray}$ 1.48$ & $0.1146^{+0.0031}_{-0.0031}$ &  $ 172.9^{+1.3}_{-1.2}$ & $ 0.58$ &&&  \\
\hline
$\yt$ & $\mtt$ &\cellcolor{Gray}$0.1171^{+0.0016}_{-0.0019} $ &\cellcolor{Gray}$ 173.1^{+0.6}_{-0.5}$ &\cellcolor{Gray}$ 1.73$ &\cellcolor{Gray}$0.1154^{+0.0021}_{-0.0021}$ &\cellcolor{Gray}$172.9^{+0.6}_{-0.5}$ &\cellcolor{Gray}$ 0.75$ &&&  \\
\hline
$\ytt$ & $\mtt$ &\cellcolor{Gray}$0.1194^{+0.0016}_{-0.0019} $ &\cellcolor{Gray}$ 173.1^{+0.6}_{-0.5}$ &\cellcolor{Gray}$ 1.57$ &\cellcolor{Gray}$0.1170^{+0.0029}_{-0.0029}$ &\cellcolor{Gray}$172.9^{+0.5}_{-0.5}$ &\cellcolor{Gray}$ 0.68$ &&&  \\
\hline
$\yt$ & $\mttu$ &\cellcolor{Gray}$0.1171^{+0.0015}_{-0.0019} $ & \cellcolor{Gray}$ 173.1^{+0.5}_{-0.5}$ &\cellcolor{Gray}$ 1.72$ & \cellcolor{Gray}$0.1154^{+0.0021}_{-0.0021}$ &\cellcolor{Gray}$172.9^{+0.5}_{-0.5}$ &\cellcolor{Gray}$ 0.74$ &&&  \\
\hline
$\ytt$ & $\mttu$ &\cellcolor{Gray}$0.1194^{+0.0015}_{-0.0019} $ & \cellcolor{Gray}$ 173.1^{+0.5}_{-0.5}$ &\cellcolor{Gray}$ 1.55$ &\cellcolor{Gray}$0.1170^{+0.0029}_{-0.0029}$ &\cellcolor{Gray}$172.9^{+0.5}_{-0.5}$ &\cellcolor{Gray}$ 0.66$ &&&  \\
\hline
$\yt$ & $\ytt$ &  $0.1188^{+0.0014}_{-0.0018} $ &  $177.9^{+1.4}_{-1.5}$ &  $ 1.85$ &  $0.1120^{+0.0018}_{-0.0018}$ &  $169.0^{+2.4}_{-2.8}$ &  $ 0.53$ &&&  \\
\hline

\end{tabular}

\endgroup

\normalsize
\caption{As in \tab{as-extr-1d-abs} but for the simultaneous extractions of $\as$ and $\mt$ from two ATLAS distributions.}
\label{tab:as-mt-extr-2d-1exp2dists-abs}
\end{table}

\begin{table}[h!]
\centering
\footnotesize

\begingroup
\setlength{\tabcolsep}{3pt}
\begin{tabular}{ |c|c|c c c|c c c|c c c| }
\hline
ATLAS & CMS & \multicolumn{3}{|c|}{CT14} & \multicolumn{3}{|c|}{NNPDF30} & \multicolumn{3}{|c|}{NNPDF31} \\
& & $\as$ & $\mt$ & $\chisqmin$   &   $\as$ & $\mt$ & $\chisqmin$   &   $\as$ & $\mt$ & $\chisqmin$ \\
\hline
$p_T^t$ & $p_T^t$ &  $0.1084^{+0.0019}_{-0.0017} $ &  $169.6^{+0.7}_{-0.7}$ &  $ 0.67$ &  $0.1115^{+0.0020}_{-0.0020}$ &  $171.6^{+0.7}_{-0.7}$ &  $ 0.71$ &  $0.1140^{+0.0017}_{-0.0017}$ &  $171.8^{+0.7}_{-0.7}$ &  $ 0.69$  \\
\hline
$p_T^t$ & $\mtt$ &  $0.1124^{+0.0016}_{-0.0016} $ &  $169.7^{+0.8}_{-0.8}$ &  $ 1.99$ &  $0.1098^{+0.0023}_{-0.0023}$ &  $170.2^{+0.8}_{-0.8}$ &  $ 1.55$ &  $0.1138^{+0.0020}_{-0.0021}$ &  $170.7^{+0.8}_{-0.8}$ &  $ 1.52$  \\
\hline
$p_T^t$ & $y_t$ &  $0.1086^{+0.0018}_{-0.0014} $ &  $170.0^{+0.8}_{-0.8}$ &  $ 1.13$ &  $0.1216^{+0.0021}_{-0.0021}$ &  $175.2^{+0.8}_{-0.8}$ &  $ 1.56$ &&&  \\
\hline
$p_T^t$ & $\ytt$ &\cellcolor{Gray}$0.1177^{+0.0012}_{-0.0014} $ &\cellcolor{Gray}$174.3^{+0.9}_{-0.8}$ &\cellcolor{Gray}$ 1.01$ &\cellcolor{Gray}$0.1160^{+0.0018}_{-0.0018}$ &\cellcolor{Gray}$173.4^{+0.9}_{-0.8}$ &\cellcolor{Gray}$ 0.59$ &&&  \\
\hline
$\mtt$ & $p_T^t$ &\cellcolor{Gray}$0.1150^{+0.0015}_{-0.0016} $ &\cellcolor{Gray}$172.5^{+0.5}_{-0.6}$ &\cellcolor{Gray}$ 1.02$ & $0.1148^{+0.0020}_{-0.0020}$ &  $ 172.7^{+0.4}_{-0.5}$ &  $ 0.98$ &\cellcolor{Gray}$0.1171^{+0.0016}_{-0.0017}$ &\cellcolor{Gray}$173.0^{+0.5}_{-0.4}$ & \cellcolor{Gray}$ 0.91$  \\
\hline
$\mtt$ & $\mtt$ &  $0.1144^{+0.0016}_{-0.0017} $ &  $171.3^{+0.8}_{-0.7}$ &  $ 2.74$ &  $0.1123^{+0.0023}_{-0.0023}$ &  $171.6^{+0.8}_{-0.7}$ &  $ 2.28$ &\cellcolor{Gray}$0.1167^{+0.0019}_{-0.0019}$ &\cellcolor{Gray}$172.3^{+0.5}_{-0.7}$ &\cellcolor{Gray}$ 2.05$  \\
\hline
$\mtt$ & $y_t$ &\cellcolor{Gray}$0.1157^{+0.0015}_{-0.0017} $ &\cellcolor{Gray}$172.9^{+0.5}_{-0.5}$ &\cellcolor{Gray}$ 1.30$ &\cellcolor{Gray}$0.1192^{+0.0019}_{-0.0019}$ &\cellcolor{Gray}$173.5^{+0.5}_{-0.5}$ &\cellcolor{Gray}$ 1.84$ &&&  \\
\hline
$\mtt$ & $\ytt$ &\cellcolor{Gray}$0.1167^{+0.0013}_{-0.0014} $ & \cellcolor{Gray}$173.1^{+0.6}_{-0.5}$ & \cellcolor{Gray}$ 1.15$ &\cellcolor{Gray}$0.1163^{+0.0017}_{-0.0017}$ &\cellcolor{Gray}$173.0^{+0.5}_{-0.5}$ &\cellcolor{Gray}$ 0.71$ &&&  \\
\hline
$y_t$ & $p_T^t$ &  $0.1124^{+0.0014}_{-0.0014} $ &  $170.9^{+0.9}_{-0.8}$ &  $ 1.71$ &  $0.1104^{+0.0018}_{-0.0018}$ &  $170.9^{+0.9}_{-0.8}$ &  $ 0.70$ &&&  \\
\hline
$y_t$ & $\mtt$ &  $0.1127^{+0.0014}_{-0.0014} $ &  $167.9^{+1.3}_{-1.4}$ &  $ 3.08$ &  $0.1101^{+0.0019}_{-0.0019}$ &  $168.1^{+1.3}_{-1.4}$ &  $ 1.34$ &&&  \\
\hline
$y_t$ & $y_t$ &\cellcolor{Gray}$0.1167^{+0.0014}_{-0.0016} $ &\cellcolor{Gray}$174.4^{+1.1}_{-1.0}$ &\cellcolor{Gray}$ 2.05$ &\cellcolor{Gray}$0.1199^{+0.0019}_{-0.0019}$ &\cellcolor{Gray}$175.7^{+1.1}_{-1.0}$ &\cellcolor{Gray}$ 2.06$ &&&  \\
\hline
$y_t$ & $\ytt$ &  $0.1193^{+0.0010}_{-0.0012} $ &  $176.7^{+1.1}_{-1.1}$ &  $ 1.68$ &  $0.1142^{+0.0016}_{-0.0016}$ &  $172.6^{+1.1}_{-1.1}$ &  $ 0.74$ &&&  \\
\hline
$\ytt$ & $p_T^t$ &  $0.1127^{+0.0015}_{-0.0015} $ &  $171.0^{+0.9}_{-0.8}$ &  $ 1.75$ &  $0.1097^{+0.0020}_{-0.0020}$ &  $170.8^{+0.8}_{-0.8}$ &  $ 0.67$ &&&  \\
\hline
$\ytt$ & $\mtt$ &  $0.1132^{+0.0016}_{-0.0015} $ &  $168.1^{+1.2}_{-1.3}$ &  $ 3.10$ &  $0.1096^{+0.0022}_{-0.0022}$ &  $168.5^{+1.2}_{-1.3}$ &  $ 1.39$ &&&  \\
\hline
$\ytt$ & $y_t$ &\cellcolor{Gray}$0.1180^{+0.0015}_{-0.0016} $ &\cellcolor{Gray}$175.0^{+1.1}_{-1.0}$ &\cellcolor{Gray}$ 2.05$ &\cellcolor{Gray}$0.1207^{+0.0021}_{-0.0021}$ &\cellcolor{Gray}$175.7^{+1.1}_{-1.0}$ &\cellcolor{Gray}$ 1.99$ &&&  \\
\hline
$\ytt$ & $\ytt$ &  $0.1197^{+0.0011}_{-0.0012} $ &  $176.8^{+1.1}_{-1.1}$ &  $ 1.66$ &  $0.1145^{+0.0017}_{-0.0017}$ &  $172.8^{+1.1}_{-1.1}$ &  $ 0.72$ &&&  \\
\hline
\end{tabular}

\endgroup

\normalsize
\caption{As in \tab{as-extr-1d-abs} but for the simultaneous extractions of $\as$ and $\mt$ from one ATLAS and one CMS distribution.}
\label{tab:as-mt-extr-2d-2exp-abs}
\end{table}

\FloatBarrier

\section{Extractions using $\mtt$ supplemented with underflow events}\label{app:C}

\begin{table}[h!]
\centering

\begin{tabular}{ |c|c c|c c|c c| }
\hline
\multicolumn{1}{|c|}{} & \multicolumn{6}{|c|}{ATLAS}\\
\hline
& \multicolumn{2}{|c|}{CT14} & \multicolumn{2}{|c|}{NNPDF30} & \multicolumn{2}{|c|}{NNPDF31} \\
& $\as$ & $\chisqmin$ & $\as$ & $\chisqmin$ & $\as$ & $\chisqmin$ \\
\hline
$\mtt$ &\cellcolor{Gray}$0.1158^{+0.0020}_{-0.0021}$ &\cellcolor{Gray}0.17 & 
        \cellcolor{Gray}$0.1157^{+0.0027}_{-0.0027}$ &\cellcolor{Gray}0.19 & 
        \cellcolor{Gray}$0.1173^{+0.0022}_{-0.0022}$ &\cellcolor{Gray}0.22  \\
\hline
$\mtt^u$ &\cellcolor{Gray}$0.1158^{+0.0020}_{-0.0021}$ &\cellcolor{Gray}0.15 & 
          \cellcolor{Gray}$0.1157^{+0.0027}_{-0.0027}$ &\cellcolor{Gray}0.17 & 
          \cellcolor{Gray}$0.1173^{+0.0022}_{-0.0022}$ &\cellcolor{Gray}0.21 \\
\hline
\end{tabular} \\[10pt]

\begin{tabular}{ |c|c c|c c|c c| }
\hline
\multicolumn{1}{|c|}{} & \multicolumn{6}{|c|}{CMS}\\
\hline
& \multicolumn{2}{|c|}{CT14} & \multicolumn{2}{|c|}{NNPDF30} & \multicolumn{2}{|c|}{NNPDF31} \\
& $\as$ & $\chisqmin$ & $\as$ & $\chisqmin$ & $\as$ & $\chisqmin$ \\
\hline
$\mtt$ & $0.1147^{+0.0014}_{-0.0014}$ & 9.68 & 
         $0.1100^{+0.0014}_{-0.0043}$ & 6.97 & 
         \cellcolor{Gray}$0.1161^{+0.0018}_{-0.0018}$ &\cellcolor{Gray}6.01  \\
\hline
$\mttu$ & $0.1147^{+0.0014}_{-0.0014}$ & 9.66 & 
           $0.1100^{+0.0014}_{-0.0043}$ & 6.94 & 
           \cellcolor{Gray}$0.1161^{+0.0018}_{-0.0018}$ &\cellcolor{Gray}5.99  \\
\hline
\end{tabular}

\caption{
As in \tab{as-extr-1d} but the case with and without underflow events is considered.
}
\label{tab:as-extr-mttunderflow}
\end{table}

\begin{table}[h!]
\centering

\begin{tabular}{ |c|c c|c c|c c| }
\hline
\multicolumn{1}{|c|}{} & \multicolumn{6}{|c|}{ATLAS}\\
\hline
& \multicolumn{2}{|c|}{CT14} & \multicolumn{2}{|c|}{NNPDF30} & \multicolumn{2}{|c|}{NNPDF31} \\
& $\mt$ & $\chisqmin$ & $\mt$ & $\chisqmin$ & $\mt$ & $\chisqmin$ \\
\hline
$\mtt$ &\cellcolor{Gray}$173.4^{+0.6}_{-0.5}$ &\cellcolor{Gray}0.35 & 
        \cellcolor{Gray}$173.4^{+0.6}_{-0.5}$ &\cellcolor{Gray}0.28 & 
        \cellcolor{Gray}$173.5^{+0.5}_{-0.5}$ &\cellcolor{Gray}0.21  \\
\hline
$\mttu$ &\cellcolor{Gray}$173.3^{+0.5}_{-0.5}$ &\cellcolor{Gray}0.33 & 
          \cellcolor{Gray}$173.4^{+0.5}_{-0.5}$ &\cellcolor{Gray}0.27 & 
          \cellcolor{Gray}$173.5^{+0.5}_{-0.5}$ &\cellcolor{Gray}0.20  \\
\hline
\end{tabular} \\[10pt]

\begin{tabular}{ |c|c c|c c|c c| }
\hline
\multicolumn{1}{|c|}{} & \multicolumn{6}{|c|}{CMS}\\
\hline
& \multicolumn{2}{|c|}{CT14} & \multicolumn{2}{|c|}{NNPDF30} & \multicolumn{2}{|c|}{NNPDF31} \\
& $\mt$ & $\chisqmin$ & $\mt$ & $\chisqmin$ & $\mt$ & $\chisqmin$ \\
\hline
$\mtt$ &\cellcolor{Gray}$170.4^{+0.6}_{-0.7}$ &\cellcolor{Gray}7.31 & 
         \cellcolor{Gray}$170.5^{+0.7}_{-0.7}$ &\cellcolor{Gray}5.98 & 
         \cellcolor{Gray}$170.8^{+0.7}_{-0.7}$ &\cellcolor{Gray}3.87  \\
\hline
$\mttu$ &\cellcolor{Gray}$170.8^{+0.6}_{-0.5}$ &\cellcolor{Gray}7.31 & 
           \cellcolor{Gray}$170.9^{+0.6}_{-0.5}$ &\cellcolor{Gray}5.98 & 
           \cellcolor{Gray}$171.2^{+0.6}_{-0.6}$ &\cellcolor{Gray}3.91  \\
\hline
\end{tabular}

\caption{
As in \tab{as-extr-mttunderflow} but for $\mt$ rather than $\as$, which has been set to the world average value of 0.118.}
\label{tab:mt-extr-mttunderflow}
\end{table}

In \tab{as-extr-mttunderflow} we show the extraction of $\as$ for a fixed value of the top mass, $\mt=173.3$~GeV, using
either the invariant mass distribution, $\mtt$, or the invariant mass distribution supplemented with underflow events, $\mtt^u$.
In \tab{mt-extr-mttunderflow} we show the corresponding table for the extraction of $\mt$ for $\as=0.118$.
In the extraction of $\as$, for all PDF sets and for both experiments it is clear that the addition of the underflow events
only slightly increases the value of $\chisqmin$ and barely affects the best-fit value and the associated uncertainties.
In the extraction of $\mt$, when using data from either experiment, the addition of underflow events again has a very
small effect on  $\chisqmin$. Interestingly however, in the case of CMS, the best-fit value of $\mt$
is shifted upwards by 0.3~GeV in all cases and the associated uncertainties decrease.

\begin{table}[h!]
\centering

\footnotesize
\begingroup
\setlength{\tabcolsep}{3pt}
\begin{tabular}{ |c|c c c|c c c|c c c| }
\hline
\multicolumn{1}{|c|}{} & \multicolumn{9}{|c|}{ATLAS}\\
\hline
& \multicolumn{3}{|c|}{CT14} & \multicolumn{3}{|c|}{NNPDF30} & \multicolumn{3}{|c|}{NNPDF31} \\
&   $\as$ & $\mt$ & $\chisqmin$   &   $\as$ & $\mt$ & $\chisqmin$   &   $\as$ & $\mt$ & $\chisqmin$ \\
\hline
$\mtt$ &\cellcolor{Gray}$0.1155^{+0.0020}_{-0.0022}$ &\cellcolor{Gray}$173.1^{+0.6}_{-0.5}$ &\cellcolor{Gray}0.16 & 
        \cellcolor{Gray}$0.1157^{+0.0027}_{-0.0027}$ &\cellcolor{Gray}$173.2^{+0.6}_{-0.5}$ &\cellcolor{Gray}0.19 & 
        \cellcolor{Gray}$0.1176^{+0.0022}_{-0.0022}$ &\cellcolor{Gray}$173.5^{+0.5}_{-0.5}$ &\cellcolor{Gray}0.21  \\
\hline
$\mttu$ &\cellcolor{Gray}$0.1156^{+0.0020}_{-0.0022}$ &\cellcolor{Gray}$173.1^{+0.5}_{-0.5}$ &\cellcolor{Gray}0.13 & 
          \cellcolor{Gray}$0.1157^{+0.0027}_{-0.0027}$ &\cellcolor{Gray}$173.2^{+0.5}_{-0.5}$ &\cellcolor{Gray}0.17 & 
          \cellcolor{Gray}$0.1176^{+0.0022}_{-0.0022}$ &\cellcolor{Gray}$173.4^{+0.5}_{-0.5}$ &\cellcolor{Gray}0.20  \\
\hline
\end{tabular} \\[10pt]

\begin{tabular}{ |c|c c c|c c c|c c c| }
\hline
\multicolumn{1}{|c|}{} & \multicolumn{9}{|c|}{CMS}\\
\hline
& \multicolumn{3}{|c|}{CT14} & \multicolumn{3}{|c|}{NNPDF30} & \multicolumn{3}{|c|}{NNPDF31} \\
&   $\as$ & $\mt$ & $\chisqmin$   &   $\as$ & $\mt$ & $\chisqmin$   &   $\as$ & $\mt$ & $\chisqmin$ \\
\hline
$\mtt$ & $0.1108^{+0.0013}_{-0.0012}$ & $168.5^{+0.8}_{-0.8}$ & 4.43 & 
         $0.1055^{+0.0021}_{-0.0020}$ & $168.8^{+0.9}_{-0.9}$ & 2.02 & 
         $0.1100^{+0.0023}_{-0.0030}$ & $169.1^{+0.8}_{-0.9}$ & 2.41  \\
\hline
$\mtt^u$ & $0.1119^{+0.0014}_{-0.0013}$ & $169.9^{+0.6}_{-0.6}$ & 5.03 & 
           $0.1069^{+0.0022}_{-0.0022}$ & $170.2^{+0.6}_{-0.6}$ & 2.71 & 
           $0.1125^{+0.0020}_{-0.0023}$ & $170.5^{+0.6}_{-0.6}$ & 2.91  \\
\hline
\end{tabular}

\endgroup
\normalsize

\caption{
As in \tab{as-extr-1d} but comparing the case with and without underflow events for the simultaneous extractions of $\mt$ and $\as$ using ATLAS (upper table) and CMS (lower table).
}
\label{tab:as-mt-extr-2d-underflow}
\end{table}

We have also investigated the effects of the addition of underflow events in simultaneous extractions
of $\as$ and $\mt$, the results of which are tabulated for each experiment in \tab{as-mt-extr-2d-underflow}.
Reflecting the results of the individual extractions, we see that adding the underflow events has
only a very small effect on the extractions when using ATLAS data.
This is not the case when using the CMS data, where we now observe significant increases in both
the best-fit values of $\as$ and $\mt$, with $\mt$ in particular being shifted by almost 1$\sigma$ for all PDFs.
These shifts are consistent within uncertainties, however it is interesting that they push the extracted
values upwards towards the respective world averages.

Perhaps the most surprising feature of the numbers presented in this section is that the change in the
extractions is not the same for both experiments.
This may suggest a different treatment by the two experiments, of underflow events either in the
direct measurements themselves or in the process of the extrapolation of direct measurements to stable tops.
Given that we have found that different treatments of underflow events can have a significant
effect on extractions of $\mt$ and $\as$, this aspect of the measurement of $\mtt$ deserves careful
consideration in future measurements. This issue has been considered with respect to resummed
calculations in ref.~\cite{Ju:2020otc}, where similar conclusions were drawn.

\section{$\as$ and $\mt$ averaging procedure}\label{app:averaging}

The averaging procedure introduced in \sec{averaging} works as follows. For the parameter $\beta=\{\as,\mt\}$ we obtain an average value as
\begin{equation}
  \bar{\beta}=\frac{\sum_i u_i \beta_i}{\sum_i u_i}\,,
\end{equation}
where $\beta_i$ is the $i$\textsuperscript{th} measurement of $\beta$. The weights $u_i$ are given by the $p$-values associated with the $\chi^2_{\mathrm{min},i}$,
\begin{equation}
  u_i=\int^\infty_{\chi^2_{\mathrm{min},i}}f(z;n_\mathrm{d})\mathrm{d}z\,,
\end{equation}
where $f(z;n_\mathrm{d})$ is the $\chi^2$ probability distribution function and $n_\mathrm{d}$ is the appropriate number of degrees of freedom. The estimated uncertainty of the average $\bar\beta$ reads
\begin{equation}
  \delta\bar\beta^{\mathrm{up/down}}=\sqrt{(\delta\beta_{\mathrm{stat}}^{\mathrm{up/down}})^2 + (\delta\beta_{\mathrm{sys}})^2}\,.
\end{equation}
It includes both statistical and systematic contributions:
\begin{eqnarray}
  \delta\beta_{\mathrm{stat}}^{\mathrm{up/down}} &=& \frac{\sqrt{\sum_i u_i^2 (\delta\beta_i^{\mathrm{up/down}})^2}}{\sum_i u_i}\,,\\
  \delta\beta_{\mathrm{sys}} &=& \frac{\sum_i u_i|\beta_i-\bar{\beta}|}{\sum_i u_i}\,,
\end{eqnarray}
where $\delta\beta_i^{\mathrm{up/down}}$ is the uncertainty on $\beta_i$ derived from the condition eq.~(\ref{eq:deltachi2}). The pairs $\left(\bar\beta_i, \delta\beta_i^{\mathrm{up/down}}\right)$ are given in the tables in \sec{results} and appendices \ref{app:A},\ref{app:B},\ref{app:C}.

\end{document}